\documentclass[twocolumn]{aastex701}

\usepackage{placeins}
\usepackage{amsmath}
\usepackage{comment}
\usepackage{chngcntr}

\definecolor{blue-violet}{rgb}{0.54, 0.17, 0.89}

\begin{document}

\title{A Highly Reflective Atmosphere on the Lava World TOI-561b Revealed by JWST/NIRSpec Phase Curve}


\author[0009-0003-5977-9581]{Samuel Boucher}
\affiliation{Département de physique and Institut Trottier de recherche sur les exoplanètes, Université de Montréal, C.P. 6128, Succ. Centre-ville,
Montréal, H3C 3J7, Québec, Canada}
\affiliation{Waterloo Centre for Astrophysics and Department of Physics and Astronomy, University of Waterloo; Waterloo, Ontario, Canada N2L 3G1}
\email{samuel.boucher.3@umontreal.ca}

\author[0000-0003-4987-6591]{Lisa Dang}
\affiliation{Waterloo Centre for Astrophysics and Department of Physics and Astronomy, University of Waterloo; Waterloo, Ontario, Canada N2L 3G1}
\affiliation{Département de physique and Institut Trottier de recherche sur les exoplanètes, Université de Montréal, C.P. 6128, Succ. Centre-ville,
Montréal, H3C 3J7, Québec, Canada}
\email{lisa.dang@uwaterloo.ca}

\author[0009-0000-8327-2631]{Alex McGinty}
\affiliation{Atmospheric, Oceanic, and Planetary Physics, Department of Physics; University of Oxford, Oxford OX1 3PU, UK}
\email{alex.mcginty@physics.ox.ac.uk}

\author[0009-0008-2801-5040]{Johanna K. Teske}
\affiliation{Earth and Planets Laboratory, Carnegie Institution for Science, 5241 Broad Branch Road, NW, Washington, DC 20015, USA}
\affiliation{The Observatories of the Carnegie Institution for Science, 813 Santa Barbara St., Pasadena, CA 91101, USA}
\email{jteske@carnegiescience.edu}

\author[0000-0002-5887-1197]{Raymond Pierrehumbert}
\affiliation{Atmospheric, Oceanic, and Planetary Physics, Department of Physics; University of Oxford, Oxford OX1 3PU, UK}
\affiliation{Department of Earth, Atmospheric and Planetary Sciences, Massachusetts Institute of Technology, 77 Massachusetts Avenue, 55-101, Cambridge, MA 02139}
\email{raymond.pierrehumbert@physics.ox.ac.uk}

\author[0000-0002-4487-5533]{Anjali A. A. Piette}
\affiliation{School of Physics \& Astronomy, University of Birmingham, Edgbaston, Birmingham, B15 2TT, UK}
\email{a.a.a.piette@bham.ac.uk}

\author[0000-0003-0354-0187]{Nicole L. Wallack}
\affiliation{Earth and Planets Laboratory, Carnegie Institution for Science, 5241 Broad Branch Road, NW, Washington, DC 20015, USA}
\email{nwallack@carnegiescience.edu}

\author[0000-0001-5341-8978]{Angharad Weeks}
\affiliation{Sydney Institute for Astronomy, School of Physics, University of Sydney, Sydney, NSW 2006, Australia}
\email{angharad.weeks@sydney.edu.au}

\author[0000-0002-3724-5728]{Neil T. Lewis}
\affiliation{Department of Physics and Astronomy, University of Exeter; Exeter, EX4 4QL, UK}
\email{n.t.lewis@exeter.ac.uk}

\author[0000-0002-3286-7683]{Tim Lichtenberg}
\affiliation{Kapteyn Astronomical Institute, University of Groningen; Groningen, The Netherlands}
\email{tim.lichtenberg@rug.nl}

\author[0000-0002-9479-2744]{Mykhaylo Plotnykov}
\affiliation{Department of Astronomy \& Astrophysics, University of Toronto; Toronto, Ontario, Canada M5S 3H4}
\email{mykhaylo.plotnykov@mail.utoronto.ca}

\author[0009-0009-5036-3049]{Emma Postolec}
\affiliation{Kapteyn Astronomical Institute, University of Groningen; Groningen, The Netherlands}
\email{e.n.postolec@rug.nl}

\author[0000-0001-8832-4488]{Daniel Huber}
\affiliation{Institute for Astronomy, University of Hawai`i, 2680 Woodlawn Drive, Honolulu, HI 96822, USA}
\email{huberd@hawaii.edu}

\author[0000-0001-8832-4488]{Timothy R. Bedding}
\affiliation{Sydney Institute for Astronomy, School of Physics, University of Sydney, Sydney, NSW 2006, Australia}
\email{tim.bedding@sydney.edu.au}

\author[0000-0002-8368-4641]{Harrison Nicholls}
\affiliation{Institute of Astronomy, University of Cambridge, Madingley Road, Cambridge, CB3 0HA, UK}
\email{harrison.nicholls@ast.cam.ac.uk}

\author[0009-0009-6098-296X]{Bo Peng}
\affiliation{Department of Earth and Planetary Sciences, Stanford University; Stanford, CA, 94305, USA}
\affiliation{Department of Astronomy \& Astrophysics, University of Toronto; Toronto, Ontario, Canada M5S 3H4}
\email{bpengeps@stanford.edu}

\author[0000-0002-6893-522X]{Mark Hammond}
\affiliation{Atmospheric, Oceanic, and Planetary Physics, Department of Physics; University of Oxford, Oxford OX1 3PU, UK}
\email{mark.hammond@physics.ox.ac.uk}

\author[0000-0003-3993-4030]{Diana Valencia}
\affiliation{Department of Astronomy \& Astrophysics, University of Toronto; Toronto, Ontario, Canada M5S 3H4}
\email{diana.valencia@utoronto.ca}

\correspondingauthor{lisa.dang@uwaterloo.ca}

\begin{abstract}
Ultra-short period (USP) rocky planets are expected to be depleted of any substantial gaseous envelope due to the intense irradiation they receive from their host star, making the recent detection of an atmosphere around TOI-561~b particularly surprising. That finding was based on the planet's bulk density and dayside emission spectrum, but full-orbit phase curve observations offer a more comprehensive way to constrain the presence and characteristics of a planet's atmosphere. In this paper, we map TOI-561~b's 3-5~$\mu$m emission using JWST/NIRSpec multi-orbit spectroscopic phase curves, explicitly accounting for the curvature of the out-of-eclipse baseline and possible hotspot offsets. From a simple energy balance argument as well as comparing to general circulation models, we find the phase curve observations are best explained with high Bond albedo and moderate global heat transfer. Our 37-hour continuous observation provides a long baseline that allows us to model stellar granulation, the planetary phase curve, and dayside emission simultaneously, effectively disentangling these signals and yielding a dayside emission spectrum that is more robust against stellar variability and consistent with previous eclipse-only fits. Using general circulation model outputs, we constrain where clouds can plausibly form and test candidate compositions, finding that silicate clouds, such as SiO$_2$ and MgSiO$_3$, can form on the dayside near the terminator and explain the observed albedo. Our findings confirm that TOI-561~b appears to have a global reflective atmosphere, suggesting exchange of volatiles with the interior to maintain it. 
\end{abstract}

\section{Introduction} 
One of the most pressing questions in exoplanet science today is whether small rocky exoplanets can host atmospheres. These atmospheres are generally expected to be ``secondary'' in the sense that they have some, perhaps total, contribution from exchange with the planet's interior and thus may leave a fingerprint of processes that are otherwise hidden from remote observations \citep{Schaefer2016ApJ,Bower2019A&A,Katyal2020,Baumeister2023,Gillmann2024,KrissansenTotton2024NatCo,Lichtenberg2025Sci}. The details of how these atmospheres form and evolve is an active area of theoretical research \citep{Kite2021ApJL,Dorn2021ApJL,Lichtenberg2021JGRE,Krissansen-Totton2022ApJ,Piette2023ApJ,Maurice2024,Seo2024ApJ,Arora2025arXiv,Ito2025ApJ,Boer2025ApJ,Gupta2025ApJL,Nicholls2025MNRAS,Nicholls2026NatAstron,Steinmeyer2026ApJ}, and the launch of JWST opened up the observational capacity to investigate them in more detail given the enhanced precision and wider wavelength coverage versus \textit{Hubble} and \textit{Spitzer} (see \citealt{Wordsworth2022,Espinoza2025,Kreidberg2025} for recent reviews). However, strong evidence for rocky planet atmospheres remains elusive, both from transmission observations \citep{Moran2023,May2023,Lim2023,Scarsdale2024,Alam2025,AdamsRedai2025} and emission observations \citep{Kreidberg2019,Crossfield2022,Greene2023,Zieba2023,Xue2024,Zhang2024,Xue2025}. 

Somewhat counter-intuitively, the most promising place to look for secondary atmospheres may be around hot rocky exoplanets, those in short, tidally-locked orbits around their host stars. These planets likely have an enduring magma ocean at least on the highly irradiated dayside, which can be in continuous exchange with and fuel an atmosphere \citep{Kite2016,Kite2020,Meier2023A&A,Meier2026MNRAS,Boukare2025NatAs}. A subset of ultra-short period planets have bulk densities that are inconsistent with a ``bare rock'' scenario, suggesting atmospheres that are composed of more than just rock vaporized from the surface, perhaps hosting volatile species like H$_2$O, CO, CO$_2$, N$_2$, or SO$_2$. Indeed, the poster child of hot rocky planets, 55 Cnc e, is believed to have a substantial atmosphere redistributing heat around the planet that likely contains abundant CO$_2$ or CO \citep{Demory2016,Hu2024}.   

\cite{EmissionPaper} added another example of a hot rocky exoplanet with evidence for a substantial atmosphere -- TOI-561~b. This ultra-short period (USP, $\sim 0.44$ day) planet has a mass of 2.24$\pm$0.20~M$_{\oplus}$ \citep{Brinkman2023} and a radius of 1.42$\pm$0.02~R$_{\oplus}$ \citep{patel2023}, amounting to a low bulk density of 4.30$\pm$0.4~g~cm$^{-3}$. This density distinguishes it from many other ultra-short period planets that cluster around an Earth-like composition \citep{Dai2019,Dai2021,Brinkman2025}. The environment in which TOI-561~b formed may also be distinct from other USPs -- the host star is a thick disk star with low metallicity ([Fe/H]=-0.41$\pm$0.05 dex), high alpha abundances ([$\alpha$/H]=0.23$\pm$0.05), and an old age ($\geq$10 Gyr) \citep{lacedelli21,weiss2021}. \cite{EmissionPaper} presented the JWST/NIRSpec 3-5 micron emission (dayside) spectrum based on four secondary eclipses, finding that it was too cool to be consistent with a bare rock or even a pure rock vapor atmosphere. They suggested instead that the most likely explanation was a thick volatile envelope on TOI-561~b.

Unlike an emission spectrum, a full phase curve -- monitoring the planet around its entire orbit -- can be used to directly constrain how heat is redistributed around the planet via the day-to-nightside temperature contrast, as well as the offset in position of the hottest part of the planet from the sub-stellar point (see \citealt{Parmentier2018,Hammond2025} and references therein). These measurements can help break degeneracies between albedo and composition from dayside emission observations alone, and thus provide strong additional evidence for the presence and nature of an atmosphere \citep[e.g.,][]{Crossfield2020, Coulombe2025}. The emission observations in \cite{EmissionPaper} were part of a full continuous ($\sim$37 hours) phase curve observation (GO program 3860), and we present the results and interpretation of those observations in this paper. Section \ref{sec:data} reviews the observations and data reduction, which were also presented in \cite{EmissionPaper} but are summarized here. Section \ref{sec:phase_curve} describes the de-trending and fitting of the white light and spectroscopic phase curves, including accounting for a possible stellar granulation signal, which is further investigated in Section \ref{sec:stellar_signals}. In Section \ref{sec:physical_parameters} we present a simple planet heat redistribution analysis, while Section \ref{sec:gcm} goes into more depth with the interpretation, presenting a comparison of the results with a general circulation model (GCM). In Section \ref{sec:discussion} we discuss the implications of our findings, and finally we summarize our conclusions in Section \ref{sec:conclusions}.
    

\section{Observations} \label{sec:data}

The data and reductions used in our analysis are the same as in \cite{EmissionPaper}, which we briefly summarize here for completeness. The full phase curve observations of TOI-561~b, covering four secondary eclipses and three transits consecutively, were taken with JWST/NIRSpec between 2024 May 1 at 11:16 UT and 2024 May 3 at 00:37 UT. The observations were conducted using the high-resolution (with a resolving power of $\sim$2700) G395H grating, providing wavelength coverage from 2.67 to 5.14 $\mu$m across the two NIRSpec detectors, including a gap between them resulting in a break in wavelength coverage between 3.72 and 3.82 $\mu$m. The full data set can be found in MAST: \dataset[10.17909/3g6t-he86]{http://dx.doi.org/10.17909/3g6t-he86}. 

We reduced the NIRSpec data with both \texttt{ExoTiC JEDI} \citep{Alderson2022JEDI,Alderson2022} and \texttt{Eureka!} \citep{Bell2022}, which both have been used extensively in the literature for JWST small planet data reduction \citep[][e.g.,]{Greene2023,Zieba2023,Moran2023,Alderson2024,Wallack2024,Scarsdale2024,Luque2025,Alderson2025}. The pipelines take as input the raw uncal files and start with the standard Stage 1 and 2 steps of the STScI \texttt{jwst} pipeline \citep{Bushouse2022}. The main differences between \texttt{ExoTiC JEDI} and \texttt{Eureka!} are in the 1/$f$ noise and background subtraction -- which \texttt{ExoTiC JEDI} performs once in Stage 1 at the group level and again in Stage 3 at the integration level, and \texttt{Eureka!} does once in Stage 1 at the group level -- and the spectral trace extraction -- \texttt{ExoTiC JEDI} fits a Gaussian to each column and then median smooths the fourth-order polynomial fit to both the trace center and width and sets the aperture to be 5$\times$ the FWHM of the trace, whereas our implementation of \texttt{Eureka!} tests different combinations of extraction parameters (aperture width, background aperture width, additional background subtractions, and the sigma threshold for the outlier rejection during optimal extraction) to produce the minimum of the median absolute deviation in the NRS1 and NRS2 white light curves. In addition to the one-dimensional spectra, \texttt{ExoTiC JEDI} extracts and reports $X$- and $Y$-position shifts as a function of time, calculated by cross-correlating the spectra, and \texttt{Eureka!} extracts and reports the same values in addition to the spread (standard deviation) of the centroid position, in both directions. As discussed in section \ref{sec:phase_curve}, our final systematic model does not include any centroid regressors. The full details of our reductions can be found in \cite{EmissionPaper}.

\section{Fitting Phase Curves} \label{sec:phase_curve}

    In this section, we present the detrending and fitting procedures used to isolate the planet's contribution to the observed flux in our observations. Different from \cite{EmissionPaper}, here we use one model for the entire time-series observation, and also include a component in the model to account for stellar granulation. A complete list of the parameters that are fitted with their respective priors and fixed orbital parameters is listed in Table \ref{tab:bestfit_eureka_vs_jedi} in Appendix \ref{sec:A_reduction_consistency}.

\subsection{Fitted Model Details}

    We fit both the \texttt{Eureka!} and \texttt{ExoTiC JEDI} reductions, but only present the \texttt{Eureka!} results in our figures and tables unless otherwise noted. We discuss the overall agreement of the two reductions in Appendix \ref{sec:A_reduction_consistency}. The datasets contain a white light curve for each detector, as well as seven binned spectroscopic light curves (three in NRS1 and four in NRS2, see Table \ref{tab:phasecurveresults}). 
    We run two separate fits, one using the white light curves (WLCs) and one using the spectroscopic light curves (SLCs). We assume a circular, tidally locked orbit and fix the period, inclination and eccentricity to the values from \cite{patel2023}.

    \subsubsection{General framework}
    
    The measured star-plus-planet signal is modeled by $F_{model} = I \times S + G$, where $I$ is the ideal light-curve, $S$ is the detector systematic model, and $G$ is a Gaussian process (GP) prediction of the correlated noise. We describe $I$ and $G$ in more detail below, and $S$ is simply the sum of a temporal polynomial and optional centroid linear regressors (see more details in \S\ref{sec:white_light_curve_fits}). We consider quadratic stellar limb-darkening in this model and compute the coefficients using \texttt{ExoTiC-LD} \citep{Grant2024}, with stellar parameters from \cite{lacedelli}. The light-curves are fitted using the affine invariant MCMC ensemble sampler \texttt{emcee} \citep{emcee}, and we select the posterior distribution median parameters as best-fit values (although the Bayesian Information Criterion [BIC] values are calculated from the maximum likelihood parameters). 
    

    \subsubsection{Planetary Signal Model \label{sec:planetary_signal_model}}

    The ideal light-curve is $I = L_E + L_T$, where $L_T$ is the transit light-curve computed by \texttt{batman} (\cite{batman}, with quadratic limb-darkening) and $L_E$ is the eclipse model, which also incorporates a planetary flux model. More precisely, we have $L_E = \Phi_P\times(E-1)$ where $E$ is the secondary eclipse light-curve computed by \texttt{batman} and $\Phi_P$ is the first-order phase variation of the planet's apparent brightness (see Equation 6 of \cite{dang} or Equation \ref{equation:phasevariation} below).

    \subsubsection{Stellar Granulation Model \label{sec:stellar_gran_model}}
    Several JWST studies have identified correlated noise in the residuals of time-series observations of exoplanet host stars \citep[e.g][]{Coulombe2025,splinter}. One possible source of this noise for Sun-like stars is stellar granulation, the turbulent motion of convective cells in subsurface stellar layers. The process manifests as a temperature fluctuation on the surfaces of stars, which can be measured as stochastic variability in brightness. Granulation occurs over a wide range of frequencies, with preferred timescales associated with granule sizes resulting in a spectrum that is well described through a combination of Lorentzian profiles \citep{Harvey1985}. The smallest scales (several minutes) are commonly referred to as granulation, intermediate scales (minutes to hours) as mesogranulation, and the largest scales (hours to days) as supergranulation \citep{Rasr2003APJ}. Granulation timescales and amplitudes closely scale with the pressure scale height and thus the surface gravity of the star \citep{Mathur2011,Kallinger2014AAP}. Granulation amplitudes furthermore vary with wavelength and can thus imprint a significant stellar spectral signature on exoplanet phase curves. Indeed, Weeks at al. (in prep) show that JWST time-series data are suitable for the detection of stellar oscillations and granulation. 
    
    To model the stellar variability, we implement a Gaussian process (GP) to model stellar granulation (see more motivation for this in Section \ref{sec:stellar_signals} below). We use the likelihoods computed by \texttt{celerite} \citep{celerite} with a `SHO' term, which models a stochastically-driven damped harmonic oscillator, as appropriate for stellar granulation. Following the approach of previous studies (e.g.,  \citealt{Coulombe2025} and \citealt{  splinter}), we fix the dampening quality factor $Q$ to $\frac{1}{\sqrt{2}}$ and use the amplitude and timescale parametrization of \cite{pereira2019}. Moreover, since our integration times ($\sim5.5$ s) are much smaller than the expected granulation timescale, we fit a jitter constant (added in quadrature to the measurement errors). Incorporating this error inflation parameter aims to ensure that the GP component of our model focuses on removing correlated noise. Without it, \texttt{celerite} might remove white noise by overfitting if the observational errors are underestimated and cannot explain the fast variability of our fluxes.
    
    This GP term is dynamically fitted during the MCMC sampling, effectively asking at each step: Given the current white noise ($\sqrt(\sigma_{obs}^2+\sigma_{jitter}^2)$), what is the probability of obtaining such residuals considering the expected effects of stellar granulation modeled by the current granulation parameters? It is therefore crucial to model this correlated noise with as much physical accuracy as possible. 

    Since we do not know \textit{a priori} the type of granulation TOI-561 exhibits in our observations, we follow the approach used in \cite{Coulombe2025} by using equations 5 and 6 of \cite{granguess} to set our initial granulation amplitude and timescale to rough estimates of 69.2 ppm and 3.1 minutes and imposing large uniform priors of $1-500$ ppm and $0.5-100$ minutes on these parameters. 
    
    \subsection{White Light Curve Fitting Scheme \label{sec:white_light_curve_fits}}
    In the WLC runs, we sample a single joint parameter space in order to impose a shared time of eclipse at every wavelength. To achieve this, we subtract the earliest timestamp across both detectors from all time values to ensure we obtain a uni-modal time-of-eclipse posterior distribution. We also normalize the fluxes and flux errors using the median flux of the corresponding detector, and we add the log-likelihoods computed for each detector during the sampling process.
 
    We run a preliminary comparison step, where we calculate the Bayesian Information Criterion

 \begin{equation}
        BIC = -2\log L+N_{param}\log N_{data}
    \label{equation:bic}
    \end{equation}
   
 \noindent   between eight different systematics models with the same priors we subsequently use in our main sampling (see Table \ref{tab:bestfit_eureka_vs_jedi} for the priors), non-binned data, $N_{burn} = 5000$, $N_{prod} = 2000$, and $N_{walkers} = 4\times N_{dim}$. Again, note that these BIC values are calculated with the posterior distribution parameters associated with the maximum likelihood, rather than the medians. These models combine a cubic or fourth-order temporal polynomial (possibly different between NRS1 and NRS2) and either all the available linear centroid regressors in each detector or none. The available regressors are $X$, $S_X$, $Y$, $S_Y$ for \texttt{Eureka!} and $X$, $Y$ for \texttt{ExoTiC JEDI}. We choose the model with the lowest BIC value for each reduction. It turns out it is the same for both \texttt{Eureka!} and \texttt{ExoTiC JEDI} --  a cubic polynomial in NRS1 and a fourth-order one in NRS2, without any centroid regressors.

\begin{figure*}[ht]
    \centering
    \includegraphics[width=0.8\linewidth]{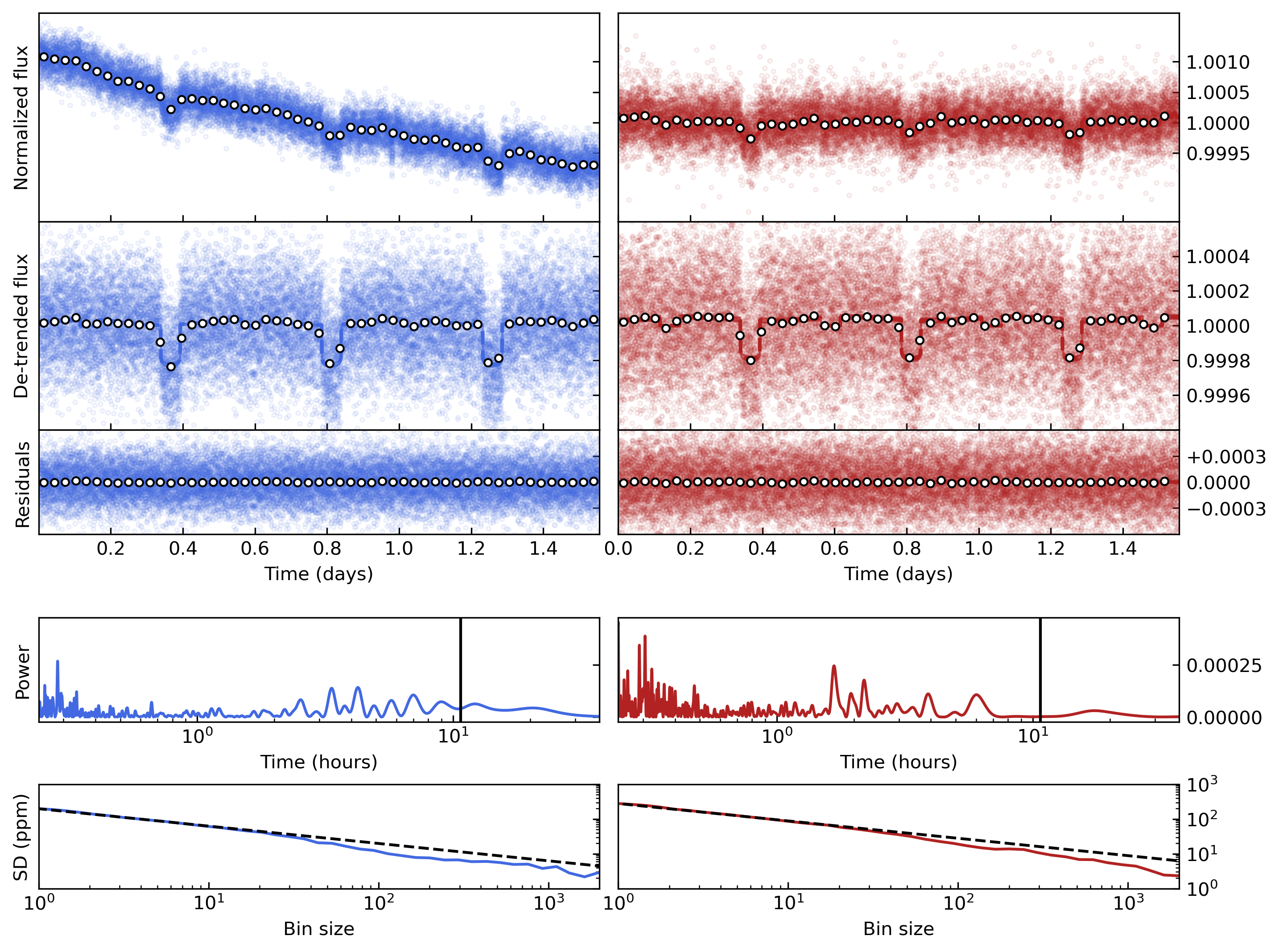}
    \caption{Data and fits corresponding to the NRS1 (left, blue) and NRS2 (right, red) white light curves. The first row shows the raw fluxes and the full ($F_{model}$, three-component) models. The second row shows the corrected fluxes and the astrophysical ($I$) models after removing the systematic ($S$) and stellar ($G$) components. The third row shows the residuals. The second to last row shows the residual periodograms, with vertical lines indicating the planetary orbital period in hours. The last row plots the standard deviation of the residuals versus bin size 
    , with a dashed line showing the white noise expectation.}
    \label{fig:wlcs}
\end{figure*}

    Once the model is selected, we run the main WLC sampling process with the chosen systematics models, non-binned data, $N_{burn} = 40000$, $N_{prod} = 10000$, and $N_{walkers} = 4\times N_{dim}$. Figure \ref{fig:wlcs} illustrates our results, showing the NRS1 and NRS2 light curves, best-fit models, residuals, and two different characterizations of the residuals (periodograms and standard deviation versus bin size). We note three things: First, we obtain $\gtrsim3\sigma$ detections of non-zero eclipse depths in both detectors. Second, the residual periodograms show no statistically significant peaks that could indicate unmodeled correlated noise. Third, after GP detrending, the binned residuals decrease somewhat faster than the canonical white-noise expectation (bottom panel of Figure \ref{fig:wlcs}). We do not interpret this as evidence for physically ``sub-white'' noise, but rather as a consequence of a correlated-noise model with a GP that can suppress the variance. In our case, the characteristic GP timescale ($\sim$25 min, see next paragraph) is shorter than the eclipse duration ($\sim$1.3 hr) and the phase curve duration ($\sim$10.7 hours), consistent with the GP modeling time-correlated systematics rather than the eclipse signal itself. This interpretation is supported by the fact that the inferred eclipse depth remains consistent within 1$\sigma$ with the results presented in \citet{EmissionPaper} (see Appendix \ref{sec:A_comparison_with_eclipse_only}).

    The GP fit to the WLCs yields granulation amplitudes of $61.9^{+3.7}_{-3.4}$~ppm and $54.6^{+4.9}_{-5.0}$~ppm, and characteristic timescales of $21.8^{+3.4}_{-2.7}$~minutes and $25.4^{+13.9}_{-6.7}$~minutes, for NRS1 and NRS2 respectively. The derived granulation amplitudes are consistent with the expected order of magnitude for this type of star \citep{KjeldsenBedding1995, Mathur2011,Kallinger2014AAP}, and the derived granulation timescales are consistent with mesogranulation. We analyze the stellar signature more in depth in the Appendix \S \ref{sec:stellar_signals} to validate our approach.

    \subsection{Spectroscopic Light Curve Fitting Scheme}
   
    \begin{deluxetable*}{cccccccccc}[hbpt!] 
    \tablecaption{Day-to-night contrast results for the \texttt{Eureka!} reduction, with temperatures in Kelvin and fluxes in ppm. The bond Albedo and heat recirculation efficiency are dimensionless values between 0 to 1. The peak-to-peak amplitude is defined as $\frac{F_{day}}{F_*}(\max\Phi_{P}-\min\Phi_{P})$ in ppm and the phase offset is in degrees. The NRS1 nightside fluxes and temperatures are given as $2\sigma$ 
    upper limits.} \label{tab:phasecurveresults}
    \tablewidth{0pt}
    \tablehead{\colhead{$\lambda$ ($\mu m$)}&\colhead{$R_P/R_*$}&\colhead{$F_{day}/F_*$}&\colhead{$F_{night}/F_*$}&\colhead{$T_{day}$}&\colhead{$T_{night}$}&\colhead{$A_{B}$}&\colhead{$\varepsilon$}&\colhead{Amplitude}&\colhead{Phase offset}}
    \startdata
    \textbf{NRS1} & ${0.0147}^{+0.0004}_{-0.0004}$ & ${34.8}^{+10.4}_{-11.6}$ & $\leq 12.1$ & $1970^{+273}_{-320}$ &  $\leq 1388$ & $0.79^{+0.10}_{-0.13}$ & $0.06^{+0.18}_{-0.06}$ & ${32.5}^{+10.4}_{-11.5}$ & ${-17.0}^{+15.5}_{-16.3}$ \\
    2.863 - 3.147 & ${0.0149}^{+0.0004}_{-0.0004}$ & ${51.2}^{+11.6}_{-11.7}$ & $\leq 10.4$ & ${2404}^{+239}_{-255}$ & $\leq 1384$ & ${0.55}^{+0.16}_{-0.20}$ & ${0.03}^{+0.09}_{-0.03}$  & ${49.5}^{+11.6}_{-11.9}$ & ${-14.8}^{+10.5}_{-11.3}$\\
    3.147 - 3.430 & ${0.0151}^{+0.0004}_{-0.0005}$ & ${31.5}^{+14.1}_{-16.5}$ & $\leq 16.5$ & ${1782}^{+377}_{-583}$ &  $\leq 1503$ & ${0.85}^{+0.10}_{-0.16}$ & ${0.13}^{+0.44}_{-0.13}$ &${27.7}^{+13.8}_{-15.4}$ & ${-13.0}^{+25.4}_{-24.0}$\\
    3.430 - 3.714 & ${0.0144}^{+0.0005}_{-0.0005}$ & ${26.5}^{+14.5}_{-16.9}$ & $\leq 24.7$ & ${1623}^{+428}_{-696}$ &  $\leq 1722$ & ${0.87}^{+0.09}_{-0.14}$ & ${0.31}^{+0.73}_{-0.31}$  & ${22.8}^{+13.2}_{-13.3}$ & ${-27.3}^{+50.0}_{-30.5}$\\
    \textbf{NRS2} & ${0.0145}^{+0.0004}_{-0.0004}$ & ${47.3}^{+11.7}_{-12.4}$ & ${33.0}^{+12.3}_{-12.3}$ & $2095^{+275}_{-291}$ & $1742^{+302}_{-359}$ & $0.52^{+0.18}_{-0.22}$ & $0.70^{+0.31}_{-0.34}$ & ${24.0}^{+11.0}_{-10.1}$& ${32.7}^{+32.1}_{-58.2}$\\
    3.820 - 4.136 & ${0.0139}^{+0.0004}_{-0.0005}$ & ${24.9}^{+11.4}_{-10.2}$ & ${15.9}^{+10.6}_{-8.0}$ & ${1618}^{+313}_{-446}$ & ${1282}^{+345}_{-632}$ & ${0.83}^{+0.10}_{-0.12}$ &${0.64}^{+0.52}_{-0.55}$  & ${20.3}^{+10.6}_{-10.0}$ & ${66.7}^{+17.1}_{-31.9}$\\
    4.136 - 4.451 & ${0.0144}^{+0.0005}_{-0.0005}$ & ${67.6}^{+14.1}_{-14.7}$ & ${29.7}^{+15.3}_{-14.3}$ & ${2578}^{+311}_{-325}$ & ${1625}^{+388}_{-561}$ & ${0.23}^{+0.28}_{-0.36}$&${0.33}^{+0.31}_{-0.26}$  & ${40.2}^{+15.1}_{-15.1}$ & ${13.2}^{+21.0}_{-17.9}$\\
    4.451 - 4.767 & ${0.0142}^{+0.0006}_{-0.0006}$ & ${52.6}^{+17.6}_{-18.6}$ & ${20.6}^{+15.9}_{-12.9}$ & ${2207}^{+434}_{-475}$ & ${1277}^{+445}_{-695}$ & ${0.59}^{+0.21}_{-0.34}$& ${0.25}^{+0.43}_{-0.24}$ & ${33.0}^{+17.7}_{-16.5}$ & ${1.0}^{+28.0}_{-26.3}$\\
    4.767 - 5.082 & ${0.0154}^{+0.0006}_{-0.0006}$ & ${51.3}^{+18.8}_{-18.7}$ & ${33.0}^{+18.1}_{-17.4}$ & ${1947}^{+378}_{-416}$ & ${1500}^{+398}_{-557}$ &${0.66}^{+0.17}_{-0.24}$ &${0.59}^{+0.48}_{-0.45}$ & ${25.9}^{+15.6}_{-12.9}$ & ${36.5}^{+36.9}_{-46.7}$ \\
    \textbf{NRS1 $\cup$ NRS2} & - & - & - & $2048^{+193}_{-206}$ & $1089^{+175}_{-379}$ & $0.74^{+0.08}_{-0.10}$ & $0.18^{+0.14}_{-0.15}$ & - & - \\
    \enddata 
    \end{deluxetable*}
    
    We fit all seven spectroscopic light curves separately. To achieve this, we subtract the same timestamp as in the WLC fits 
    from all spectroscopic time values 
    and fix $t_0$ to the obtained posterior value. Moreover, we now use detector-specific and shared Gaussian priors for the granulation amplitude (that is, one prior for the NRS1 bins and another for the NRS2 bins), but Gaussian priors on the granulation timescale derived from NRS1 for all bins, since the timescale should be similar at all wavelengths and it is better constrained at these shorter wavelengths. Note that we use the largest error-bar as $\sigma$ for these $\mathcal{N}(\mu,\sigma)$ priors, and we also impose that the amplitude and timescale remain positive. Then, we run the sampling process with the chosen systematics models, non-binned data, $N_{burn} = 20000$, $N_{prod} = 5000$, and $N_{walkers} = 4\times N_{dim}$. Note that instead of only imposing a strictly positive planetary flux prior, $\Phi_P > 0$, we add a stricter prior of $0 < A < 0.5$ on the phase curve parameter $A$ (see equation \ref{equation:phasevariation}) in the fourth and seventh bins where the planetary signal is particularly small to avoid overfitting the noise resulting in a brighter nightside than dayside. Indeed, we deem such a heat distribution not plausible given the results in NRS1 (see below). 

\section{Deriving Physical Parameters from the Phase Curve \label{sec:physical_parameters}}

Now that we are confident in our treatment of stellar granulation -- having used a GP in our overall model to disentangle mesogranulation from planetary flux variation -- we can robustly interpret the phase curve signal as planetary in origin. The complete NRS1 and NRS2 best-fit parameters are presented in Table \ref{tab:phasecurveresults} (and Table \ref{tab:bestfit_eureka_vs_jedi} for both reductions). Here we take the empirical results of the phase curve fitting above and use them as inputs into physical (analytic) models to infer brightness temperatures and longitudinal heat maps, which are subsequently used to obtain rough estimates of TOI-561~b's reflectivity and heat transport efficiency. Finally, we discuss the consistency of our spectroscopic eclipse depths with those obtained using secondary eclipses only in our previous emission-only paper \citep{EmissionPaper}.

    \subsection{Heat Distribution Calculation}
    
        All the equations in this paragraph are particular cases of the more general method described in \cite{dang}. Note that the orbital phase is defined as $\xi(t) = 2\pi (t-t_e)/P$, where $t_e$ is the time of eclipse and $P$ is the period of revolution, $\phi$ is the longitude from the sub-stellar point, and $\theta$ is the latitude.

        From \S\ref{sec:planetary_signal_model}, we have $\Phi_P$
        
        \begin{equation}
            \Phi_P(\xi) = 1+A(\cos(\xi)-1) + B\sin(\xi)
        \label{equation:phasevariation}
        \end{equation}

\begin{figure*}
            \centering
            \includegraphics[width=1.6\columnwidth]{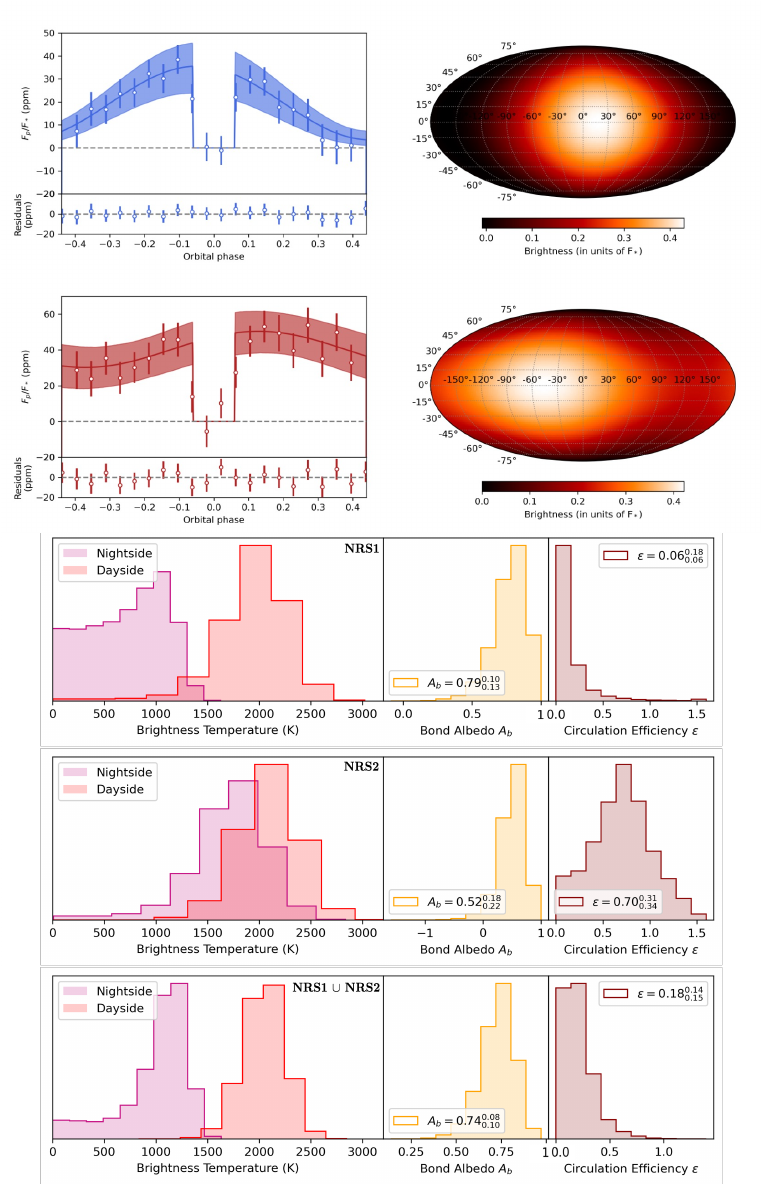}
            \caption{Summary of the white light curve fits for the TOI-561~b phase curve. The top two rows (NRS1 at the first and NRS2 in the second) display the phase-folded WLCs along with 16-84$^{th}$ percentile contours from the posterior distribution and error-bars corresponding to the errors on each bin's mean, and on the right are the relative brightness maps. The bottom three rows show the posterior distributions of the brightness temperatures, Bond albedo estimates, and circulation efficiency estimates for NRS1, NRS2 and NRS1 $\cup$ NRS2, respectively. Note that the data used to create this figure come from the \texttt{Eureka!} reduction.}
            \label{fig:summary}
        \end{figure*}

\noindent For each detector, we convert the phase curve coefficients $A$ and $B$ 
to Fourier series coefficients via 
       \begin{equation}
            F_P(\xi) = F_P(0)\Phi_P(\xi) = F_0+C_1\cos(\xi) + C_2 \sin(\xi)
        \label{equation:fourier}
        \end{equation}
\noindent and then to longitudinal brightness map coefficients valid for tidally-locked planets via 
        \begin{equation}
            J(\phi) = \frac{F_0}{2} + \frac{2}{\pi}C_1\cos(\phi)- \frac{2}{\pi}D_1\sin(\phi)
        \label{equation:longitude}
        \end{equation}
\noindent Since no latitudinal information is constrained by the observations ($\sim$ an edge-on orbit), these maps simply follow Equation \ref{equation:maps}: 

          \begin{equation}
            I(\phi,\theta) = \frac{3}{4}J(\phi)\sin(\theta)
        \label{equation:maps}
        \end{equation}    
\noindent  assuming that the brightness drops as the cosine of latitude. The maps derived from the WLC fit are included in Figure \ref{fig:summary} in the two top right panels.

        Then, posterior distributions of the dayside and nightside temperature are obtained using \texttt{emcee} with a blackbody model through the following equations: 
    
        \begin{equation}
            B_\lambda(\lambda,T)=\frac{2hc^2}{\lambda^5}\Big[e^{\frac{hc}{\lambda kT}}-1\Big]^{-1}
        \label{equation:blackbody}
        \end{equation}
    
        \begin{equation}
            \Big[ \frac{F_P}{F_*} \Big](\lambda,T) = \Big[\frac{R_P}{R_*}\Big]^2\frac{B_{\lambda} (\lambda,T)}{B_{\lambda}(\lambda,T_{*})}
        \label{equation:fluxratio}
        \end{equation}
    
        \begin{equation}
            \Big[\frac{F_P}{F_*}\Big](T) = \frac{\int^{\lambda_{max}}_{\lambda_{min}} \tau(\lambda) \Big[\frac{F_P}{F_*}\Big](\lambda,T) B_{\lambda} (\lambda,T_{*})  \lambda d\lambda}{\int^{\lambda_{max}}_{\lambda_{min}} \tau(\lambda) B_{\lambda} (\lambda,T_{*}) \lambda d\lambda}
        \label{equation:integrated_blackbody}
        \end{equation}
    
        \noindent Here, $\tau$ represents the throughput NIRSpec/G395H, computed by \texttt{ExoTiC-LD} \citep{Grant2024} with the MPS-ATLAS-1 stellar grid. The $\lambda$ thresholds correspond to the minimum and maximum wavelength of each bin/detector.
        
        We derive the dayside and nightside brightness temperatures by comparing the observed eclipse depth to the one calculated at a given brightness temperature $T_{P}$. We first generate a grid of blackbody spectra computed via Planck's law over a planet temperature range of 10 to 4000 K (Equation \ref{equation:blackbody}). For each $T_{P}$ considered, we compute the expected planet flux ratio $F_P/F_\star$ by integrating over the NIRSpec/G395H bandpass using the instrumental throughput $\tau$ (Equation \ref{equation:integrated_blackbody}). We then construct an interpolating function mapping $F_p/F_\star$ to a brightness temperature, and we use this function to translate our posterior distributions of the dayside and nightside fluxes obtained from our phase curve fit, using \texttt{emcee} again. This method provides a posterior distribution of the dayside and nightside temperatures. We report the median temperatures, with uncertainties defined by the 16th and 84th percentiles.
        
        Finally, we compute estimates of the Bond albedo and circulation efficiency with Equations \ref{equation:irrad_temp} through \ref{equation:boundalbedo}, which are derived from Equations 4 and 5 of \cite{cowanalbedo}.

        \begin{equation}
            T_0=\frac{T_{*}}{\sqrt(a/R_*)}
        \label{equation:irrad_temp}
        \end{equation}
        \begin{equation}
             \varepsilon= \frac{8}{3}\Big[\Big(\frac{T_{day}}{T_{night}}\Big)^4 + \frac{5}{3}\Big]^{-1}
        \label{equation:epsilon}
        \end{equation}
        \begin{equation}
            A_B = 1-\frac{4}{\varepsilon}\Big[\frac{T_{night}}{T_0}\Big]^4
        \label{equation:boundalbedo}
        \end{equation}
    \noindent Note that these conversions assume a surface emissivity of unity.
 
        \subsection{Results}
\begin{figure*}[ht]
        \centering
        \includegraphics[width=1\linewidth]
        {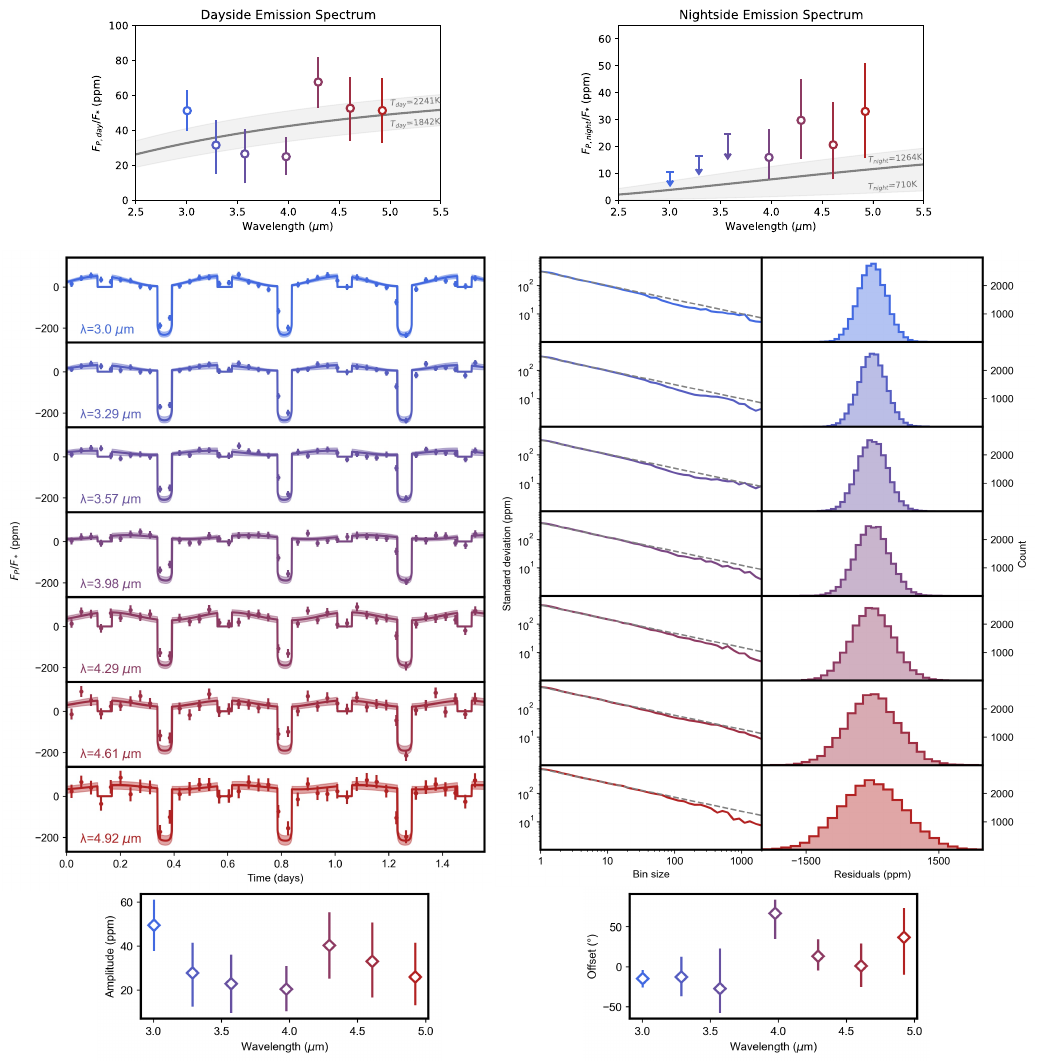}
        \caption{\texttt{Eureka!} dayside-nightside emission spectra at the top, with the NRS1 $\cup$ NRS2 blackbody predictions using $R_P/R_* = 0.0146$ (the average between the posteriors of each detector) as grey lines with shading. Note that on the nightside spectrum, only the 2$\sigma$ upper-limits are reported for the three NRS1 spectroscopic channels. Our spectroscopic light curves are shown in the middle left panel, with the 16-84$^{th}$ percentile contours of our best-fit models overplotted. In the middle right panel, we show the standard deviation of the residuals versus bin size and the distributions of the data-model fit residuals in ppm. The bottom row shows our measured peak to peak amplitude (in ppm) and phase offset of the planet's apparent brightness variation (in degrees).}
        \label{fig:spectroscopic_results}
    \end{figure*}

         Overall, our results suggest a global atmosphere on TOI-561~b. The broadband phase folded light curves, relative brightness maps, and $T_{day}$, $T_{night}$, $A_{B}$, and $\varepsilon$ distributions obtained during the WLC fitting are shown in 
         Figure \ref{fig:summary}. We note that the strict positivity prior on the planetary flux truncates the NRS1 phase curve posteriors, leading to an underestimated nightside flux uncertainty in this channel; we therefore report the NRS1 nightside flux as a 2$\sigma$ upper-limit rather than an absolute measurement. This effect does not affect the NRS2 channel.
         The NRS1 and NRS2 channels yield dayside fluxes of $34.8^{+10.4}_{-11.6}$~ppm and $47.3^{+11.7}_{-12.4}$~ppm, and nightside fluxes of $\leq 12.1$~ppm at 2$\sigma$ and $33.0^{+12.3}_{-12.3}$~ppm, respectively.
         The combined NRS1~$\cup$~NRS2 fit yields a dayside temperature of $2048^{+193}_{-206}$~K and a nightside temperature of $1089^{+175}_{-379}$~K, with a Bond albedo of $A_B = 0.74^{+0.08}_{-0.10}$ and a heat redistribution efficiency of $\varepsilon = 0.18^{+0.14}_{-0.15}$. The full spectroscopic phase curve results, including the planet's radius $R_{P}/R_{\star}$, dayside flux $F_{day}/F_{\star}$, nightside flux $F_{night}/F_{\star}$, dayside temperature $T_{day}$, nightside temperature $T_{night}$, Bond albedo $A_B$, heat redistribution efficiency $\varepsilon$, and the phase curve peak-to-peak amplitude and phase offset, are all reported in Table~\ref{tab:phasecurveresults}. To clarify, the reported offsets are orbital and not longitudinal. A negative value indicates that the brightness peak occurs when the planet has not yet reached the eclipse point, i.e., an eastern offset if we assume a counter-clockwise revolution as seen from above the north pole. In Figure~\ref{fig:spectroscopic_results} we also visualize the results of the spectroscopic light curve fitting, including the dayside and nightside emission spectra, spectroscopic light curve fits and residuals, and phase curve amplitudes and offsets as a function of wavelength. 

         We find some differences between our results for NRS1 versus NRS2, some of which (e.g., the nightside brightness temperatures) are more significant than others (e.g., the phase offset values) when considering the 1$\sigma$ errors. We discuss these differences and their potential origin(s) further in \S\ref{sec:limitations_and_future_obs}, after also considering a more detailed modeling analysis that takes into account hyper-illumination and non-grey effects in \S\ref{sec:gcm}. 
         
         As a cross-check, in Appendix \ref{sec:A_reduction_consistency} we compare our results from fitting the \texttt{Eureka!} and \texttt{ExoTiC-JEDI} reductions and find that they are generally consistent. In Appendix \ref{sec:A_comparison_with_eclipse_only} we show the emission spectra derived here from the \texttt{Eureka!} and \texttt{ExoTiC-JEDI} reductions compared to those presented in \cite{EmissionPaper} and also find general agreement.

\section{General circulation modelling of phase curves \label{sec:gcm}}

In addition to our analytic estimates of the temperatures, albedo, and heat redistribution for TOI-561~b, we wish to take advantage of our phase curve observations to gain deeper insight into the possible atmospheric composition, structure, and circulation for this planet. For this investigation, we turn to a general circulation model (GCM), which can simulate the whole-atmosphere dynamics of a planet self-consistently and is particularly useful for tidally locked planets that can have large day-to-nightside temperature contrasts resulting in complex climate patterns (\citealt{Pierrehumbert2019} and references therein).  

Since the spin period of a synchronously rotating tidally-locked planet is identical to its orbital period, a USP planet like TOI-561~b is a rapid rotator. The effect of this rapid rotation on the general circulation, and its ability to redistribute heat from dayside to nightside, can be characterized by the Weak Temperature Gradient (WTG) parameter \citep{Pierrehumbert2019} 
\begin{equation}
\Lambda \equiv \frac{\sqrt{R^*T/\mu}}{\Omega a}  
\end{equation}
where $R^*$ is the universal gas constant, $T$ is a characteristic temperature of the dynamically active part of the atmosphere, $\mu$ is its mean molecular weight, $\Omega$ is the angular velocity of the planet's spin, and $a$ is the planet's radius. This parameter is the ratio of the radius of deformation of the atmosphere relative to the size of the planet, and characterizes the ability of the pressure gradients arising from strong temperature gradients to be balanced by the Coriolis acceleration. The local projection of the Coriolis
parameter is proportional to the sine of the latitude, so the local Coriolis parameter is always weak near the equator, leading to weak temperature gradients there -- a phenomenon central to the conceptual understanding of Earth's tropics. 

When $\Lambda \gg 1$, temperature gradients are weak globally (the temperature is more homogeneous), whereas when $\Lambda \ll 1$ day/night temperature homogenization is only guaranteed in a thin strip near the equator, and the rest of the planet can support strong day/night temperature contrast.  This is particularly so for 
ultra-hot planets, because the short radiative damping time makes it hard for midlatitude jets to transport heat to the nightside, since it tends to get radiated away before it gets very far. Near the equator, heat transport is due to wave dynamics in addition to the typical super-rotating jet. Earth has $\Lambda \approx 0.6$, and the WTG approximation applies over roughly half the planet extending around the equator. TOI-561~b has $\Lambda \approx 2 $ for a pure $\mathrm{H_2}$ atmosphere, $\Lambda \approx 0.66 $ for pure 
$\mathrm{H_2O}$, and  $\Lambda \approx 0.42 $ for pure $\mathrm{CO_2}$. Note that the effect of high temperatures on $\Lambda$ partly compensates for the rapid spin
of TOI-561~b.  Except in the case of a very low molecular weight atmosphere, we conclude that TOI-561~b is in a roughly Earth-like circulation regime, with weak temperature gradients in a substantial band in the tropics, but high temperature gradients in the midlatitudes. The radiative damping time is much smaller than Earth's, though, on account of the
high temperature, and this would tend to further enhance strong temperature gradients away from the tropics.  We will see these effects at play in the general circulation model
simulations. 

\subsection{Methods}\label{section:GCM_method}
For interpreting the observed phase curves, we run the GCM  \texttt{Isca} \citep{Vallis2018}, a flexible framework that has been used to study Earth and other planets \citep{Thomson2019, Lewis2022}. 
We run \texttt{Isca} assuming a dry atmosphere (i.e. no condensible species) configured with a suite of idealized parameters. The model is run at 
roughly 5.6$^o$ resolution in latitude and longitude (T21 in \texttt{Isca}). In the model, there are 30 vertical layers distributed accordingly to $\sigma =\exp [-5(0.05 x+0.95x^3)]$, where $\sigma=p/p_s$ and $x$ is evenly spaced on the unit interval. The GCM simulations were performed using planetary parameters representative of TOI-561~b. We set the radius of the planet $R_p = 1.42$ R$_{\oplus}$ \citep{patel2023},the mass of the planet $M_p = 2.24$ M$_{\oplus}$ \citep{Brinkman2023} giving a surface gravity of $g=10.91$ ms$^{-2}$. The orbital period $P$=0.447 days \citep{patel2023} orbiting a star with radius $R_s$ = 0.843 R$_\odot$ \citep{lacedelli21} and with stellar flux $S_0=6.42\times10^6$ W/m$^2$ at a distance of $a$=0.011 AU.  All simulations are run for 310 orbits with time averaging over the last 12 orbits used for analysis. To ensure the GCMs have reached steady-state, we verified that the global average of the sum of radiative fluxes at the top of the atmosphere is relatively constant. Most GCM models have some energy leakage even if converged. The top-of-atmosphere energy leakage for the 1 bar H$_2$O is 5.24x10$^3$ W/m$^2$ which represents 0.6 \% of the total energy output

We consider the three possible atmosphere compositions that were investigated in \cite{EmissionPaper} -- pure H$_2$O, 50\% H$_2$O + 50\% CO$_2$ and  99.9\% O$_2$ + 0.01\% H$_2$O. These compositions were chosen simply to show a few example cases and do not represent a comprehensive exploration of parameter space, which is beyond the scope of this first phase curve comparison and we leave for future work. 
Our idealized GCM uses a semi-grey gas scheme with prescribed shortwave and longwave opacity (see, e.g., \citealp{pierrehumbert2010principles}). 
To determine the semi-opacities, we fit an analytical grey gas solution from \cite{Guillot2010} to the self consistent real-gas 1D results presented in \cite{EmissionPaper}.
Aside from computational efficiency, the semi-grey model allows us to be somewhat agnostic about the precise source of the opacity. The assumption is that if the longwave and shortwave opacities are tuned to give a vertical structure and overall temperature approximately matching 1D globally averaged climate computations, the 3D dynamics and atmospheric structure will indirectly account for spectroscopic radiative effects. To take into account the real gas opacity effect on the observed phase curve, we perform post processing with real gas opacity.
The composition of the atmosphere mainly impacts the dynamics through the heat capacity of the atmosphere and the effect of mean molecular weight on wave speeds \citep{Hammond2017}, although composition also effects the fitted semi-grey opacities. The heat capacity for our models was determined by computing the Shomate equation assuming a temperature of 2000~K using coefficents from \citep{chase1998j} and \citep{garvin1986codata}.  

For each composition, we consider five possible surface pressures, $P_s$ =0.1,1, 3, 5 and 10 bar. The actual atmosphere of TOI-561~b may be thicker than 10 bar, but at this pressure the amplitude of the phase curve does not substantially change while the runtime of the GCM does, and we cannot study the heat transport of such a thick atmosphere within the GCM. Choosing 10 bar as the maximum surface pressure is also well-motivated since, as discussed in \S\ref{sec:gcm}, TOI-561~b is expected to have substantial temperature gradients in the midlatitudes due to the fast rotation period. We provide a complete list of our experiments in Table \ref{tab:toi561b-gcm-runs}. 

Finally, for planets such as TOI-561~b having extremely small orbital semi-major axes, the host star occupies a larger proportion of the sky and the terminator broadens. For TOI-561~b, up to two-thirds of the surface will be illuminated by the star. We therefore take nightside hyper-illumination in account by following the method outlined in \cite{Kang2023} (see their Appendix A).

In the radiation scheme, the longwave (planetary thermal emission) and shortwave (stellar spectrum) opacities are
set as constants independent of pressure. Scattering is not explicitly taken into account in the radiation
scheme, but an effective planetary albedo is imposed by reducing the net incoming stellar flux at the top of the
atmosphere relative to the incident flux. The shortwave surface albedo is set to zero; any flux that reaches the surface is assumed to be absorbed there \citep{fortin_lava_2024}. 
Imposing a top-of-atmosphere albedo this way more closely represents the situation in which the albedo is controlled by reflective high haze or clouds, which reflect back stellar radiation before it has a chance to penetrate into the deeper atmosphere and become absorbed there.

The planetary surface, turbulence layer, and sub-scale convection are parametrized in a similar way to the simulations shown in \cite{Lewis2022}. The surface is modeled as a static slab whose temperature evolves according to
\begin{equation}
\label{Eqn:Surface_eqn }
    C \frac{\partial T_s}{\partial t} = S+I^{\downarrow}_{lw} -\sigma T_s^4 -H
\end{equation}
where $C$=$2.1\times 10^7$ JK$^{-1}$ m$^{-2}$ is the surface heat capacity (for context, this value for $C$ is equivalent that of a liquid water ocean with a depth of 5~m). We assume no time variation in the radiative forcing, and dynamical heat redistribution within day-side magma oceans is known to be negligible \citep{lai_three_2026}, so a shallow ocean layer parametrization is appropriate for modeling TOI-561~b. The variables $S$ and $I^\downarrow_{lw}$ are the downward stellar and longwave fluxes of radiation, respectively. The variable $H$ is the sensible heat flux, which we calculate using a turbulent kinetic energy scheme as
\begin{equation}
\label{Eqn:Heat_flux}
    H = \rho_a c_p \zeta |\boldsymbol{u}_a| (T_s-T_a).
\end{equation}
The subscript $a$ indicates quantities at the lowest model level, and $\zeta$ is a dimensionless bulk transfer coefficient that damps the formation of large temperature discontinuities between the surface and the lowest model level ($T_s-T_a$). We set $\zeta=0.1$, which is greater than that in \cite{Lewis2022}. This choice was necessary to prevent the formation of large near-surface temperature discontinuities. 

The turbulent vertical mixing of heat and momentum in the near-surface atmospheric boundary layer is parameterized using eddy diffusion as
\begin{equation}
\label{Eqn:Turbulent_mixing}
    \frac{\partial X}{\partial t} = ... + \frac{1}{\rho} \frac{\partial}{\partial z} K \rho \frac{\partial X}{\partial z}
\end{equation}
where $X$ is a placeholder for either the dry static energy or the velocity, and $z$ is the height. This is a standard diffusion equation where we have neglected to write the advection and source terms.  The diffusion coefficient $K$ is calculated according to
\begin{equation}
\label{Eqn:Diff coef}
K=
    \begin{cases} 
      \zeta |\boldsymbol{u}_a| z_a, & \text{for } P > P_{bl} \\ 
      \zeta |\boldsymbol{u}_a|\exp \left(\left( -\frac{P_{bl}-P}{P_{strat}} \right)^2\right), & \text{for}  \ P\leq P_{bl}
    \end{cases}
\end{equation}
We set the pressure at the top of the boundary layer, $P_{bl}$, equal to $0.85~P_s$. Above $P_{bl}$, there is an exponential decay in the diffusivity, controlled by $P_{strat}$=0.1 $P_s$.This is not the stratosphere pressure but a model parameter that controls the damping of turbulence as we approach free troposphere. Finally, the model includes a dry convection scheme, which instantaneously
restores the temperature structure to the dry adiabat whenever it is convectively unstable.

\subsection{Grey Gas Fit}
We tuned longwave and shortwave grey opacities by performing an analytical grey gas fit of the analytic grey gas solution in \cite{Guillot2010} to  \texttt{GENESIS} real-gas radiative convective simulations \citep{EmissionPaper} for the compositions of interest. Details are provided in Appendix \ref{sec:A_full_gcm_experiment}. The Guillot solution takes longwave opacity $\kappa_{lw}$ and the ratio of shortwave to longwave opacity $\gamma = \frac{\kappa_{sw}}{\kappa_{lw}}$ as inputs; these are the same opacity parameters we use in the grey radiation model for GCM simulations.  From these fits, we found a Root Mean Square Error (RMSE) of 112 for H$_2$O, 106 for 50\% H$_2$O+50\% CO$_2$ and 116 for 99.9\% O$2$+0.1\% H$_2$O. These RMSE values were computed for pressures between 10 bar and 1 mbar and correspond to relative errors on the Guillot fits of around 3.3\% to 9\%, which we argue are small enough for us to proceed. For these best fits we derive the following opacities $\kappa_{lw}=[193.1,60.5,7.94]$ cm$^2$/kg and $\kappa_{sw}=[30.1,8.6,1.04] $ cm$^2$/kg for H$_2$O, 50\% H$_2$O+50\% CO$_2$ and 99.9\% O$2$+0.1\% H$_2$O respectively. 

\begin{deluxetable*}{lccccc}
\tablecaption{List of GCM model parameters. Specific heat capacity is computed using the Shomate Equation at 2000\,K.\label{tab:toi561b-gcm-runs}}
\tablewidth{0pt}
\tablehead{
\colhead{Composition} &
\colhead{$P_s$ [bar]} &
\colhead{$R_d$ [J\,kg$^{-1}$\,K$^{-1}$]} &
\colhead{$c_p$ [J\,kg$^{-1}$\,K$^{-1}$]} &
\colhead{$\kappa_{lw}$ [cm$^2$\,kg$^{-1}$]} &
\colhead{$\kappa_{sw}$ [cm$^2$\,kg$^{-1}$]}
}
\startdata
100\% H$_2$O  & [0.1,1,3,5,10]  & 462 & 2842 & 193.1  & 30.1 \\
50\% CO$_2$ + 50\% H$_2$O & [0.1,1,3,5,10]  & 323 & 2107 & 60.5 & 8.6 \\
99.9\% O$_2$ + 0.1\% H$_2$O & [0.1,1,3,5,10]  & 260 & 1181 & 7.94 & 1.04 \\
\enddata
\end{deluxetable*}

\begin{figure}
            \centering
            \includegraphics[width=\columnwidth]{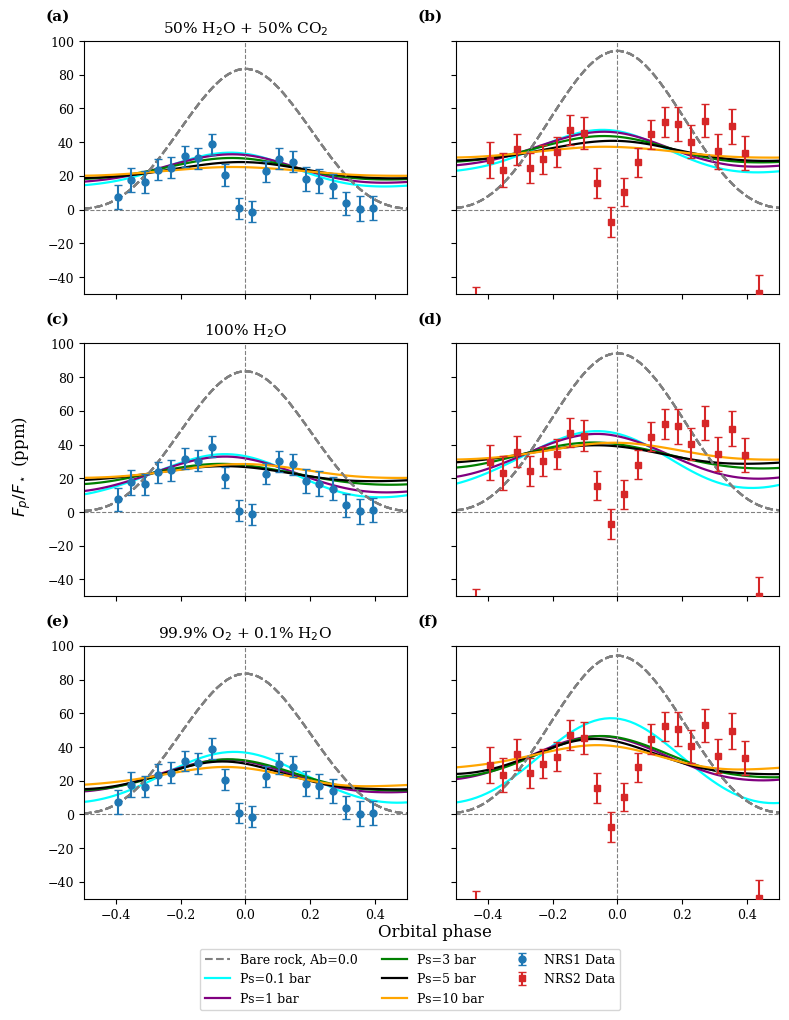}
            \caption{The phase curves from the GCM that best match TOI-561~b are those at $A_B=0.6$, as shown here (compared to Figure \ref{fig:all_comp_phasecurves}). The left column shows the NRS1 data (blue points) and simulated phase curves (lines) with each row representing a different composition, and the right column is the same for NRS2 (data as red points). We show phase curves for $P_s$ = [0.1,1,3,5,10] bar. The expected phase curve for a zero-albedo, unit emissivity bare rock surface (dashed grey line) is inconsistent with our observations. 
            The NRS1 data show a larger phase-curve amplitude, whereas the NRS2 data exhibit a smaller day-to-night variation. This contrast presents a tension with the predictions of idealized GCMs that neglect effects such as non-grey clouds; possible explanations are discussed later in Section \ref{sec:discussion}. More importantly, we do not suggest a ranking among the different composition or the different pressures scenarios, only that those with high albedo are preferred. The opacity values used in these simulations are provided in Table \ref{tab:toi561b-gcm-runs}. }
            \label{fig:Ab_06_phasecurves}
\end{figure}

\subsection{GCM Results \label{sec:gcm_results}}
To produce synthetic phase curves for comparison with our TOI-561~b observations, we start with the T-P profile for each atmosphere column, and use this in \texttt{petitRADTRANS} \citep{Molliere2019} to produce emission spectra of the composition of interest.
Our molecular opacities are H$_2$O \citep{2010JQSRT.111.2139R}, CO$_2$ \citep{2020MNRAS.496.5282Y} and O$_2$ \citep{2022JQSRT.27707949G}. This becomes our outgoing flux $F_{out}$ in equation

\begin{equation}
    I_p(\xi) = \frac{\int_{-\pi/2}^{\pi/2}\int^{-\xi+\pi/2}_{-\xi-\pi/2}F_{out} \cos(\lambda+\xi)\cos^2(\theta) d\lambda d\theta}{\int_{-\pi/2}^{\pi/2}\int^{-\xi+\pi/2}_{-\xi-\pi/2}\cos(\lambda+\xi)\cos^2(\theta) d\lambda d\theta}
\end{equation}
where $\xi$ is the phase angle, longitude $\lambda$ and latitude $\theta$. The planetary flux $F_p$ is compared to the stellar flux $F_s$ through this equation
\begin{equation}
    \frac{F_p}{F_{s}} = \frac{I_p}{I_s} \left(\frac{r_p}{r_s}\right)^2
\end{equation}
\noindent where $r_p$ and $r_s$ are the planetary and stellar radii, respectively. The stellar flux spectral energy distribution was obtained from PHOENIX stellar models at an effective temperature $T_{\rm{eff}} = 5372$~K \citep{husser_a_2013,lacedelli}. 

We explore a range of albedo values in our comparisons to the phase curve data. The best-matching cases -- those with high albedo -- are in Figure \ref{fig:Ab_06_phasecurves}, which shows the measured data for NRS1 (blue points) and NRS2 (red points) and simulated phase curves assuming $A_B = 0.6$ (lines). A bare rock phase curve, assuming a local energy budget without heat transport following the hyper-illumination distribution given in \cite{Kang2023} for $A_B=0.0$ with unit emissivity, is included in this figure (dashed grey line), illustrating that the phase curve amplitude for this case is substantially larger than the observations, providing a poor fit. For completeness, Figure \ref{fig:Ab_0_phasecurves} shows the zero albedo cases across all of the compositions and surface pressures considered (rows), again indicating that the planet's dayside simulated by the GCM is too hot compared to the observed dayside flux ratio. 
Further comparisons across a wider range of albedos are shown in Figure \ref{fig:all_comp_phasecurves}. We find that a Bond albedo greater than 0.5 is needed to match the observed dayside flux in the NRS1 band, but increasing the Bond albedo still cannot match the low nightside emission, with the best fits being around 5 to 10 ppm higher than the observed nightside fluxes. In the NRS2 band a Bond albedo of around 0.4 is needed to match the dayside emission, although at some pressures this is too high to match the nightside flux. We see similar mismatches for the 50\% CO$_2+$50\% H$_2$O and 99.9\% O$_2+$0.1\% H$_2$O cases. 

To explore more quantitative comparisons between the modeled and observed phase curves, we compute a reduced $\chi^2$ value for each composition across five different pressures and three different albedo values, and report these $\chi^2$ fitting coefficients in Table \ref{tab:toi561b-x^2/N-gcm}. We caution against reading too much into these values due to the small number of data points in our observations and other limitations of the data discussed in \S\ref{sec:limitations_and_future_obs}, but report them here for completeness. 
Our GCM phase curve models do not capture the in-eclipse and in-transit behavior, so we ignore these parts of the phase curve when computing the reduced $\chi^2/N$ ($N=16$ is the number of data points). For the NRS2 phase curve, the best-matching atmospheric composition is 50\% CO$_2+$50\% H$_2$O ($\mu=31.0$ g/mol). For this composition under a high Bond albedo scenario, the $\chi^2/N$ ranges from 1.13 to 1.36 depending on the surface pressure. For the NRS1 phase curve, the best-matching composition is 99.9\% O$_2+$0.1\% H$_2$O ($\mu=31.9$ g/mol), with $\chi^2/N$ ranging from 0.4 to 1.51. However, we do not suggest that one composition is preferred over another, indeed Figure \ref{fig:Ab_06_phasecurves} shows how similar the simulated phase curves are across compositions. As mentioned above, these compositions are example cases only; a wider exploration is not merited for our analysis here. 

That being said, it is clear that across all pressures and compositions, phase curves produced from simulations with a Bond albedo $A_B = 0.6$ are generally preferred, as shown in Figure \ref{fig:Ab_06_phasecurves} compared to Figure \ref{fig:all_comp_phasecurves} and the bare rock case. 
While some thin ($\leq10$-bar) atmosphere models produce good fits to our observed phase curves, we disfavor such a thin atmosphere as the most plausible explanation because the low bulk density of TOI-561~b favors a volatile envelope \citep{Plotnykov2024} and thin volatile atmospheres are unstable over Gyr timescales due to strong stellar irradiation that govern atmospheric escape \citep{Owen2019AREPS}.
Instead, a substantially thick atmosphere can lower the amplitude of the phase curve because, all other things being equal, a deeper atmosphere has greater heat capacity and can redistribute more energy for a given wind speed. The higher heat capacity also increases the radiative damping time, and allows high temperatures to extend towards the nightside of TOI-561~b. This is why the amplitude of the phase curves decreases in all cases with increased surface pressure (see the example GCM outputs in Figure \ref{fig:gcm-windvectors-temp}). 
However, TOI-561~b is likely within the fast rotator-regime, so given a high molecular weight atmosphere, temperature gradients are only weak in a band surrounding the equator (though the tropical regions cover sufficient area to also meaningfully contribute to the disk-averaged phase curve). 

\begin{table*}[ht] \centering \small \setlength{\tabcolsep}{5pt} \renewcommand{\arraystretch}{1} \begin{tabular}{|l|c|c|c|c||c|c|c|} \hline \textbf{Composition} & $P_s$ [bar] & \shortstack{$\chi^2/N$ NRS1 \\ $A_B$=0.0} & \shortstack{$\chi^2/N$ NRS1 \\ $A_B$=0.3} & \shortstack{$\chi^2/N$ NRS1 \\ $A_B$=0.6} & \shortstack{$\chi^2/N$ NRS2 \\ $A_B$=0.0} & \shortstack{$\chi^2/N$ NRS2 \\ $A_B$=0.3} & \shortstack{$\chi^2/N$ NRS2 \\ $A_B$=0.6} \\ \hline Bare rock & 0 & 12.1 & 5.67 & 1.26 & 5.54 & 3.70 & 3.28 \\ \hline 100\% H$_2$O & 0.1 & 7.10 & 2.51 & 0.48 & 4.39 & 3.09 & 2.73 \\ 100\% H$_2$O & 1 & 8.38 & 2.11 & 0.76 & 4.74 & 2.19 & 1.97 \\ 100\% H$_2$O & 3 & 9.70 & 2.96 & 1.47 & 4.57 & 1.55 & 1.52 \\ 100\% H$_2$O & 5 & 11.17 & 3.43 & 1.81 & 4.47 & 1.39 & 1.46 \\ 100\% H$_2$O & 10 & 12.55 & 4.39 & 2.21 & 4.90 & 1.30 & 0.98 \\ \hline 50\% CO$_2$ + 50\% H$_2$O & 0.1 & 8.50 & 3.62 & 0.92 & 4.69 & 2.30 & 1.66 \\ 50\% CO$_2$ + 50\% H$_2$O & 1 & 8.65 & 2.58 & 1.27 & 4.67 & 1.88 & 1.36 \\ 50\% CO$_2$ + 50\% H$_2$O & 3 & 9.74 & 3.06 & 1.62 & 4.83 & 1.59 & 1.21 \\ 50\% CO$_2$ + 50\% H$_2$O & 5 & 11.73 & 3.97 & 1.89 & 5.00 & 1.40 & 1.13 \\ 50\% CO$_2$ + 50\% H$_2$O & 10 & 11.72 & 3.86 & 2.29 & 4.92 & 1.16 & 1.25 \\ \hline 99.9\% O$_2$ + 0.1\% H$_2$O & 0.1 & 7.06 & 2.67 & 0.40 & 4.73 & 2.97 & 2.57 \\ 99.9\% O$_2$ + 0.1\% H$_2$O & 1 & 9.06 & 2.64 & 0.88 & 5.45 & 1.90 & 1.89 \\ 99.9\% O$_2$ + 0.1\% H$_2$O & 3 & 8.96 & 2.63 & 1.01 & 5.72 & 2.73 & 1.55 \\ 99.9\% O$_2$ + 0.1\% H$_2$O & 5 & 8.94 & 2.50 & 1.09 & 4.83 & 1.62 & 1.70 \\ 99.9\% O$_2$ + 0.1\% H$_2$O & 10 & 8.85 & 2.12 & 1.51 & 4.59 & 1.61 & 1.62 \\ \hline \end{tabular} \caption{List of $\chi^2/N$ values for low, intermediate, and high albedo model runs in NRS1 and NRS2. Note that did not simulate pressures above 10 bars as the majority of the incoming flux is absorbed at pressures below 10 bars, therefore a GCM run at higher surface pressures than 10 bars would resemble that of a 10 bar atmosphere.} \label{tab:toi561b-x^2/N-gcm} \end{table*}


\section{Discussion} \label{sec:discussion}

\subsection{Consistency with a Global Atmosphere}
We took two approaches to investigating the evidence for an atmosphere on TOI-561~b from our phase curve observations. 
In \S\ref{sec:physical_parameters}, we performed fits to the NRS1 and NRS2 light curves, accounting for differences in systematics between the detectors. 
We then combined both channels using \texttt{emcee} to jointly infer the dayside and nightside temperatures, Bond albedo, and heat recirculation efficiency. The combined dayside temperature is significantly lower than the predicted maximum dayside temperature for a planet with zero albedo and no heat redistribution to the nightside ($\sim$2050~K versus $\sim$3200K), and the nightside temperature ($\sim$1100~K) is also suggestive of non-negligible nightside emission. Note that these values are best-estimates based on the very limited wavelength range probed by NIRSpec/G395H, hence non-grey effects could be important since our observations do not cover the peak of the planetary blackbody. Using the day-to-night contrast, we inferred a high Bond albedo and a moderate heat redistribution efficiency, the combination of which is consistent with the presence of a global atmosphere driving large-scale circulation. 

Our second approach was to compare our measured white light phase curves with those produced synthetically via a GCM for the atmosphere (\S\ref{sec:gcm}), which also takes into account for non-grey effects during a post-processing step. We conclude from the GCM simulations that heat redistribution alone cannot match the observed dayside fluxes or amplitude of the phase curve, and a high albedo source ($A_b = 0.6$) indicative of clouds or haze \citep{gao_aerosols_2021,Lee2025,Janssen2026} is suggested, although these scenarios still overestimate the NRS1 nightside emission (e.g. Figure \ref{fig:Ab_06_phasecurves}a). For completeness, we also examined thin 0.1 bar atmosphere cases with surface shortwave albedo of 0.6, finding that it provides a better match to the NRS1 phase curve but is inconsistent with the NRS2 phase curve's high nightside emission, and would be incompatible with the low observed bulk density \citep{Plotnykov2024} and expected escape rates based on the high stellar irradiation the planet has received over its Gyr lifetime \citep{Owen2019AREPS}.
A high-albedo bare rock is also implausible because (1) lab measurements indicate that lava surfaces have low-reflective properties \citep{2020Essack,fortin_lava_2024} (2) such a hot surface would be evaporating, and the resulting rock vapor atmosphere would be optically thick at the visible wavelengths where the stellar flux peaks, meaning the relevant surface would be effectively hidden from view, and (3) again, it would be inconsistent with the low bulk density of the planet.

\subsection{Day and Night Emission Spectra of TOI-561~b \label{subsec: daynight_emission_spectra}}
The dayside spectrum we retrieved from our full phase curve analysis is largely consistent with the dayside spectrum presented in \cite{EmissionPaper}, see Figure \ref{fig:comparison_with_emission_paper_spectrum}. In the dayside spectrum presented here, where we have fit the full phase curve and accounted for stellar granulation noise, we see more of a downward trend from $\sim$2.8~$\mu$m to $\sim$4$~\mu$m, followed by a significant increase between 4.0 and 4.5~$\mu$m. 
Since this pattern spans the two detectors, 
we suspect molecular emission features, possibly muted-down by clouds, might be the cause \citep{gao_aerosols_2021,Janssen2026}. In particular, we note that this pattern is consistent with the CO$_2$ opacity function (see the fourth panel of Figure 1 from \cite{Hammond2025}). Therefore, our $\sim$2.8~$\mu$m and $\sim$4.3~$\mu$m spikes could be caused by CO$_2$ emission features in the scenario of an inverted T-P profile caused by clouds \citep{nguyen_clouds_2024}. An in-depth analysis of these spectra will be presented in future work.

We also present the first nightside spectrum of TOI-561~b in Figure \ref{fig:spectroscopic_results} and highlight the low nightside emission in NRS1 (2.67-3.72 $\mu$m). Notably, the GCM simulations overpredict the nightside flux in NRS1, suggesting that an additional cooling mechanism could be at play on the nightside that the models do not capture. This could be indicative of nightside clouds suppressing thermal emission at the wavelengths covered by NRS1 that is optically thin at the wavelengths covered by NRS2. Indeed, multiple species could condensate and form on the nightside as discussed in the next section. 

\subsection{Clouds in the Atmosphere of TOI-561~b} \label{sec: discussion clouds}
Laboratory measurements indicate that lava planet surfaces are poor reflectors \citep{2020Essack,fortin_lava_2024}. However, radiative-convective models show that atmospheres outgassed from magma ocean planets can reach high Bond albedos through Rayleigh/Mie scattering from gases/clouds, respectively \citep{2019Pluriel,nicholls_magma_2024}. The high Bond albedo inferred from our white light curve analysis and the GCM fits is suggestive of reflective clouds in the atmosphere of TOI-561~b. 
The dayside temperature of TOI-561~b is sufficient to vaporize rock that would be optically thick at UV and optical wavelengths, and while a pure rock vapor atmosphere is inconsistent with our NIR observations, a volatile-rich atmosphere could host mineral clouds condensed from rock vapor that has evaporated from the magma ocean 
\citep[e.g.,][]{Lee2025,Janssen2026,nguyen_clouds_2024}. The dominant condensate species are sensitive to both background atmospheric composition and the enrichment of heavier elements such as Si, Mg and Ti. For example, \citet{Janssen2026} find that TiO$_2$ is a major cloud component in O-rich atmospheres with solar refractory abundances, while \citet{Lee2025} consider the evaporation of minerals from a sub-Neptune's magma ocean surface and find that SiO$_2$ and MgSiO$_3$ are important condensates close to the surface, while Na$_2$S condenses at higher altitudes. The locations of condensates also depend on vertical and horizontal mixing as well as detailed microphysical processes, while thermal inversions in the upper atmosphere can potentially vaporize high-altitude clouds \citep[e.g.,][]{Lee2025,Janssen2026}.

To explore whether clouds could exist in the atmosphere of TOI-561~b, we used \texttt{FASTCHEM COND} \citep{2024Kitzmann} to compute condensation of outgassed rock vapors for hemisphere-averaged temperature-pressure profiles on the dayside and nightside of the planet. \texttt{FASTCHEM COND} was coupled to \texttt{vaporock} \citep{2023Wolf} to compute the abundance of chemical species outgassed from the magma ocean surface where we considered a bulk silicate Earth mantle with oxygen fugacity of $\Delta$IW=$\mp 2$ as proxies for a reduced and an oxidized mantle, respectively. We use the GCM to compute a dayside mean surface temperature of 2600 K.
We apply this model to case of a 1 100\% H$_2$O atmosphere as an illustrative example. 
\begin{figure*}
            \centering
            \includegraphics[width=\linewidth]{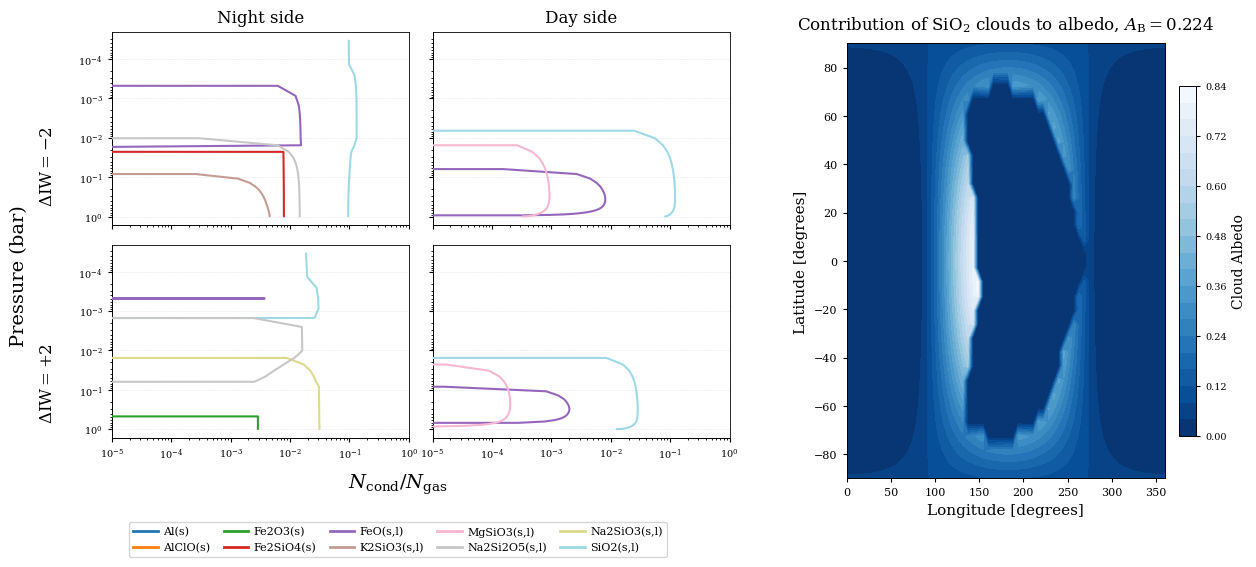}
            \caption{On the left we show condensate mixing ratio profiles for a 1 bar 100 \% H$_2$O atmosphere on TOI-561~b. The left column shows the nightside and the middle column is the dayside. The top row represents a reduced mantle with f$O_2$ $\Delta$IW=$-2$ and the bottom row is an oxidised mantle with f$O_2$ $\Delta$IW=$+2$. On the right we show a representative albedo contribution map of SiO$_2$ clouds based on the H$_2$O $P_s$=1 bar $A_b$=0.0 GCM run. We compute condensation temperatures of SiO$_2$ using the cloud model \texttt{Virga} at the 0.1 bar pressure level. Where the air temperature prediction by the GCM is less than this condensation temperature, we set the albedo to be equal to 0.9. This is then scaled by the shortwave flux at the top of the atmosphere so only cloud regions that receive incoming stellar radiation contribute to the globally averaged albedo of 0.224, thereby cooling the planet, triggering further cloud formation, and further increasing the albedo.}
            \label{fig:toi561b_O2_cloud}
\end{figure*}
As shown in Figure \ref{fig:toi561b_O2_cloud}, multiple possible cloud species can form on the dayside including MgSiO$_3$ and SiO$_2$. In particular, the SiO$_2$ single scattering albedo in the NIRSpec range is very high \citep{Kitzmann2018} and could help to explain the high albedo required by the GCM to match the observed dayside emission. To further investigate whether silicate clouds could explain the observed albedo, we conservatively used the GCM output of the $A_b$=0.0 case to identify regions where the atmospheric temperature drops below the condensation temperature of SiO$_2$. We then recalculate the Bond albedo contribution from SiO$_2$ clouds assuming a single scattering albedo of 0.9 in the NIRSpec wavelength range based on where SiO$_2$ condensates can form and where the planet actually receives incoming shortwave stellar flux and find an albedo of $A_b$=0.24. This demonstrates that even in the most conservative estimate -- where the planet's dayside is hottest -- we can get cloud formation on the morning limb. This, in turn, can help to cool the planet and trigger further cloud formation, further increasing the albedo.

On the nightside of the planet we get an extended cloud deck of SiO$_2$ forming in the reduced case, versus in the oxidized case where the SiO$_2$ deck is constrained to the upper atmosphere with sodium clouds like Na$_2$SiO$_3$ forming in the deeper atmosphere. Depending on the particle-size distribution, these clouds could contribute substantially more opacity at shorter wavelengths, muting the emission in NRS1 while having a weaker effect in NRS2.

Moreover, the presence of highly reflective clouds could be part of a feedback loop aiding the retention of the atmosphere of TOI-561~b. A similar mechanism has been proposed for the hot Neptune LTT~9779~b, whose high geometric albedo ($A_g\sim$0.8) is attributed to reflective silicate clouds that reduce the net stellar flux absorbed by the planet, limiting atmospheric escape \citep{2023Hoyer, 2024Radica, Coulombe2025}. Further exploration of the impact of different cloud species on atmosphere evolution is beyond the scope of this work, but would be a fruitful direction for future studies.

\subsection{Limitations and Future Observations \label{sec:limitations_and_future_obs}}
Our derived stellar granulation parameters are well-constrained for NRS1, giving us confidence that stellar variability has been effectively disentangled from the planetary flux variation. By comparison, the NRS2 stellar granulation parameters are less tightly constrained, meaning that residual stellar granulation signals could still be contributing differences in our results between the detectors. 
In particular, the westward hot-spot offset measured from NRS2 (\S\ref{sec:physical_parameters} and shown Figure \ref{fig:summary}) cannot be explained by our GCM models. While the phase offset values are consistent with zero within 3$\sigma$ (Table \ref{tab:phasecurveresults}), the discrepancy between the two channels is notable and potentially hints at unmodeled stellar granulation in NRS2.

On the other hand, NRS1 is also known to exhibit larger long-timescale detector systematics than NRS2 in G395H observations \citep[e.g.,][]{Espinoza2023,Luque2025,Gordon2026}, likely related to the different properties of the two detectors \citep{Alam2026}. We see in our comparison to GCM simluations that the NRS1 dayside flux is consistent with some of the scenarios, but the measured nightside flux is lower (\S\ref{sec:gcm_results} and Figure \ref{fig:Ab_06_phasecurves}). 
We therefore cannot rule out that residual detector systematics in NRS1 are responsible for part or all of the anomalous nightside flux in NRS1 and offset discrepancy between the two detectors. 

A definitive test to resolve the discrepancies between detectors -- are they due to the planet atmosphere, stellar atmosphere, or instrument? -- will require observations at longer wavelengths. 
Although a MIRI eclipse could corroborate the low dayside flux, a MIRI phase curve would be particularly useful as it would independently constrain the dayside and nightside temperatures and any hot-spot offset, and be more robust against stellar noise (as stellar variability signals are much smaller in amplitude in the mid-IR). Moreover, combining NIRSpec with MIRI LRS would probe complementary pressure levels and molecular bands, placing stronger simultaneous constraints on the thermal structure and composition on the dayside and the nightside.
Future observations with MIRI could also provide powerful additional constraints on the atmospheric composition of TOI-561~b. In particular, the 15 $\mu$m CO$_2$ band falls in the F1500W filter, so mid-infrared phase curves in and out of the CO$_2$ band could distinguish a CO$_2$-rich and CO$_2$-poor atmosphere \citep{Hammond2025}. On the dayside, our updated emission spectrum shows a tentative feature that could be consistent with a CO$_2$ emission feature if the dayside exhibits an inverted T-P profile. On the nightside, where we expect a non-inverted T-P profile, CO$_2$ would appear as an absorption feature further decreasing the nightside flux at 15 $\mu$m. 


\section{Conclusions} \label{sec:conclusions}

In this paper, we present and analyze a phase curve of the ultra-hot super-Earth planet TOI-561~b, covering four full orbits of the planet, taken with JWST NIRSpec/G395H from 3-5~$\mu$m. By analyzing the continuous 37-hr light curve, we simultaneously capture both the day-to-night brightness variation of TOI-561~b and the brightness variation of the host star due to stellar granulation. The duration of the observation is long enough to comprehensively disentangle the granulation signal from the planetary phase curve by fitting both signals simultaneously, and minimize the contamination from stellar variability in our isolated phase curve. Thus, the spectra and phase curves we produce in this paper represent the most robust analysis of this TOI-561~b JWST/NIRSpec data set. 

We find that the dayside and nightside temperatures, a high Bond albedo, and moderate heat redistribution efficiency for TOI-561~b are collectively inconsistent with a bare rock, and most naturally explained by a global atmosphere driving large-scale circulation. This explanation is supported by comparing our phase curves to general circulation models, in which we find that an atmosphere with 1-10 bars of surface pressure is necessary to come close to matching the observed phase curves. In these observations we also obtained a transit spectrum of TOI-561~b, which will be analyzed in detail in an upcoming manuscript to place complementary constraints on the atmospheric scale height, metallicity, and presence of clouds in its atmosphere. The discrepancy in the GCM fits between NRS1 and NRS2 is still somewhat mysterious, and could be due to detector systematics, wavelength-dependent stellar variation, or (more excitingly) wavelength-dependent cloud or composition variation. 

More data, in particular a MIRI phase curve, could help resolve this degenerate explanation by probing in and out of the 15~$\mu$m CO$_2$ band across the full orbit. 

Our confirmation of \citet{EmissionPaper}'s finding of a significant global volatile atmosphere on TOI-561~b still presents a puzzle to our current understanding of planetary evolution. Atmospheric escape theory suggests that strongly irradiated super-Earths should be devoid of any volatile envelopes \citep{Owen2019AREPS,Rogers2021MNRAS}, which is in contrast with our findings. This is particularly astonishing given that the current JWST census of rocky exoplanets has been interpreted as suggestive of a population of bare-rock planets (\citealt{Zhang2024,Xue2024,WeinerMansfield2024,Xue2025,Kreidberg2025}; although we note that most of these planets are hosted by M-dwarfs, while TOI-561 is a late G-/early K-dwarf). If rocky exoplanets at intermediate irradiation are completely desiccated, even-more irradiated exoplanets like TOI-561~b should be, too. A few possible options to explain this mismatch between our observations and theoretical expectations are: (a) The planet formed with an enormous amount of atmospheric volatiles at birth, i.e., it started out as a volatile-enriched sub-Neptune \citep[e.g.][]{Venturini2024A&A,Burn2024NatAs,Nicholls2026NatAstron} (b) Atmospheric escape efficiency of secondary atmospheres (at elevated metallicity) is substantially decreased relative to primary, H$_2$/He dominated atmospheres \citep{Ji2025ApJ,Yoshida2025A&A,Chatterjee2026ApJ}, (c) The planet formed much further outwards and migrated to its current orbit at relatively late times, thus escaping the most aggressive XUV radiation from its host star, or (d) The originally accreted atmospheric volatiles were sequestered into the interior and later outgassed \citep{Dorn2021ApJL,Lichtenberg2025TrGeo}. Why TOI-561~b has an atmosphere may be due to a combination of more than one of these options. 

It is an exciting prospect that constraining the composition of a rocky planet's secondary atmosphere could provide insights into its geochemical evolution (option d), such as the redox state of the magma ocean \citep{nicholls_magma_2024}. The outlined mechanisms above, and the sensitivity of various compounds to be retained in the interior and then degassed, 
offer ways to explain a potential origin of the H$_2$O, CO$_2$, and O$_2$-enriched atmospheric scenarios explored in this work. However, as of yet it is unclear if these scenarios can self-consistently explain the present-day atmosphere. Thus, in upcoming work we will evaluate the coupled interior-atmosphere evolution of TOI-561~b to estimate whether decreased escape efficiency, interior volatile sequestration, orbital migration, other mechanisms, or a combination of these processes can best explain its atmospheric signals. 

\clearpage

\appendix
\counterwithin{figure}{section} 
\counterwithin{table}{section} 
\twocolumngrid
\section{Stellar Signatures Analysis \label{sec:stellar_signals}}

In Weeks at al. (in prep), we show that JWST time-series data are suitable for the detection of oscillations and granulation in Sun-like stars. Here, we validate the approach from \S\ref{sec:stellar_gran_model} by separately analyzing the residuals from a fit to the observations that does not include the GP in \S\ref{sec:stellar_gran_model}, i.e. only the instrumental and planetary components are modeled out.

Similar to above, we fit a Gaussian process model using \texttt{celerite} \citep{celerite} to the \texttt{Eureka!} residuals with a Simple Harmonic Oscillator term, which has a power spectral density of

\begin{equation}
    S(\omega) = \sqrt{\frac{2}{\pi}}
    \frac{S_0\,\omega_0^4}{(\omega^2-\omega_0^2)^2 + \omega_0^2\omega^2 + \omega_0^2\omega^2/Q^2},
\end{equation}

\noindent where $S_{0}$ describes the amplitude, $\omega_0$ the characteristic (angular) frequency, and $Q$ the quality factor. We fix $Q=\frac{1}{\sqrt{2}}$, and set a prior on $S_0$ that corresponds to a prior on the granulation amplitude $a_{gran}$ of $\ln (\mathcal{N}(\sigma_y^{2},10))$, where $\sigma_y$ is the variance of the flux. We use the empirical relation from \citet{Kallinger2014AAP},

\begin{equation}
    \nu_{\mathrm{gran}} = 0.317\nu_{\max}^{0.970},
\end{equation}

\noindent to predict the characteristic (linear) frequency of mesogranulation, where $\nu_{\max}$ is the predicated frequency of solar-like oscillations, given by \citep{brown1991, ulrich1986, KjeldsenBedding1995}
\begin{equation}
    \frac{\nu_{\max}}{\nu_{\max_{\odot}}} = \frac{g / g_{\odot}}{\sqrt{T_{\rm{eff}} / T_{\rm{eff}_{\odot}}}}.
\end{equation}
 Here, $g$ is the mean stellar density, and $T_{\rm{eff}}$ the effective temperature, which we adopt from \citet{lacedelli}, and find a predicted mesogranulation frequency of
 926 $\mu$Hz. This empirical prediction is made assuming no metallicity dependence, however TOI-561 is relatively metal poor in comparison to the Sun, which these relations are scaled to. Therefore, we set a normal prior on $\omega_0$ around this value with a standard deviation of $ln(10)$. We use the \texttt{PyMC3} Maximum A Posteriori optimiser \citep{pymc2023} to generate initial guesses for the sampling, fitting values for the mean and white noise of the data before fitting the granulation component. We then use these values to generate posterior distributions with the the No-U Turn Sampler from \texttt{PyMC3}. We sample the models for 2000 tuning and 5000 draw steps across two chains. 
\begin{figure}[ht]
    \centering
    \includegraphics[width=\linewidth]{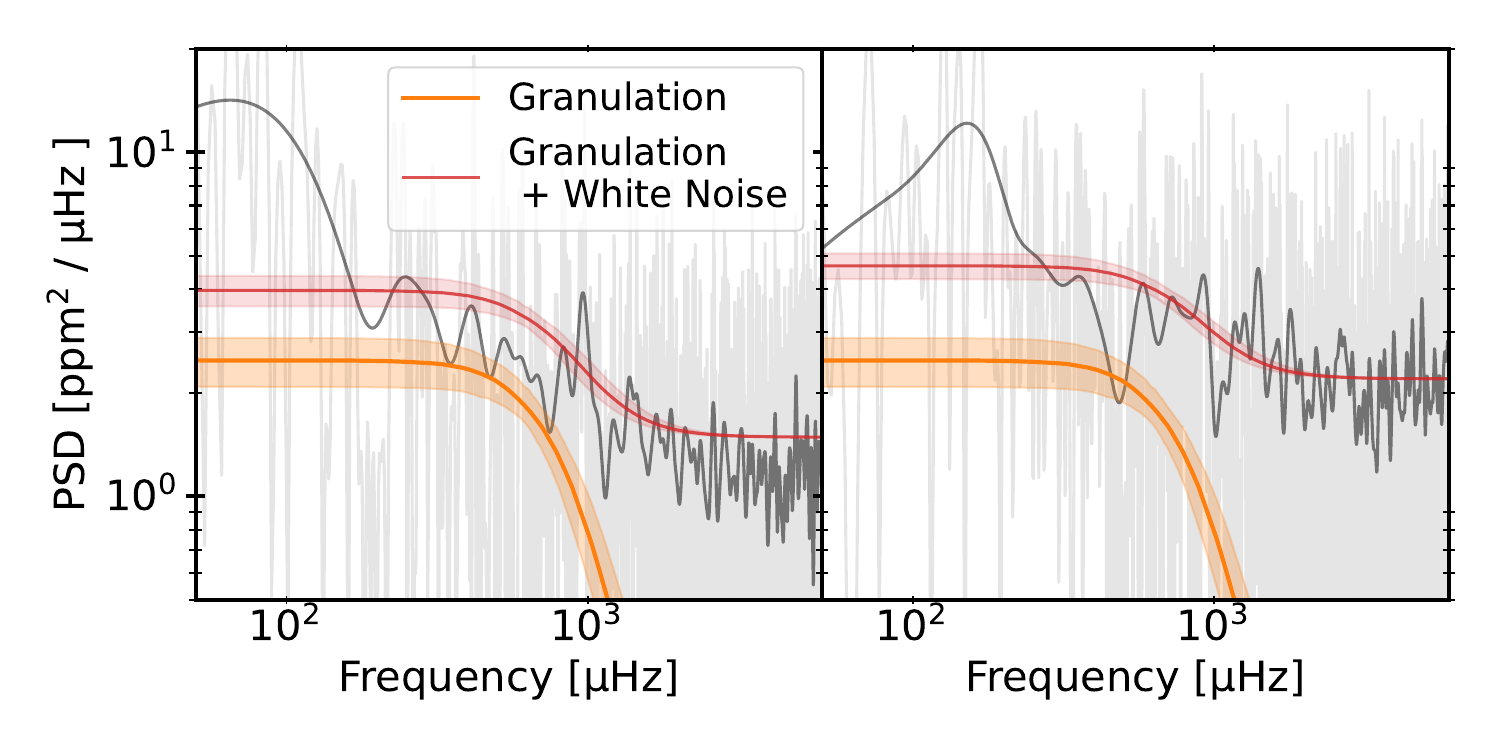}
    \caption{Gaussian process fits for TOI-561 (colored lines), with the power density spectrum of data from NRS1 (left panel) and NRS2 (right panel) in grey; a smoothed version is also overplotted in black. The plots show the Power Spectral Density 
    manifestation of the GP in frequency space, with the granulation component shown in orange, and the granulation with white noise included shown in red. The shaded regions show the 68$\%$ confidence of the sampled fits.}
    \label{fig:gp_results_sampled_PSD}
\end{figure}

\begin{deluxetable}{cccc}[ht]
\tablecaption{Granulation parameters measured for TOI-561, in both NRS1 and NRS2.\label{tab:granulation}}
\tablehead{
  \multicolumn{2}{c}{$\nu_{char}$ [$\mu$Hz]} &
  \multicolumn{2}{c}{$a_{gran}$ [ppm]} \\
  \colhead{NRS1} &
  \colhead{NRS2} &
  \colhead{NRS1} &
  \colhead{NRS2}
}
\startdata
$828.3 \pm 111.8$ & $842.4 \pm 213.2$ & $60.18 \pm 3.39$ & $54.4 \pm 4.70$ \\
\enddata
\end{deluxetable}

The results from our GP fits to the phase curve residuals are shown in Table \ref{tab:granulation} and Figure \ref{fig:gp_results_sampled_PSD}. We observe the expected red noise from granulation in both detectors. 
Reassuringly, we find that our results are consistent with those found in \S\ref{sec:stellar_gran_model} (where the GP is included in the full model), and the model converges on a frequency that is consistent with the predicted frequency from \citet{Kallinger2014AAP}. 
In addition, we calculate the predicted amplitude of the granulation signal with
\begin{equation}
    A_{\rm{gran}} \propto \frac{R}{M^{0.75}}
\end{equation}
\citep{Kallinger2014AAP}. Scaling this to the mesogranulation signal of the sun gives 
\begin{equation}
    A_{\rm{gran}} = A_{\rm{gran}_{\odot}}\left(\frac{(R/R_{\odot})}{(M/M_{\odot})^{0.75}}\right)
\end{equation}

\noindent where $A_{\rm{gran}_{\odot}}$ = 41.6 ppm \citep{Kallinger2014AAP}, and $M_{\odot}$ and $R_{\odot}$ are the solar mass and radius. Adopting stellar parameters from \citep{lacedelli} gives a predicted amplitude of mesogranulation in TOI-561 of 40.6 ppm. This is slightly lower than the amplitudes in Table \ref{tab:granulation}, likely a result of the intrinsic scatter in the scaling relations, and the different amplitude measurement technique used. The predicted amplitude is in a range that is nevertheless consistent with the finding that semi-periodic signals in the residual data are the result of stellar mesogranulation.

We obtain a slightly smaller granulation amplitude in NRS2 than in NRS1, in agreement with observations in Weeks et al. (in prep), who recommend that the mesogranulation for detectors with different spectral response functions should be fit individually in order to determine the amplitude of the signal in each wavelength band. Indeed, the granulation amplitude is expected to decrease with wavelength \citep{Lund2019mnras}, because the effects of temperature gradients should have less impact on our observations at longer wavelengths due to the shape of the Planck function.

\section{Consistency between the reductions \label{sec:A_reduction_consistency}}

\FloatBarrier
To ensure robustness, we fit the NRS1 and NRS2 white-light curves from both the \texttt{Eureka!} and \texttt{ExoTIC JEDI} reductions. Reassuringly,
our results do not significantly differ, as shown in Table \ref{tab:bestfit_eureka_vs_jedi}. Indeed, the most notable discrepancies are between the NRS1 granulation timescales ($\tau_{gran}$ in Table \ref{tab:bestfit_eureka_vs_jedi}) and the second spectroscopic bin dayside fluxes (see solid diamonds and triangles in Figure \ref{fig:comparison_with_emission_paper_spectrum}). However, the discrepancies are smaller than 1-$\sigma$. 
\onecolumngrid
\begin{deluxetable*}{cccccc}[b!]
\tablecaption{Results and priors for the WLCs fits. We assume a circular, tidally locked orbit and fix the period, inclination and eccentricity to the values from J. A. Patel et al. (2023). }\label{tab:bestfit_eureka_vs_jedi}
\tablehead{
  \colhead{Parameter} &
  \colhead{Prior} &
  \multicolumn{2}{c}{\texttt{Eureka!}} &
  \multicolumn{2}{c}{\texttt{ExoTiC JEDI}} \\
  \colhead{} &
  \colhead{} &
  \colhead{NRS1} &
  \colhead{NRS2} &
  \colhead{NRS1} &
  \colhead{NRS2}
}
\startdata
$t_0$ & $\mathcal{U}(0.3498,0.3798)$ days & $0.36579^{+0.00010}_{-0.00010}$ & $0.36579^{+0.00010}_{-0.00010}$ & $0.365788^{+0.000094}_{-0.000090}$ & $0.365788^{+0.000094}_{-0.000090}$ \\
\hline
$R_P/R_*$ & $\mathcal{U}(0.00,0.05)$ & $0.01474^{+0.00038}_{-0.00039}$ & $0.01454^{+0.00042}_{-0.00042}$ &  $0.01477^{+0.00040}_{-0.00039}$ & $0.01447^{+0.00040}_{-0.00041}$ \\
$F_{day}/F_*$ & $\mathcal{U}(1,500)$ ppm & $35^{+10}_{-12}$ & $47^{+12}_{-12}$ &  $36^{+11}_{-13}$ & $53^{+11}_{-11}$ \\
$A$ & $0 \leq \Phi_p$ & $0.439^{+0.039}_{-0.078}$ & $0.15^{+0.12}_{-0.15}$ &  $0.432^{+0.042}_{-0.088}$ & $0.21^{+0.10}_{-0.11}$ \\
$B$ & $0 \leq \Phi_p$ & $-0.14^{+0.12}_{-0.14}$ & $0.15^{+0.14}_{-0.11}$ &  $-0.16^{+0.12}_{-0.15}$ & $0.159^{+0.100}_{-0.087}$ \\
$\sigma_{jitter}$ & $\mathcal{U}(0,300)$ ppm & $153.9^{+1.2}_{-1.3}$ & $225.9^{+1.7}_{-1.8}$  &  $96.6^{+1.9}_{-2.0}$ & $116.4^{+2.8}_{-2.8} $ \\
$a_{gran}$ & $\mathcal{U}(1,500)$ ppm & $61.9^{+3.7}_{-3.4}$ & $54.6^{+4.9}_{-5.0}$ &  $58.2^{+3.9}_{-3.7}$ & $52.5^{+4.3}_{-4.3}$  \\
$\tau_{gran}$ & $\mathcal{U}(0.5,100)$ minutes & $21.8^{+3.4}_{-2.7}$ & $25.4^{+13.9}_{-6.7}$ & $29.3^{+5.7}_{-4.8}$ & $25.4^{+7.7}_{-5.1}$ \\
$c_0$ & - &  $1.001122^{+0.000014}_{-0.000014}$ & $1.000094^{+0.000019}_{-0.000019}$ &  $1.001057^{+0.000016}_{-0.000015}$ & $1.000041^{+0.000019}_{-0.000018}$ \\
$c_1$ & - & $-0.002068^{+0.000075}_{-0.000071}$ & $-0.00102^{+0.00016}_{-0.00016}$ &  $-0.001882^{+0.000077}_{-0.000077}$ & $-0.00065^{+0.00014}_{-0.00015}$\\
$c_2$ & - & $0.00105^{+0.00011}_{-0.00011}$ & $0.00249^{+0.00040}_{-0.00042}$ &  $0.00088^{+0.00012}_{-0.00011}$ & $0.00164^{+0.00038}_{-0.00037}$ \\
$c_3$ & - & $-0.000330^{+0.000045}_{-0.000044}$ & $-0.00225^{+0.00040}_{-0.00039}$ &  $-0.000271^{+0.000047}_{-0.000049}$ & $-0.00149^{+0.00036}_{-0.00036}$  \\
$c_4$ & - & - & $0.00068^{+0.00013}_{-0.00013}$ & - & $0.00046^{+0.00012}_{-0.00011}$ \\
\enddata 
\end{deluxetable*}
\twocolumngrid

\section{Dayside Spectrum Comparison with Eclipse-only analysis \label{sec:A_comparison_with_eclipse_only}}

\begin{figure}[h]
        \centering
        \includegraphics[width=0.96\columnwidth]{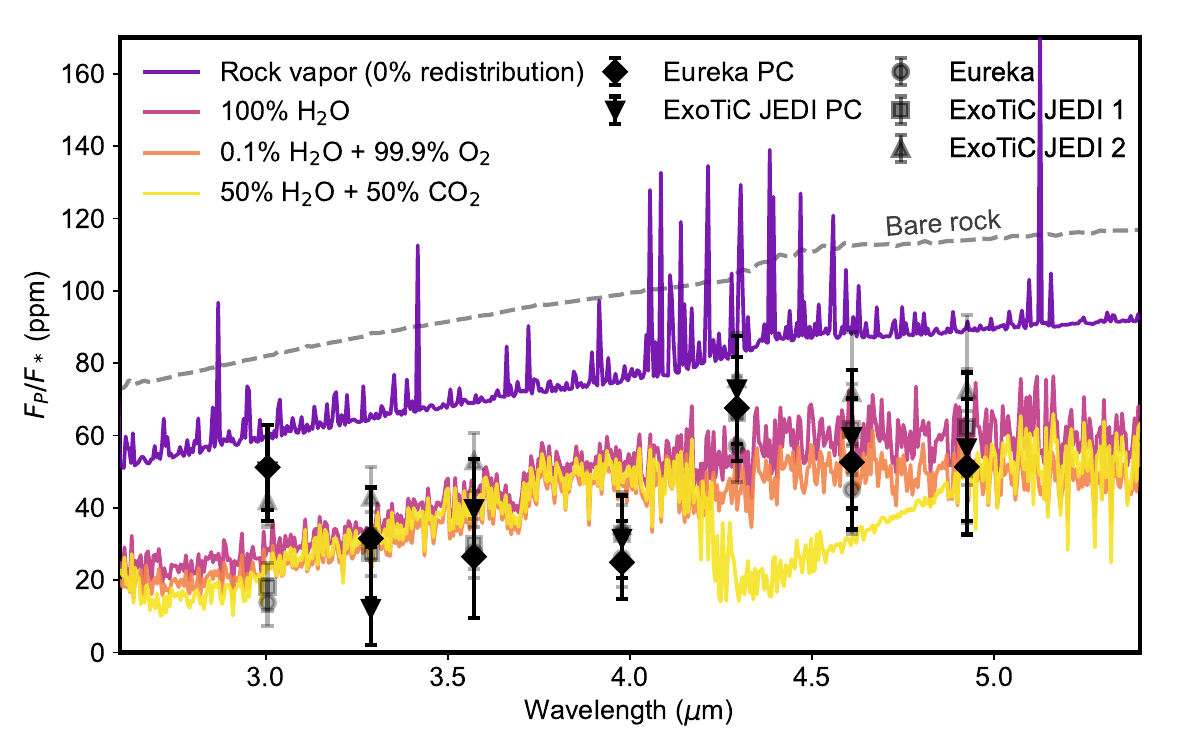}
        \caption{Adaptation of the emission spectrum plotted in \cite{EmissionPaper}.}
        \label{fig:comparison_with_emission_paper_spectrum}
        \vspace{-5mm}
\end{figure}

We produce a variation of Figure 2 from \cite{EmissionPaper} in Figure \ref{fig:comparison_with_emission_paper_spectrum} to show that we obtain consistent results in this work, fitting the full phase curve and including a stellar granulation component (solid diamonds and triangles, versus lighter symbols from \cite{EmissionPaper}). These low eclipse depths corroborate our previous claim that TOI-561~b hosts a thick volatile atmosphere. 

\section{Full GCM Experiment} \label{sec:A_full_gcm_experiment}

To set the opacity parameters of the grey gas model used in the GCM simulations, a fit of the analytic Guillot grey model to \texttt{GENESIS} real-gas column radiative-convective model results was carried out. These fits were done reproducing conditions used in the \texttt{GENESIS} simulations, with the aim of determining grey opacities which mimic the real gas results; it is only the resulting opacities that are used in the GCM. The fits were done over the full 100 bar depth of the \texttt{GENESIS} atmosphere, though the GCM runs were carried out for surface pressures of 10 bars or less. Given that the deep atmosphere participates little in the energy budget of the upper atmosphere, this has little effect on the most appropriate opacities used in the GCM. The fit was carried out assuming an equilibrium temperature (T$_{eq}$) of 2310~K similar to that found in \citep{lacedelli21} for the the 100\% H$_2$O and 50\% H$_2$O + 50\% CO$_2$ case and a slightly lower 1900~K. for the 99.9\% O$_2$ + 0.1\% H$_2$O  case. In all cases an intrinsic temperature (T$_{int}$) of 150~K is assumed, but at the high temperatures of lava planets, intrinsic temperatures of this magnitude are of little consequence over the depth of atmosphere we have considered and do not greatly change the best fitting grey opacities. 
\begin{figure}[b!]
            \centering
            \includegraphics[width=\columnwidth]{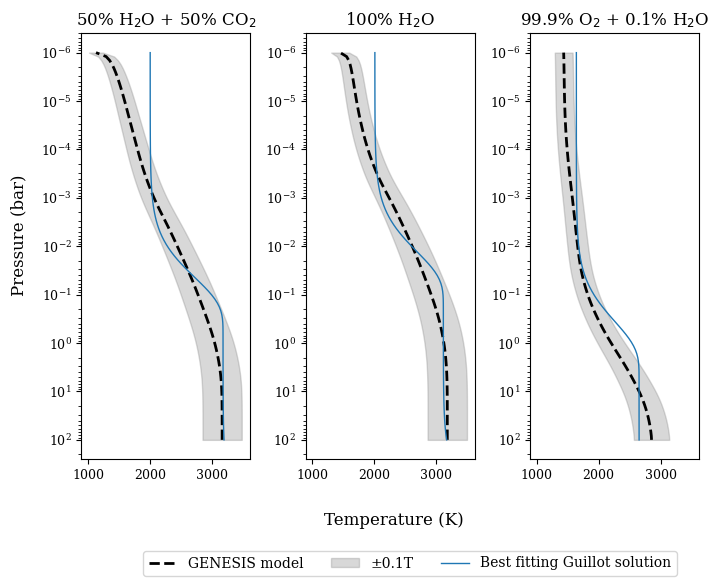}
            \caption{Best fitting of the Guillot solutions to GENESIS real gas computations for the stated compositions. The grey shaded regions indicate the error of the fit, and demarcate the region with under 10\% error relative to the GENESIS model. }
            \label{fig:GuillotFits}
\end{figure}
In Figure \ref{fig:GuillotFits} we show the best fits of the analytical solution. For most pressures, the analytical Guillot solution is in good agreement with the \texttt{GENESIS} prediction. In the upper atmosphere, for pressures less than 1 mbar, the isothermal layer is too hot in comparison to the \texttt{GENESIS} model, but this thin portion of the atmosphere has too little mass to transport much heat, and has a limited contribution to emission. The O$_2$-dominated case fit has an upper atmosphere isothermal layer with similar temperatures to the GENESIS model. A thick atmosphere can lower the amplitude of the phase curve because a deeper atmosphere has greater heat capacity and can redistribute more energy for a given wind speed. To demonstrate the difference in atmospheric circulation regime between a 10- and 1-bar atmosphere, we show the wind vectors and temperature maps in the upper atmosphere (at $P=$1 mbar) and the deep atmosphere (at $P=$0.1 bar) for the pure H$_2$O case in Figure \ref{fig:gcm-windvectors-temp}. As shown, the day-to-night temperature contrast is weaker in the $P_s=$10-bar case than the $P_s=$1 bar case. 

\begin{figure}[h!]
            \centering
            \includegraphics[width=\columnwidth]{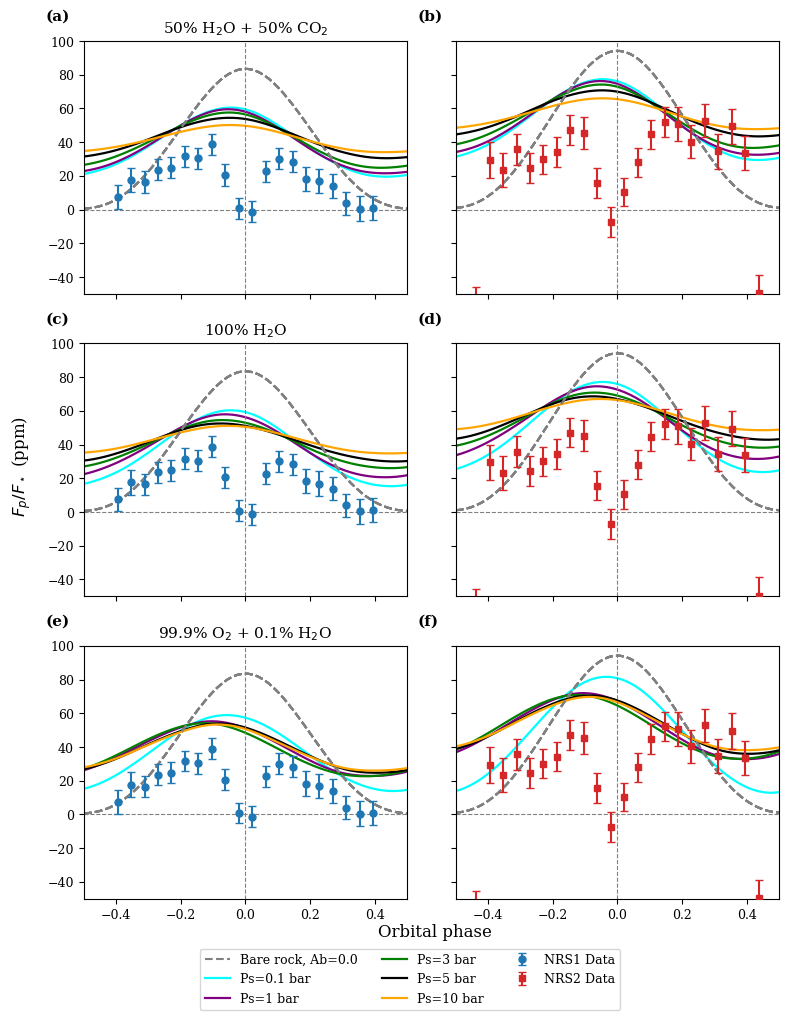}
            \caption{Phase curves for $A_b$=0 cases. The left hand column shows the NRS1 phase curves with each row representing a different composition, the middle column is the NRS2 phase curves. Within each phase curve we show phase curves for $P_s$ = [0.1,1,3,5,10] bar and a dark bare rock case in dashed grey line.}
            \label{fig:Ab_0_phasecurves}
\end{figure}

\begin{figure}
            \centering
            \includegraphics[width=1.\columnwidth]{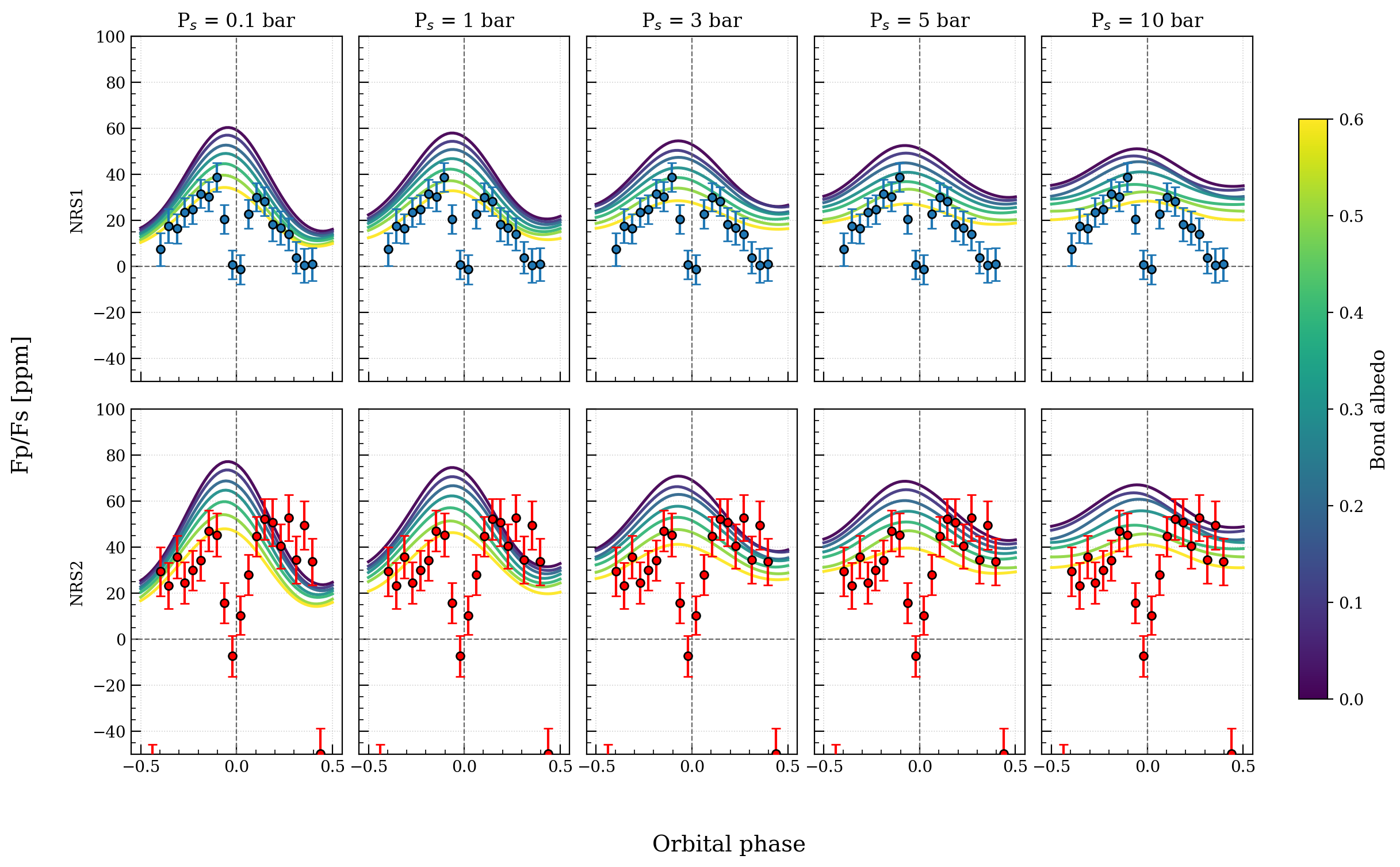}
            \includegraphics[width=1.\columnwidth]{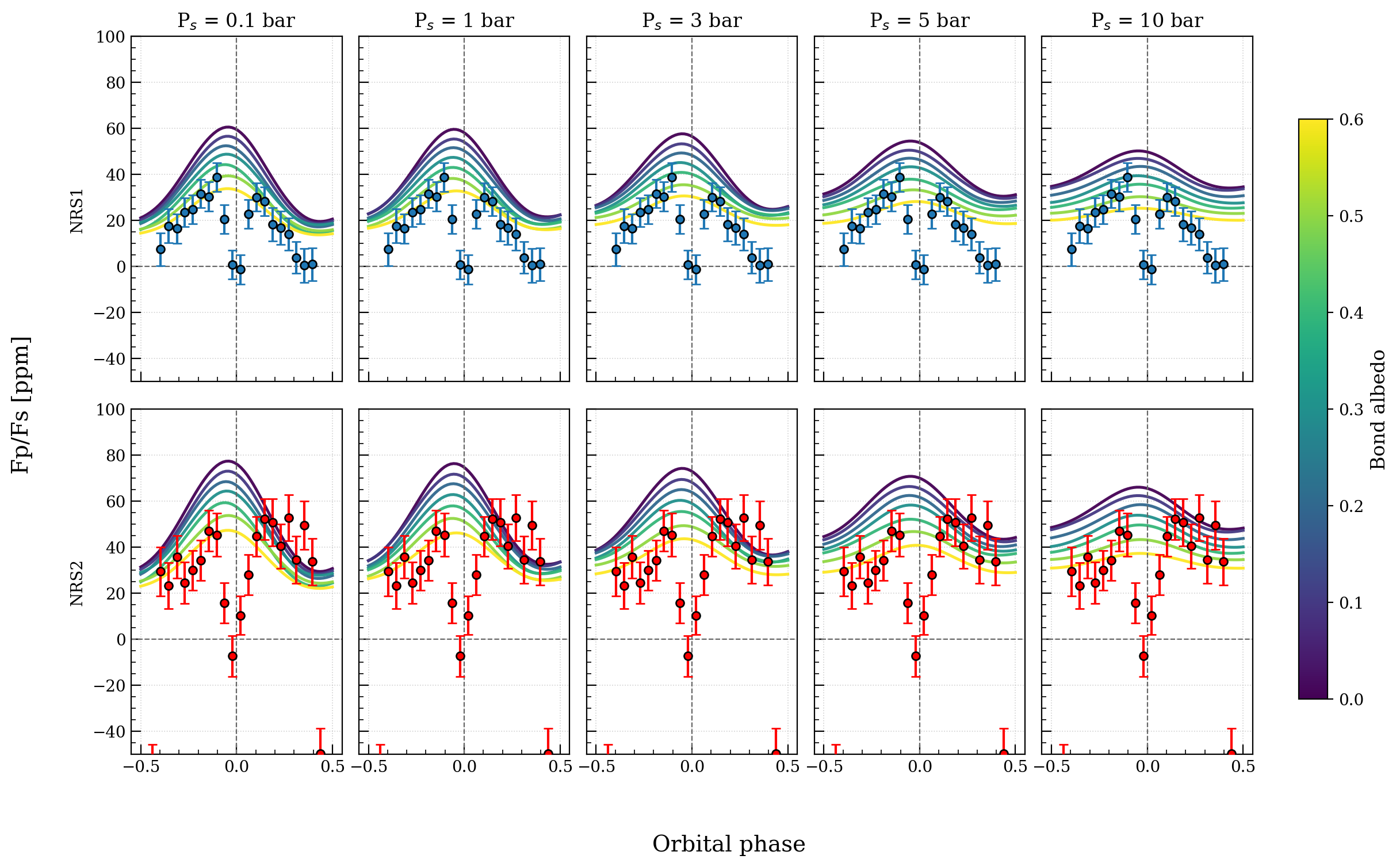}
            \includegraphics[width=1.\columnwidth]{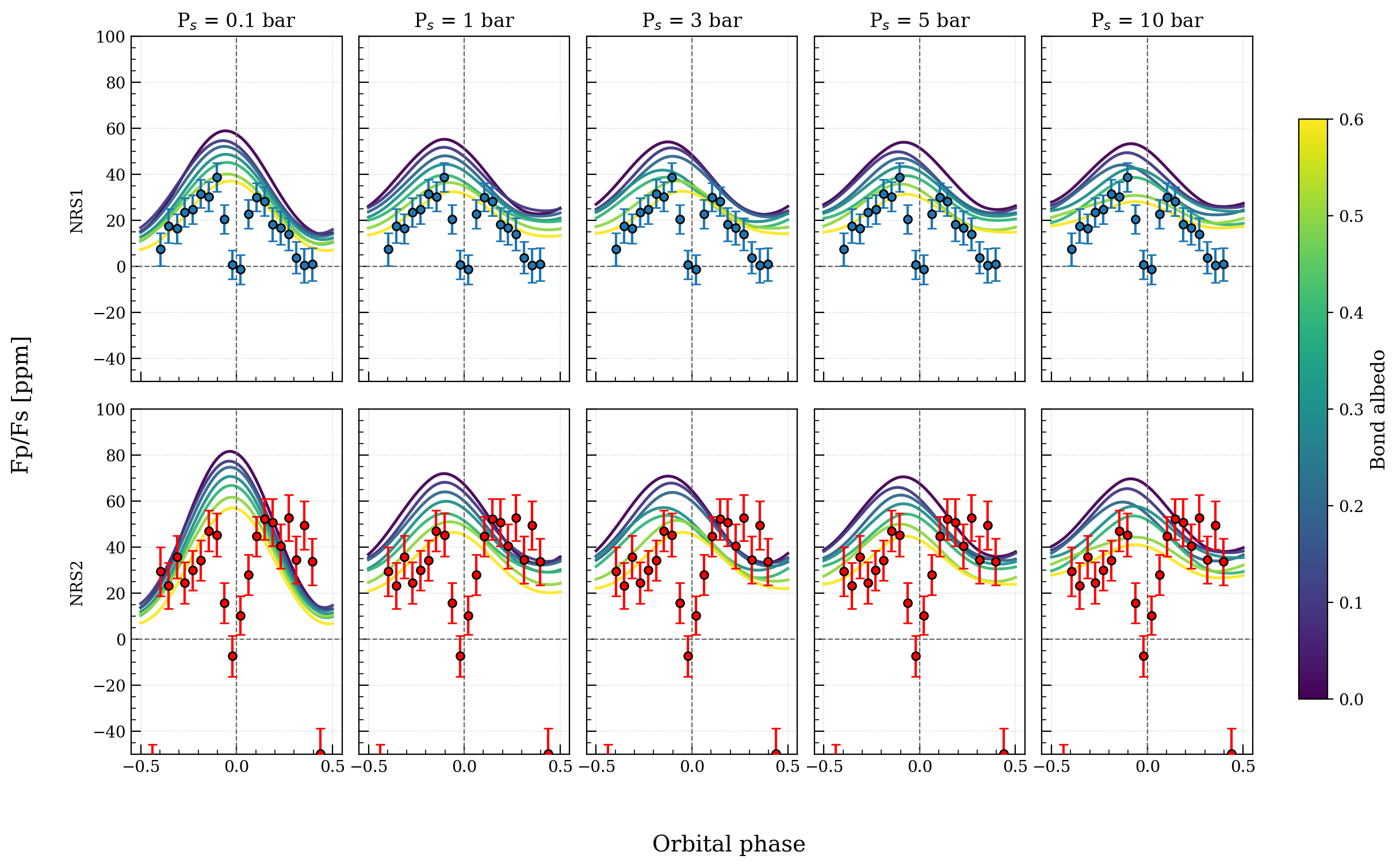}
            \caption{Varying Bond albedo phase curves for a 100\% H$_2$O (top panel), a 50\% CO$_2$ + 50\% H$_2$O (middle panel), a 99.9\% O$_2$ + 0.1\% H$_2$O (bottom panel),  TOI-561~b atmosphere. The top row of each panel shows the phase curves in the NRS1 band compared to the NRS1 observed phase curve (blue) and the bottom in the NRS2 band compared to the NRS1 observed phase curve (red). From left to right the surface pressure increases from 0.1 to 10 bar. The colour of each phase curve represents the Bond albedo used in the GCM.}
            \label{fig:all_comp_phasecurves}
\end{figure}

\begin{figure}
            \centering
            \includegraphics[width=\columnwidth]{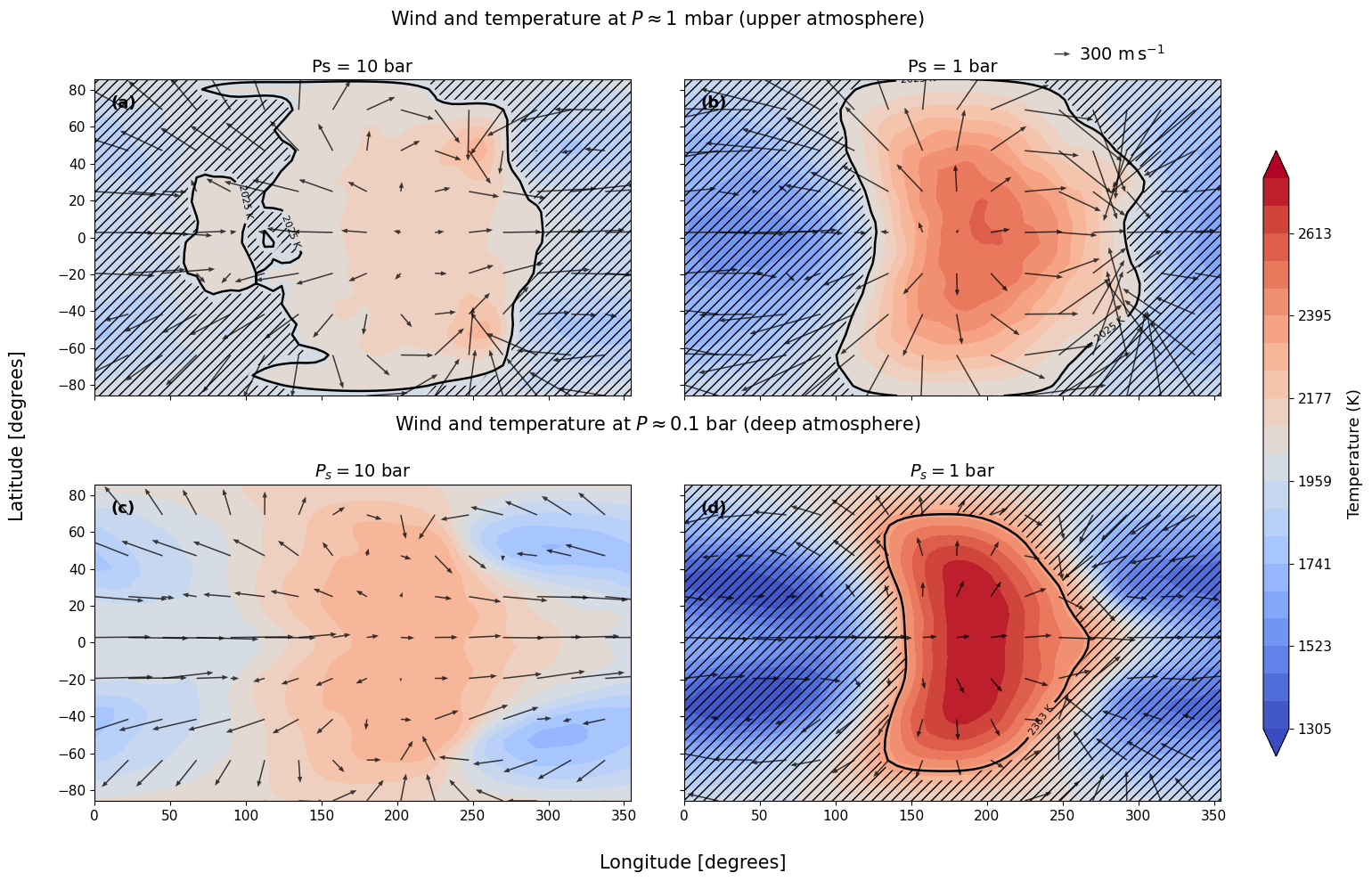}
            \caption{Wind vectors and temperature maps for the 10- and 1- bar surface pressure pure H$_2$O atmospheres at the 1 mbar pressure level (top row) and 0.1 bar (bottom row). The dashed regions represents parts of the planet where the air temperature is colder than the condensation temperature of SiO$_2$ for metallicity of 161 times solar. This was based on \texttt{VapoRock} calculations of 10\% rock vapor atmosphere with a reduced mantle ($\Delta IW=$-2).}
            \label{fig:gcm-windvectors-temp}
\end{figure}

As shown in Figure \ref{fig:Ab_0_phasecurves}, our GCM simulations of $A_B=0$ atmospheres is inconsistent with our observed phase curves, for all the composition and pressured explored in this GCM experiment. We include varying Bond albedo phase curves for three different compositions : 100\% H$_2$O, 50\% CO$_2$ + 50\% H$_2$O and 99.9\% O$_2$ + 0.1\% H$_2$O (see figures \ref{fig:all_comp_phasecurves}). All the GCM phase curves seem to underpredict the nightside flux in NRS1 and high albedo GCM best match the observed NRS1 and NRS2 phase curves.  



\begin{acknowledgments}

This research has made use of the NASA Exoplanet Archive, which is operated by the California Institute of Technology, under contract with the National Aeronautics and Space Administration under the Exoplanet Exploration Program.

This work is based on observations made with the NASA/ESA/CSA James Webb Space Telescope. The data were obtained from the Mikulski Archive for Space Telescopes at the Space Telescope Science Institute, which is operated by the Association of Universities for Research in Astronomy, Inc., under NASA contract NAS 5-03127 for JWST. These observations are associated with program 3860. Support for program 3860 was provided by NASA through a grant from the Space Telescope Science Institute, which is operated by the Association of Universities for Research in Astronomy, Inc., under NASA contract NAS5-03127. Support for program 3860 was provided by the Canadian Space Agency under contract 23JWGO2B06. This work has been partially funded by the Natural Sciences and Engineering Research Council of Canada (grant RGPIN-2021-02706). We would like to acknowledge that our work was performed on land traditionally inhabited by the Wendat, the Anishnaabeg, Haudenosaunee, Metis, and the Mississaugas of the New Credit First Nation.
T.L. was supported by the Branco Weiss Foundation, the Alfred P. Sloan Foundation (AEThER project, G202114194), NASA's Nexus for Exoplanet System Science research coordination network (Alien Earths project, 80NSSC21K0593), and the European Research Council (ERC) under the European Union's Horizon Europe research and innovation programme (101219807, MagmaWorlds).
J.T., T.L., A.P., N.W., and R.P. thank the AEThER project, funded by the Alfred P. Sloan Foundation (G202114194), for the opportunity to discuss ideas related to this manuscript. R.P. and A. M. additionally received support from the U.K. Science and Technology Facilities Council consolidated grant ST/W000903/1. A.M. and R.P. would like to acknowledge the use of the University of Oxford Advanced Research Computing (ARC) facility in carrying out this work. https://doi.org/10.5281/zenodo.22558 
L.D. and S.B. acknowledge support from the Natural Sciences and Engineering Research Council (NSERC), the Trottier Family Foundation and the Waterloo Centre for Astrophysics.
A.P. acknowledges funding from a UK Science and Technology Facilities Council (STFC) Small Award, grant number UKRI/ST/B001171/1, and the Alfred P. Sloan Foundation AEThER project, grant number G202114194.
S.B. thanks L.D. for their guidance during and following his undergraduate research internship with the Trottier Institute for Research on Exoplanets (IREX). S.B. also acknowledges the use of large language model tools (Microsoft Copilot and ChatGPT) for assistance with coding tasks (e.g., debugging and figure production) and as a learning aid for methods and underlying theory involved in this work.
H.N. acknowledges support from STFC grant UKRI1184.
A.W., D.H. and T.R.B. are supported by the Australian Research Council (DP 250102562).

\textbf{Author Contribution} S.B. performed the simultaneous fits of the phase curve, systematics and stellar variability and contributed heavily in the writing of the manuscript. L.D. participated in the conceptualization of the project, provided initial code for phase curve and systematic model fitting, residuals diagnostics, and brightness temperature calculation, contributed to the paper writing, and general advising of S.B. A.M. led the theoretical atmospheric dynamics and cloud modelling and wrote the associated description in the manuscript. J.T. oversaw the entire program and contributed heavily to the writing of this manuscript. A.W. contributed the stellar variability analysis to validate our approach to stellar variability modelling. N.W. provided the reduction of the JWST data. R.P, A.P, N.T contributed to the interpretation of the results. All co-authors have read and provided feedback on the manuscript.

\end{acknowledgments}

\bibliography{references}{}

@ARTICLE{Molliere2019,
       author = {{Molli{\`e}re}, P. and {Wardenier}, J.~P. and {van Boekel}, R. and {Henning}, Th. and {Molaverdikhani}, K. and {Snellen}, I.~A.~G.},
        title = "{petitRADTRANS. A Python radiative transfer package for exoplanet characterization and retrieval}",
      journal = {\aap},
         year = 2019,
        month = jul,
       volume = {627},
          eid = {A67},
        pages = {A67},
          doi = {10.1051/0004-6361/201935470},
archivePrefix = {arXiv},
       eprint = {1904.11504},
 primaryClass = {astro-ph.EP},
       adsurl = {https://ui.adsabs.harvard.edu/abs/2019A&A...627A..67M}
}

@ARTICLE{Lee2025,
       author = {{Lee}, Elspeth K.~H. and {Werlen}, Aaron and {Dorn}, Caroline},
        title = "{Mineral Cloud Formation above Magma Oceans in Sub-Neptune Atmospheres}",
      journal = {\apjl},
         year = 2025,
        month = sep,
       volume = {990},
       number = {2},
          eid = {L43},
        pages = {L43},
          doi = {10.3847/2041-8213/adfe62},
archivePrefix = {arXiv},
       eprint = {2508.15097},
 primaryClass = {astro-ph.EP},
       adsurl = {https://ui.adsabs.harvard.edu/abs/2025ApJ...990L..43L}
}

@INPROCEEDINGS{Harvey1985,
       author = {{Harvey}, J.},
        title = "{High-Resolution Helioseismology}",
    booktitle = {Future Missions in Solar, Heliospheric \& Space Plasma Physics},
         year = 1985,
       editor = {{Rolfe}, Erica and {Battrick}, Bruce},
       series = {ESA Special Publication},
       volume = {235},
        month = jun,
        pages = {199},
       adsurl = {https://ui.adsabs.harvard.edu/abs/1985ESASP.235..199H}
}

@article{pymc2023,
  title = {{PyMC}: A Modern and Comprehensive Probabilistic Programming Framework in {P}ython},
  author = {Oriol Abril-Pla and Virgile Andreani and Colin Carroll and Larry Dong and Christopher J. Fonnesbeck and Maxim Kochurov and Ravin Kumar and Junpeng Lao and Christian C. Luhmann and Osvaldo A. Martin and Michael Osthege and Ricardo Vieira and Thomas Wiecki and Robert Zinkov },
  journal = {{PeerJ} Computer Science},
  volume = {9},
  number = {e1516},
  doi = {10.7717/peerj-cs.1516},
  year = {2023}
}

@book{pierrehumbert2010principles,
  title={Principles of planetary climate},
  author={Pierrehumbert, Raymond T},
  year={2010},
  publisher={Cambridge University Press}
}

@article{nicholls_magma_2024,
	author = {Nicholls, Harrison and Lichtenberg, Tim and Bower, Dan J. and Pierrehumbert, Raymond},
	title = {{Magma Ocean Evolution at Arbitrary Redox State}},
	journal = {J. Geophys. Res. Planets},
	volume = {129},
	number = {12},
	pages = {e2024JE008576},
	year = {2024},
	month = dec,
	issn = {2169-9097},
	publisher = {John Wiley {\&} Sons, Ltd},
	doi = {10.1029/2024JE008576}
}

@ARTICLE{Janssen2026,
       author = {{Janssen}, L.~J. and {Miguel}, Y. and {Min}, M. and {Huang}, H. and {Zilinskas}, M. and {van Buchem}, C.~P.~A.},
        title = "{Hot and cloudy: high temperature clouds in super-Earths and sub-Neptunes}",
      journal = {\mnras},
         year = 2026,
        month = mar,
       volume = {546},
       number = {4},
          eid = {stag180},
        pages = {stag180},
          doi = {10.1093/mnras/stag180},
archivePrefix = {arXiv},
       eprint = {2601.15927},
 primaryClass = {astro-ph.EP},
       adsurl = {https://ui.adsabs.harvard.edu/abs/2026MNRAS.546ag180J}
}

@ARTICLE{Pierrehumbert2019,
       author = {{Pierrehumbert}, Raymond T. and {Hammond}, Mark},
        title = "{Atmospheric Circulation of Tide-Locked Exoplanets}",
      journal = {Annual Review of Fluid Mechanics},
         year = 2019,
        month = jan,
       volume = {51},
       number = {1},
        pages = {275-303},
          doi = {10.1146/annurev-fluid-010518-040516},
       adsurl = {https://ui.adsabs.harvard.edu/abs/2019AnRFM..51..275P}
}

@INCOLLECTION{Parmentier2018,
       author = {{Parmentier}, Vivien and {Crossfield}, Ian J.~M.},
        title = "{Exoplanet Phase Curves: Observations and Theory}",
    booktitle = {Handbook of Exoplanets},
         year = 2018,
       editor = {{Deeg}, Hans J. and {Belmonte}, Juan Antonio},
          eid = {116},
        pages = {116},
          doi = {10.1007/978-3-319-55333-7_116},
       adsurl = {https://ui.adsabs.harvard.edu/abs/2018haex.bookE.116P}
}

@TECHREPORT{Alam2026,
       author = {{Alam}, Munazza K. and {Ubeda}, Leonardo and {Lu}, Qinyan ''Apple'' and {Espinoza}, N{\'e}stor and {Nikolov}, Nikolay},
        title = "{Charge Migration and Residual Non-Linearity in NIRSpec BOTS Observations}",
  institution = {STScI},
         year = 2026,
       number = {Technical Report JWST-STScI-009234},
 howpublished = {Technical Report JWST-STScI-009234, 10 pages},
          doi = {10.48550/arXiv.2601.04255},
       adsurl = {https://ui.adsabs.harvard.edu/abs/2026jwst.rept.9234A}
}

@ARTICLE{Espinoza2023,
       author = {{Espinoza}, N{\'e}stor and {{\'U}beda}, Leonardo and {Birkmann}, Stephan M. and {Ferruit}, Pierre and {Valenti}, Jeff A. and {Sing}, David K. and {Rustamkulov}, Zafar and {Regan}, Michael and {Kendrew}, Sarah and {Sabbi}, Elena and {Schlawin}, Everett and {Beatty}, Thomas and {Albert}, Lo{\"\i}c and {Greene}, Thomas P. and {Nikolov}, Nikolay and {Karakla}, Diane and {Keyes}, Charles and {Alves de Oliveira}, Catarina and {B{\"o}ker}, Torsten and {Pena-Guerrero}, Maria and {Giardino}, Giovanna and {Kumari}, Nimisha and {Manjavacas}, Elena and {Proffitt}, Charles and {Rawle}, Timothy},
        title = "{Spectroscopic Time-series Performance of JWST/NIRSpec from Commissioning Observations}",
      journal = {\pasp},
         year = 2023,
        month = jan,
       volume = {135},
       number = {1043},
          eid = {018002},
        pages = {018002},
          doi = {10.1088/1538-3873/aca3d3},
archivePrefix = {arXiv},
       eprint = {2211.01459},
 primaryClass = {astro-ph.EP},
       adsurl = {https://ui.adsabs.harvard.edu/abs/2023PASP..135a8002E}
}

@ARTICLE{Gordon2026,
       author = {{Gordon}, Tyler A. and {Batalha}, Natalie M. and {Batalha}, Natasha E. and {Aguichine}, Artyom and {Gagnebin}, Anna and {Kirk}, James and {L{\'o}pez-Morales}, Mercedes and {Meech}, Annabella and {Scarsdale}, Nicholas and {Teske}, Johanna and {Wallack}, Nicole L. and {Wogan}, Nicholas},
        title = "{JWST COMPASS: Insights into the Systematic Noise Properties of NIRSpec/G395H from a Uniform Reanalysis of Seven Transmission Spectra}",
      journal = {\aj},
         year = 2026,
        month = mar,
       volume = {171},
       number = {3},
          eid = {178},
        pages = {178},
          doi = {10.3847/1538-3881/ae3de9},
archivePrefix = {arXiv},
       eprint = {2511.18196},
 primaryClass = {astro-ph.EP},
       adsurl = {https://ui.adsabs.harvard.edu/abs/2026AJ....171..178G}
}

@ARTICLE{Kite2021ApJL,
       author = {{Kite}, Edwin S. and {Schaefer}, Laura},
        title = "{Water on Hot Rocky Exoplanets}",
      journal = {\apjl},
         year = 2021,
        month = mar,
       volume = {909},
       number = {2},
          eid = {L22},
        pages = {L22},
          doi = {10.3847/2041-8213/abe7dc},
archivePrefix = {arXiv},
       eprint = {2103.07753},
 primaryClass = {astro-ph.EP},
       adsurl = {https://ui.adsabs.harvard.edu/abs/2021ApJ...909L..22K}
}

@ARTICLE{Dorn2021ApJL,
       author = {{Dorn}, Caroline and {Lichtenberg}, Tim},
        title = "{Hidden Water in Magma Ocean Exoplanets}",
      journal = {\apjl},
         year = 2021,
        month = nov,
       volume = {922},
       number = {1},
          eid = {L4},
        pages = {L4},
          doi = {10.3847/2041-8213/ac33af},
archivePrefix = {arXiv},
       eprint = {2110.15069},
 primaryClass = {astro-ph.EP},
       adsurl = {https://ui.adsabs.harvard.edu/abs/2021ApJ...922L...4D}
}

@ARTICLE{Gupta2025ApJL,
       author = {{Gupta}, Akash and {Stixrude}, Lars and {Schlichting}, Hilke E.},
        title = "{The Miscibility of Hydrogen and Water in Planetary Atmospheres and Interiors}",
      journal = {\apjl},
         year = 2025,
        month = apr,
       volume = {982},
       number = {2},
          eid = {L35},
        pages = {L35},
          doi = {10.3847/2041-8213/adb631},
archivePrefix = {arXiv},
       eprint = {2407.04685},
 primaryClass = {astro-ph.EP},
       adsurl = {https://ui.adsabs.harvard.edu/abs/2025ApJ...982L..35G}
}

@ARTICLE{Lichtenberg2025Sci,
       author = {{Lichtenberg}, Tim and {Shorttle}, Oliver and {Teske}, Johanna and {Kempton}, Eliza M.-R.},
        title = "{Constraining exoplanet interiors using observations of their atmospheres}",
      journal = {Science},
         year = 2025,
        month = oct,
       volume = {390},
       number = {6769},
          eid = {eads3660},
        pages = {eads3660},
          doi = {10.1126/science.ads3360},
archivePrefix = {arXiv},
       eprint = {2510.08844},
 primaryClass = {astro-ph.EP},
       adsurl = {https://ui.adsabs.harvard.edu/abs/2025Sci...390S3660L}
}

@ARTICLE{Nicholls2026NatAstron,
       author = {{Nicholls}, Harrison and {Lichtenberg}, Tim and {Chatterjee}, Richard D. and {Guimond}, Claire Marie and {Postolec}, Emma and {Pierrehumbert}, Raymond T.},
        title = "{Volatile-rich evolution of molten super-Earth L 98-59 d}",
      journal = {Nature Astronomy},
         year = 2026,
        month = mar,
          eid = {arXiv:2507.02656},
        pages = {arXiv:2507.02656},
          doi = {10.1038/s41550-026-02815-8},
archivePrefix = {arXiv},
       eprint = {2507.02656},
 primaryClass = {astro-ph.EP},
       adsurl = {https://ui.adsabs.harvard.edu/abs/2025arXiv250702656N}
}

@ARTICLE{Meier2023A&A,
       author = {{Meier}, Tobias G. and {Bower}, Dan J. and {Lichtenberg}, Tim and {Hammond}, Mark and {Tackley}, Paul J.},
        title = "{Interior dynamics of super-Earth 55 Cancri e}",
      journal = {\aap},
         year = 2023,
        month = oct,
       volume = {678},
          eid = {A29},
        pages = {A29},
          doi = {10.1051/0004-6361/202346950},
archivePrefix = {arXiv},
       eprint = {2308.00592},
 primaryClass = {astro-ph.EP},
       adsurl = {https://ui.adsabs.harvard.edu/abs/2023A&A...678A..29M}
}

@ARTICLE{Meier2026MNRAS,
       author = {{Meier}, Tobias G. and {Guimond}, Claire Marie and {Pierrehumbert}, Raymond T. and {Birkby}, Jayne and {Chatterjee}, Richard D. and {Fisher}, Chloe E. and {Golabek}, Gregor J. and {Hammond}, Mark and {Komacek}, Thaddeus D. and {Lichtenberg}, Tim and {McGinty}, Alex and {Vald{\'e}s}, Erik Meier and {Nicholls}, Harrison and {Parker}, Luke T. and {Spaargaren}, Rob J. and {Tackley}, Paul J.},
        title = "{Mantle Convection and Nightside Volcanism on Lava World K2-141 b}",
      journal = {\mnras},
         year = 2026,
        month = feb,
          doi = {10.1093/mnras/stag390},
archivePrefix = {arXiv},
       eprint = {2603.02408},
 primaryClass = {astro-ph.EP},
       adsurl = {https://ui.adsabs.harvard.edu/abs/2026MNRAS.tmp..377M}
}

@MISC{Bushouse2022,
       author = {{Bushouse}, Howard and {Eisenhamer}, Jonathan and {Dencheva}, Nadia and {Davies}, James and {Greenfield}, Perry and {Morrison}, Jane and {Hodge}, Phil and {Simon}, Bernie and {Grumm}, David and {Droettboom}, Michael and {Slavich}, Edward and {Sosey}, Megan and {Pauly}, Tyler and {Miller}, Todd and {Jedrzejewski}, Robert and {Hack}, Warren and {Davis}, David and {Crawford}, Steven and {Law}, David and {Gordon}, Karl and {Regan}, Michael and {Cara}, Mihai and {MacDonald}, Ken and {Bradley}, Larry and {Shanahan}, Clare and {Jamieson}, William and {Teodoro}, Mairan and {Williams}, Thomas},
        title = "{JWST Calibration Pipeline}",
 howpublished = {Zenodo},
         year = 2022,
        month = oct,
          eid = {10.5281/zenodo.7229890},
          doi = {10.5281/zenodo.7229890},
      version = {1.8.2},
    publisher = {Zenodo},
       adsurl = {https://ui.adsabs.harvard.edu/abs/2022zndo...7229890B}
}

@ARTICLE{Bell2022,
       author = {{Bell}, Taylor and {Ahrer}, Eva-Maria and {Brande}, Jonathan and {Carter}, Aarynn and {Feinstein}, Adina and {Caloca}, Giannina and {Mansfield}, Megan and {Zieba}, Sebastian and {Piaulet}, Caroline and {Benneke}, Bj{\"o}rn and {Filippazzo}, Joseph and {May}, Erin and {Roy}, Pierre-Alexis and {Kreidberg}, Laura and {Stevenson}, Kevin},
        title = "{Eureka!: An End-to-End Pipeline for JWST Time-Series Observations}",
      journal = {The Journal of Open Source Software},
         year = 2022,
        month = nov,
       volume = {7},
       number = {79},
          eid = {4503},
        pages = {4503},
          doi = {10.21105/joss.04503},
archivePrefix = {arXiv},
       eprint = {2207.03585},
 primaryClass = {astro-ph.IM},
       adsurl = {https://ui.adsabs.harvard.edu/abs/2022JOSS....7.4503B}
}

@ARTICLE{2019Pluriel,
       author = {{Pluriel}, W. and {Marcq}, E. and {Turbet}, M.},
        title = "{Modeling the albedo of Earth-like magma ocean planets with H$_{2}$O-CO$_{2}$ atmospheres}",
      journal = {\icarus},
         year = 2019,
        month = jan,
       volume = {317},
        pages = {583-590},
          doi = {10.1016/j.icarus.2018.08.023},
archivePrefix = {arXiv},
       eprint = {1809.02036},
 primaryClass = {astro-ph.EP},
       adsurl = {https://ui.adsabs.harvard.edu/abs/2019Icar..317..583P}
}

@ARTICLE{2024Kitzmann,
       author = {{Kitzmann}, Daniel and {Stock}, Joachim W. and {Patzer}, A. Beate C.},
        title = "{FASTCHEM COND: equilibrium chemistry with condensation and rainout for cool planetary and stellar environments}",
      journal = {\mnras},
         year = 2024,
        month = jan,
       volume = {527},
       number = {3},
        pages = {7263-7283},
          doi = {10.1093/mnras/stad3515},
archivePrefix = {arXiv},
       eprint = {2309.02337},
 primaryClass = {astro-ph.EP},
       adsurl = {https://ui.adsabs.harvard.edu/abs/2024MNRAS.527.7263K}
}

@ARTICLE{2023Wolf,
       author = {{Wolf}, Aaron S. and {J{\"a}ggi}, Noah and {Sossi}, Paolo A. and {Bower}, Dan J.},
        title = "{VapoRock: Thermodynamics of Vaporized Silicate Melts for Modeling Volcanic Outgassing and Magma Ocean Atmospheres}",
      journal = {\apj},
         year = 2023,
        month = apr,
       volume = {947},
       number = {2},
          eid = {64},
        pages = {64},
          doi = {10.3847/1538-4357/acbcc7},
archivePrefix = {arXiv},
       eprint = {2208.09582},
 primaryClass = {astro-ph.EP},
       adsurl = {https://ui.adsabs.harvard.edu/abs/2023ApJ...947...64W}
}

@ARTICLE{2023Hoyer,
       author = {{Hoyer}, S. and {Jenkins}, J.~S. and {Parmentier}, V. and {Deleuil}, M. and {Scandariato}, G. and {Wilson}, T.~G. and {D{\'\i}az}, M.~R. and {Crossfield}, I.~J.~M. and {Dragomir}, D. and {Kataria}, T. and {Lendl}, M. and {Ramirez}, R. and {Pe{\~n}a Rojas}, P.~A. and {Vin{\'e}s}, J.~I.},
        title = "{The extremely high albedo of LTT 9779 b revealed by CHEOPS. An ultrahot Neptune with a highly metallic atmosphere}",
      journal = {\aap},
         year = 2023,
        month = jul,
       volume = {675},
          eid = {A81},
        pages = {A81},
          doi = {10.1051/0004-6361/202346117},
       adsurl = {https://ui.adsabs.harvard.edu/abs/2023A&A...675A..81H}
}

@ARTICLE{2024Radica,
       author = {{Radica}, Michael and {Coulombe}, Louis-Philippe and {Taylor}, Jake and {Albert}, Loic and {Allart}, Romain and {Benneke}, Bj{\"o}rn and {Cowan}, Nicolas B. and {Dang}, Lisa and {Lafreni{\`e}re}, David and {Thorngren}, Daniel and {Artigau}, {\'E}tienne and {Doyon}, Ren{\'e} and {Flagg}, Laura and {Johnstone}, Doug and {Pelletier}, Stefan and {Roy}, Pierre-Alexis},
        title = "{Muted Features in the JWST NIRISS Transmission Spectrum of Hot Neptune LTT 9779b}",
      journal = {\apjl},
         year = 2024,
        month = feb,
       volume = {962},
       number = {1},
          eid = {L20},
        pages = {L20},
          doi = {10.3847/2041-8213/ad20e4},
archivePrefix = {arXiv},
       eprint = {2401.15548},
 primaryClass = {astro-ph.EP},
       adsurl = {https://ui.adsabs.harvard.edu/abs/2024ApJ...962L..20R}
}

@ARTICLE{2020Essack,
       author = {{Essack}, Zahra and {Seager}, Sara and {Pajusalu}, Mihkel},
        title = "{Low-albedo Surfaces of Lava Worlds}",
      journal = {\apj},
         year = 2020,
        month = aug,
       volume = {898},
       number = {2},
          eid = {160},
        pages = {160},
          doi = {10.3847/1538-4357/ab9cba},
archivePrefix = {arXiv},
       eprint = {2008.02789},
 primaryClass = {astro-ph.EP},
       adsurl = {https://ui.adsabs.harvard.edu/abs/2020ApJ...898..160E}
}

@MISC{Alderson2022JEDI,
       author = {{Alderson}, Lili and {Grant}, David and {Wakeford}, Hannah},
        title = "{Exo-TiC/ExoTiC-JEDI: v0.1-beta-release}",
 howpublished = {Zenodo},
         year = 2022,
        month = oct,
          eid = {10.5281/zenodo.7185855},
          doi = {10.5281/zenodo.7185855},
      version = {v0.1},
    publisher = {Zenodo},
       adsurl = {https://ui.adsabs.harvard.edu/abs/2022zndo...7185855A}
}

@ARTICLE{Luque2025,
       author = {{Luque}, Rafael and {Coy}, Brandon Park and {Xue}, Qiao and {Feinstein}, Adina D. and {Ahrer}, Eva-Maria and {Changeat}, Quentin and {Zhang}, Michael and {Moran}, Sarah E. and {Bean}, Jacob L. and {Kite}, Edwin and {Weiner Mansfield}, Megan and {Pall{\'e}}, Enric},
        title = "{A Dark, Bare Rock for TOI-1685 b from a JWST NIRSpec G395H Phase Curve}",
      journal = {\aj},
         year = 2025,
        month = jul,
       volume = {170},
       number = {1},
          eid = {49},
        pages = {49},
          doi = {10.3847/1538-3881/addb40},
archivePrefix = {arXiv},
       eprint = {2412.03411},
 primaryClass = {astro-ph.EP},
       adsurl = {https://ui.adsabs.harvard.edu/abs/2025AJ....170...49L}
}

@ARTICLE{Alderson2024,
       author = {{Alderson}, Lili and {Batalha}, Natasha E. and {Wakeford}, Hannah R. and {Wallack}, Nicole L. and {Aguichine}, Artyom and {Teske}, Johanna and {Adams Redai}, Jea and {Alam}, Munazza K. and {Batalha}, Natalie M. and {Gao}, Peter and {Kirk}, James and {L{\'o}pez-Morales}, Mercedes and {Moran}, Sarah E. and {Scarsdale}, Nicholas and {Wogan}, Nicholas F. and {Wolfgang}, Angie},
        title = "{JWST COMPASS: NIRSpec/G395H Transmission Observations of the Super-Earth TOI-836b}",
      journal = {\aj},
         year = 2024,
        month = may,
       volume = {167},
       number = {5},
          eid = {216},
        pages = {216},
          doi = {10.3847/1538-3881/ad32c9},
archivePrefix = {arXiv},
       eprint = {2404.00093},
 primaryClass = {astro-ph.EP},
       adsurl = {https://ui.adsabs.harvard.edu/abs/2024AJ....167..216A}
}

@ARTICLE{Moran2023,
       author = {{Moran}, Sarah E. and {Stevenson}, Kevin B. and {Sing}, David K. and {MacDonald}, Ryan J. and {Kirk}, James and {Lustig-Yaeger}, Jacob and {Peacock}, Sarah and {Mayorga}, L.~C. and {Bennett}, Katherine A. and {L{\'o}pez-Morales}, Mercedes and {May}, E.~M. and {Rustamkulov}, Zafar and {Valenti}, Jeff A. and {Adams Redai}, J{\'e}a I. and {Alam}, Munazza K. and {Batalha}, Natasha E. and {Fu}, Guangwei and {Gonzalez-Quiles}, Junellie and {Highland}, Alicia N. and {Kruse}, Ethan and {Lothringer}, Joshua D. and {Ortiz Ceballos}, Kevin N. and {Sotzen}, Kristin S. and {Wakeford}, Hannah R.},
        title = "{High Tide or Riptide on the Cosmic Shoreline? A Water-rich Atmosphere or Stellar Contamination for the Warm Super-Earth GJ 486b from JWST Observations}",
      journal = {\apjl},
         year = 2023,
        month = may,
       volume = {948},
       number = {1},
          eid = {L11},
        pages = {L11},
          doi = {10.3847/2041-8213/accb9c},
archivePrefix = {arXiv},
       eprint = {2305.00868},
 primaryClass = {astro-ph.EP},
       adsurl = {https://ui.adsabs.harvard.edu/abs/2023ApJ...948L..11M}
}

@ARTICLE{Alderson2025,
       author = {{Alderson}, Lili and {Moran}, Sarah E. and {Wallack}, Nicole L. and {Batalha}, Natasha E. and {Wogan}, Nicholas F. and {Dattilo}, Anne and {Wakeford}, Hannah R. and {Redai}, Jea Adams and {Alam}, Munazza K. and {Aguichine}, Artyom and {Batalha}, Natalie M. and {Gagnebin}, Anna and {Gao}, Peter and {Kirk}, James and {L{\'o}pez-Morales}, Mercedes and {Meech}, Annabella and {Teske}, Johanna and {Wolfgang}, Angie},
        title = "{JWST COMPASS: NIRSpec/G395H Transmission Observations of the Super-Earth TOI-776 b}",
      journal = {\aj},
         year = 2025,
        month = mar,
       volume = {169},
       number = {3},
          eid = {142},
        pages = {142},
          doi = {10.3847/1538-3881/adad64},
archivePrefix = {arXiv},
       eprint = {2501.14596},
 primaryClass = {astro-ph.EP},
       adsurl = {https://ui.adsabs.harvard.edu/abs/2025AJ....169..142A}
}

@ARTICLE{Alderson2022,
       author = {{Alderson}, Lili and {Wakeford}, Hannah R. and {Alam}, Munazza K. and {Batalha}, Natasha E. and {Lothringer}, Joshua D. and {Adams Redai}, Jea and {Barat}, Saugata and {Brande}, Jonathan and {Damiano}, Mario and {Daylan}, Tansu and {Espinoza}, N{\'e}stor and {Flagg}, Laura and {Goyal}, Jayesh M. and {Grant}, David and {Hu}, Renyu and {Inglis}, Julie and {Lee}, Elspeth K.~H. and {Mikal-Evans}, Thomas and {Ramos-Rosado}, Lakeisha and {Roy}, Pierre-Alexis and {Wallack}, Nicole L. and {Batalha}, Natalie M. and {Bean}, Jacob L. and {Benneke}, Bj{\"o}rn and {Berta-Thompson}, Zachory K. and {Carter}, Aarynn L. and {Changeat}, Quentin and {Col{\'o}n}, Knicole D. and {Crossfield}, Ian J.~M. and {D{\'e}sert}, Jean-Michel and {Foreman-Mackey}, Daniel and {Gibson}, Neale P. and {Kreidberg}, Laura and {Line}, Michael R. and {L{\'o}pez-Morales}, Mercedes and {Molaverdikhani}, Karan and {Moran}, Sarah E. and {Morello}, Giuseppe and {Moses}, Julianne I. and {Mukherjee}, Sagnick and {Schlawin}, Everett and {Sing}, David K. and {Stevenson}, Kevin B. and {Taylor}, Jake and {Aggarwal}, Keshav and {Ahrer}, Eva-Maria and {Allen}, Natalie H. and {Barstow}, Joanna K. and {Bell}, Taylor J. and {Blecic}, Jasmina and {Casewell}, Sarah L. and {Chubb}, Katy L. and {Crouzet}, Nicolas and {Cubillos}, Patricio E. and {Decin}, Leen and {Feinstein}, Adina D. and {Fortney}, Joanthan J. and {Harrington}, Joseph and {Heng}, Kevin and {Iro}, Nicolas and {Kempton}, Eliza M. -R. and {Kirk}, James and {Knutson}, Heather A. and {Krick}, Jessica and {Leconte}, J{\'e}r{\'e}my and {Lendl}, Monika and {MacDonald}, Ryan J. and {Mancini}, Luigi and {Mansfield}, Megan and {May}, Erin M. and {Mayne}, Nathan J. and {Miguel}, Yamila and {Nikolov}, Nikolay K. and {Ohno}, Kazumasa and {Palle}, Enric and {Parmentier}, Vivien and {Petit dit de la Roche}, Dominique J.~M. and {Piaulet}, Caroline and {Powell}, Diana and {Rackham}, Benjamin V. and {Redfield}, Seth and {Rogers}, Laura K. and {Rustamkulov}, Zafar and {Tan}, Xianyu and {Tremblin}, P. and {Tsai}, Shang-Min and {Turner}, Jake D. and {de Val-Borro}, Miguel and {Venot}, Olivia and {Welbanks}, Luis and {Wheatley}, Peter J. and {Zhang}, Xi},
        title = "{Early Release Science of the Exoplanet WASP-39b with JWST NIRSpec G395H}",
      journal = {arXiv e-prints},
         year = 2022,
        month = nov,
          eid = {arXiv:2211.10488},
        pages = {arXiv:2211.10488},
          doi = {10.48550/arXiv.2211.10488},
archivePrefix = {arXiv},
       eprint = {2211.10488},
 primaryClass = {astro-ph.EP},
       adsurl = {https://ui.adsabs.harvard.edu/abs/2022arXiv221110488A}
}

@ARTICLE{Wallack2024,
       author = {{Wallack}, Nicole L. and {Batalha}, Natasha E. and {Alderson}, Lili and {Scarsdale}, Nicholas and {Adams Redai}, Jea I. and {Aguichine}, Artyom and {Alam}, Munazza K. and {Gao}, Peter and {Wolfgang}, Angie and {Batalha}, Natalie M. and {Kirk}, James and {L{\'o}pez-Morales}, Mercedes and {Moran}, Sarah E. and {Teske}, Johanna and {Wakeford}, Hannah R. and {Wogan}, Nicholas F.},
        title = "{JWST COMPASS: A NIRSpec/G395H Transmission Spectrum of the Sub-Neptune TOI-836c}",
      journal = {\aj},
         year = 2024,
        month = aug,
       volume = {168},
       number = {2},
          eid = {77},
        pages = {77},
          doi = {10.3847/1538-3881/ad3917},
archivePrefix = {arXiv},
       eprint = {2404.01264},
 primaryClass = {astro-ph.EP},
       adsurl = {https://ui.adsabs.harvard.edu/abs/2024AJ....168...77W}
}

@ARTICLE{Coulombe2025,
       author = {{Coulombe}, Louis-Philippe and {Radica}, Michael and {Benneke}, Bj{\"o}rn and {D'Aoust}, {\'E}lyse and {Dang}, Lisa and {Cowan}, Nicolas B. and {Parmentier}, Vivien and {Albert}, Lo{\"\i}c and {Lafreni{\`e}re}, David and {Taylor}, Jake and {Roy}, Pierre-Alexis and {Pelletier}, Stefan and {Allart}, Romain and {Artigau}, {\'E}tienne and {Doyon}, Ren{\'e} and {Jayawardhana}, Ray and {Johnstone}, Doug and {Kaltenegger}, Lisa and {Langeveld}, Adam B. and {MacDonald}, Ryan J. and {Rowe}, Jason F. and {Turner}, Jake D.},
        title = "{Highly reflective white clouds on the western dayside of an exo-Neptune}",
      journal = {Nature Astronomy},
         year = 2025,
        month = apr,
       volume = {9},
        pages = {512-525},
          doi = {10.1038/s41550-025-02488-9},
archivePrefix = {arXiv},
       eprint = {2501.14016},
 primaryClass = {astro-ph.EP},
       adsurl = {https://ui.adsabs.harvard.edu/abs/2025NatAs...9..512C}
}

@ARTICLE{Crossfield2020,
       author = {{Crossfield}, Ian J.~M. and {Dragomir}, Diana and {Cowan}, Nicolas B. and {Daylan}, Tansu and {Wong}, Ian and {Kataria}, Tiffany and {Deming}, Drake and {Kreidberg}, Laura and {Mikal-Evans}, Thomas and {Gorjian}, Varoujan and {Jenkins}, James S. and {Benneke}, Bj{\"o}rn and {Collins}, Karen A. and {Burke}, Christopher J. and {Henze}, Christopher E. and {McDermott}, Scott and {Mireles}, Ismael and {Watanabe}, David and {Wohler}, Bill and {Ricker}, George and {Vanderspek}, Roland and {Seager}, Sara and {Jenkins}, Jon M.},
        title = "{Phase Curves of Hot Neptune LTT 9779b Suggest a High-metallicity Atmosphere}",
      journal = {\apjl},
         year = 2020,
        month = nov,
       volume = {903},
       number = {1},
          eid = {L7},
        pages = {L7},
          doi = {10.3847/2041-8213/abbc71},
archivePrefix = {arXiv},
       eprint = {2010.12745},
 primaryClass = {astro-ph.EP},
       adsurl = {https://ui.adsabs.harvard.edu/abs/2020ApJ...903L...7C}
}

@ARTICLE{Kite2016,
       author = {{Kite}, Edwin S. and {Fegley}, Jr., Bruce and {Schaefer}, Laura and {Gaidos}, Eric},
        title = "{Atmosphere-interior Exchange on Hot, Rocky Exoplanets}",
      journal = {\apj},
         year = 2016,
        month = sep,
       volume = {828},
       number = {2},
          eid = {80},
        pages = {80},
          doi = {10.3847/0004-637X/828/2/80},
archivePrefix = {arXiv},
       eprint = {1606.06740},
 primaryClass = {astro-ph.EP},
       adsurl = {https://ui.adsabs.harvard.edu/abs/2016ApJ...828...80K}
}

@ARTICLE{Kite2020,
       author = {{Kite}, Edwin S. and {Barnett}, Megan N.},
        title = "{Exoplanet secondary atmosphere loss and revival}",
      journal = {Proceedings of the National Academy of Science},
         year = 2020,
        month = jul,
       volume = {117},
        pages = {18264-18271},
          doi = {10.1073/pnas.2006177117},
archivePrefix = {arXiv},
       eprint = {2006.02589},
 primaryClass = {astro-ph.EP},
       adsurl = {https://ui.adsabs.harvard.edu/abs/2020PNAS..11718264K}
}

@ARTICLE{Maurice2024,
       author = {{Maurice}, M. and {Dasgupta}, R. and {Hassanzadeh}, P.},
        title = "{Volatile atmospheres of lava worlds}",
      journal = {\aap},
         year = 2024,
        month = aug,
       volume = {688},
          eid = {A47},
        pages = {A47},
          doi = {10.1051/0004-6361/202347749},
archivePrefix = {arXiv},
       eprint = {2405.09284},
 primaryClass = {astro-ph.EP},
       adsurl = {https://ui.adsabs.harvard.edu/abs/2024A&A...688A..47M}
}

@ARTICLE{Gillmann2024,
       author = {{Gillmann}, Cedric and {Hakim}, Kaustubh and {Louren{\c{c}}o}, Diogo and {Quanz}, Sascha P. and {Sossi}, Paolo A.},
        title = "{Interior Controls on the Habitability of Rocky Planets}",
      journal = {Space: Science and Technology},
         year = 2024,
        month = feb,
       volume = {4},
          eid = {0075},
        pages = {0075},
          doi = {10.34133/space.0075},
archivePrefix = {arXiv},
       eprint = {2403.17630},
 primaryClass = {astro-ph.EP},
       adsurl = {https://ui.adsabs.harvard.edu/abs/2024SpScT...4...75G}
}

@ARTICLE{Katyal2020,
       author = {{Katyal}, Nisha and {Ortenzi}, Gianluigi and {Lee Grenfell}, John and {Noack}, Lena and {Sohl}, Frank and {Godolt}, Mareike and {Garc{\'\i}a Mu{\~n}oz}, Antonio and {Schreier}, Franz and {Wunderlich}, Fabian and {Rauer}, Heike},
        title = "{Effect of mantle oxidation state and escape upon the evolution of Earth's magma ocean atmosphere}",
      journal = {\aap},
         year = 2020,
        month = nov,
       volume = {643},
          eid = {A81},
        pages = {A81},
          doi = {10.1051/0004-6361/202038779},
archivePrefix = {arXiv},
       eprint = {2009.14599},
 primaryClass = {astro-ph.EP},
       adsurl = {https://ui.adsabs.harvard.edu/abs/2020A&A...643A..81K}
}

@ARTICLE{Baumeister2023,
       author = {{Baumeister}, Philipp and {Tosi}, Nicola and {Brachmann}, Caroline and {Grenfell}, John Lee and {Noack}, Lena},
        title = "{Redox state and interior structure control on the long-term habitability of stagnant-lid planets}",
      journal = {\aap},
         year = 2023,
        month = jul,
       volume = {675},
          eid = {A122},
        pages = {A122},
          doi = {10.1051/0004-6361/202245791},
archivePrefix = {arXiv},
       eprint = {2301.03466},
 primaryClass = {astro-ph.EP},
       adsurl = {https://ui.adsabs.harvard.edu/abs/2023A&A...675A.122B}
}

@ARTICLE{Brinkman2025,
       author = {{Brinkman}, Casey L. and {Weiss}, Lauren M. and {Huber}, Daniel and {Lee}, Rena A. and {Kolecki}, Jared and {Tenn}, Gwyneth and {Zhang}, Jingwen and {Narayanan}, Suchitra and {Polanski}, Alex S. and {Dai}, Fei and {Bean}, Jacob L. and {Beard}, Corey and {Brady}, Madison and {Brodheim}, Max and {Brown}, Matt and {Chontos}, Ashley and {Deich}, William and {Edelstein}, Jerry and {Fulton}, Benjamin J. and {Giacalone}, Steven and {Gibson}, Steven R. and {Gilbert}, Gregory J. and {Halverson}, Samuel and {Handley}, Luke and {Hill}, Grant M. and {Holcomb}, Rae and {Holden}, Bradford and {Householder}, Aaron and {Howard}, Andrew W. and {Isaacson}, Howard and {Kaye}, Stephen and {Laher}, Russ R. and {Lanclos}, Kyle and {Ong}, J.~M. Joel and {Payne}, Joel and {Petigura}, Erik A. and {Pidhorodetska}, Daria and {Poppett}, Claire and {Roy}, Arpita and {Rubenzahl}, Ryan and {Saunders}, Nicholas and {Schwab}, Christian and {Seifahrt}, Andreas and {Shaum}, Abby P. and {Sirk}, Martin M. and {Smith}, Chris and {Smith}, Roger and {Stef{\'a}nsson}, Gu{\dj}mundur and {St{\"u}rmer}, Julian and {Thorne}, Jim and {Turtelboom}, Emma V. and {Tyler}, Dakotah and {Valliant}, John and {Van Zandt}, Judah and {Walawender}, Josh and {Yee}, Samuel W. and {Yeh}, Sherry and {Zink}, Jon},
        title = "{The Compositions of Rocky Planets in Close-in Orbits Tend to Be Earth-like}",
      journal = {\aj},
         year = 2025,
        month = aug,
       volume = {170},
       number = {2},
          eid = {109},
        pages = {109},
          doi = {10.3847/1538-3881/ade677},
archivePrefix = {arXiv},
       eprint = {2410.00213},
 primaryClass = {astro-ph.EP},
       adsurl = {https://ui.adsabs.harvard.edu/abs/2025AJ....170..109B}
}

@ARTICLE{Brinkman2023,
       author = {{Brinkman}, Casey L. and {Weiss}, Lauren M. and {Dai}, Fei and {Huber}, Daniel and {Kite}, Edwin S. and {Valencia}, Diana and {Bean}, Jacob L. and {Beard}, Corey and {Behmard}, Aida and {Blunt}, Sarah and {Brady}, Madison and {Fulton}, Benjamin and {Giacalone}, Steven and {Howard}, Andrew W. and {Isaacson}, Howard and {Kasper}, David and {Lubin}, Jack and {MacDougall}, Mason and {Akana Murphy}, Joseph M. and {Plotnykov}, Mykhaylo and {Polanski}, Alex S. and {Rice}, Malena and {Seifahrt}, Andreas and {Stef{\'a}nsson}, Gu{\dh}mundur and {St{\"u}rmer}, Julian},
        title = "{TOI-561 b: A Low-density Ultra-short-period ``Rocky'' Planet around a Metal-poor Star}",
      journal = {\aj},
         year = 2023,
        month = mar,
       volume = {165},
       number = {3},
          eid = {88},
        pages = {88},
          doi = {10.3847/1538-3881/acad83},
archivePrefix = {arXiv},
       eprint = {2210.06665},
 primaryClass = {astro-ph.EP},
       adsurl = {https://ui.adsabs.harvard.edu/abs/2023AJ....165...88B}
}

@ARTICLE{Dai2021,
       author = {{Dai}, Fei and {Howard}, Andrew W. and {Batalha}, Natalie M. and {Beard}, Corey and {Behmard}, Aida and {Blunt}, Sarah and {Brinkman}, Casey L. and {Chontos}, Ashley and {Crossfield}, Ian J.~M. and {Dalba}, Paul A. and {Dressing}, Courtney and {Fulton}, Benjamin and {Giacalone}, Steven and {Hill}, Michelle L. and {Huber}, Daniel and {Isaacson}, Howard and {Kane}, Stephen R. and {Lubin}, Jack and {Mayo}, Andrew and {Mo{\v{c}}nik}, Teo and {Akana Murphy}, Joseph M. and {Petigura}, Erik A. and {Rice}, Malena and {Robertson}, Paul and {Rosenthal}, Lee and {Roy}, Arpita and {Rubenzahl}, Ryan A. and {Weiss}, Lauren M. and {Zandt}, Judah Van and {Beichman}, Charles and {Ciardi}, David and {Collins}, Karen A. and {Gonzales}, Erica and {Howell}, Steve B. and {Matson}, Rachel A. and {Matthews}, Elisabeth C. and {Schlieder}, Joshua E. and {Schwarz}, Richard P. and {Ricker}, George R. and {Vanderspek}, Roland and {Latham}, David W. and {Seager}, Sara and {Winn}, Joshua N. and {Jenkins}, Jon M. and {Caldwell}, Douglas A. and {Colon}, Knicole D. and {Dragomir}, Diana and {Lund}, Michael B. and {McLean}, Brian and {Rudat}, Alexander and {Shporer}, Avi},
        title = "{TKS X: Confirmation of TOI-1444b and a Comparative Analysis of the Ultra-short-period Planets with Hot Neptunes}",
      journal = {\aj},
         year = 2021,
        month = aug,
       volume = {162},
       number = {2},
          eid = {62},
        pages = {62},
          doi = {10.3847/1538-3881/ac02bd},
archivePrefix = {arXiv},
       eprint = {2105.08844},
 primaryClass = {astro-ph.EP},
       adsurl = {https://ui.adsabs.harvard.edu/abs/2021AJ....162...62D}
}

@ARTICLE{Dai2019,
       author = {{Dai}, Fei and {Masuda}, Kento and {Winn}, Joshua N. and {Zeng}, Li},
        title = "{Homogeneous Analysis of Hot Earths: Masses, Sizes, and Compositions}",
      journal = {\apj},
         year = 2019,
        month = sep,
       volume = {883},
       number = {1},
          eid = {79},
        pages = {79},
          doi = {10.3847/1538-4357/ab3a3b},
archivePrefix = {arXiv},
       eprint = {1908.06299},
 primaryClass = {astro-ph.EP},
       adsurl = {https://ui.adsabs.harvard.edu/abs/2019ApJ...883...79D}
}

@ARTICLE{Hu2024,
       author = {{Hu}, Renyu and {Bello-Arufe}, Aaron and {Zhang}, Michael and {Paragas}, Kimberly and {Zilinskas}, Mantas and {van Buchem}, Christiaan and {Bess}, Michael and {Patel}, Jayshil and {Ito}, Yuichi and {Damiano}, Mario and {Scheucher}, Markus and {Oza}, Apurva V. and {Knutson}, Heather A. and {Miguel}, Yamila and {Dragomir}, Diana and {Brandeker}, Alexis and {Demory}, Brice-Olivier},
        title = "{A secondary atmosphere on the rocky exoplanet 55 Cancri e}",
      journal = {\nat},
         year = 2024,
        month = jun,
       volume = {630},
       number = {8017},
        pages = {609-612},
          doi = {10.1038/s41586-024-07432-x},
archivePrefix = {arXiv},
       eprint = {2405.04744},
 primaryClass = {astro-ph.EP},
       adsurl = {https://ui.adsabs.harvard.edu/abs/2024Natur.630..609H}
}

@ARTICLE{Demory2016,
       author = {{Demory}, Brice-Olivier and {Gillon}, Michael and {de Wit}, Julien and {Madhusudhan}, Nikku and {Bolmont}, Emeline and {Heng}, Kevin and {Kataria}, Tiffany and {Lewis}, Nikole and {Hu}, Renyu and {Krick}, Jessica and {Stamenkovi{\'c}}, Vlada and {Benneke}, Bj{\"o}rn and {Kane}, Stephen and {Queloz}, Didier},
        title = "{A map of the large day-night temperature gradient of a super-Earth exoplanet}",
      journal = {\nat},
         year = 2016,
        month = apr,
       volume = {532},
       number = {7598},
        pages = {207-209},
          doi = {10.1038/nature17169},
archivePrefix = {arXiv},
       eprint = {1604.05725},
 primaryClass = {astro-ph.EP},
       adsurl = {https://ui.adsabs.harvard.edu/abs/2016Natur.532..207D}
}

@ARTICLE{Scarsdale2024,
       author = {{Scarsdale}, Nicholas and {Wogan}, Nicholas and {Wakeford}, Hannah R. and {Wallack}, Nicole L. and {Batalha}, Natasha E. and {Alderson}, Lili and {Aguichine}, Artyom and {Wolfgang}, Angie and {Teske}, Johanna and {Moran}, Sarah E. and {L{\'o}pez-Morales}, Mercedes and {Kirk}, James and {Gordon}, Tyler and {Gao}, Peter and {Batalha}, Natalie M. and {Alam}, Munazza K. and {Adams Redai}, Jea},
        title = "{JWST COMPASS: The 3{\textendash}5 {\ensuremath{\mu}}m Transmission Spectrum of the Super-Earth L 98-59 c}",
      journal = {\aj},
         year = 2024,
        month = dec,
       volume = {168},
       number = {6},
          eid = {276},
        pages = {276},
          doi = {10.3847/1538-3881/ad73cf},
archivePrefix = {arXiv},
       eprint = {2409.07552},
 primaryClass = {astro-ph.EP},
       adsurl = {https://ui.adsabs.harvard.edu/abs/2024AJ....168..276S}
}

@ARTICLE{AdamsRedai2025,
       author = {{Adams Redai}, Jea and {Wogan}, Nicholas and {Wallack}, Nicole L. and {Alam}, Munazza K. and {Aguichine}, Artyom and {Wolfgang}, Angie and {Wakeford}, Hannah R. and {Teske}, Johanna and {Scarsdale}, Nicholas and {Moran}, Sarah E. and {L{\'o}pez-Morales}, Mercedes and {Meech}, Annabella and {Gao}, Peter and {Gagnebin}, Anna and {Batalha}, Natasha E. and {Batalha}, Natalie M. and {Alderson}, Lili},
        title = "{JWST COMPASS: A NIRSpec G395H Transmission Spectrum of the Super-Earth GJ 357 b}",
      journal = {\aj},
         year = 2025,
        month = oct,
       volume = {170},
       number = {4},
          eid = {219},
        pages = {219},
          doi = {10.3847/1538-3881/adee92},
archivePrefix = {arXiv},
       eprint = {2507.07165},
 primaryClass = {astro-ph.EP},
       adsurl = {https://ui.adsabs.harvard.edu/abs/2025AJ....170..219A}
}

@ARTICLE{Alam2025,
       author = {{Alam}, Munazza K. and {Gao}, Peter and {Adams Redai}, Jea and {Wallack}, Nicole L. and {Wogan}, Nicholas F. and {Aguichine}, Artyom and {Dattilo}, Anne and {Alderson}, Lili and {Batalha}, Natasha E. and {Batalha}, Natalie M. and {Kirk}, James and {L{\'o}pez-Morales}, Mercedes and {Meech}, Annabella and {Moran}, Sarah E. and {Teske}, Johanna and {Wakeford}, Hannah R. and {Wolfgang}, Angie},
        title = "{JWST COMPASS: The First Near- to Mid-infrared Transmission Spectrum of the Hot Super-Earth L 168-9 b}",
      journal = {\aj},
         year = 2025,
        month = jan,
       volume = {169},
       number = {1},
          eid = {15},
        pages = {15},
          doi = {10.3847/1538-3881/ad8eb5},
archivePrefix = {arXiv},
       eprint = {2411.03154},
 primaryClass = {astro-ph.EP},
       adsurl = {https://ui.adsabs.harvard.edu/abs/2025AJ....169...15A}
}

@ARTICLE{May2023,
       author = {{May}, E.~M. and {MacDonald}, Ryan J. and {Bennett}, Katherine A. and {Moran}, Sarah E. and {Wakeford}, Hannah R. and {Peacock}, Sarah and {Lustig-Yaeger}, Jacob and {Highland}, Alicia N. and {Stevenson}, Kevin B. and {Sing}, David K. and {Mayorga}, L.~C. and {Batalha}, Natasha E. and {Kirk}, James and {L{\'o}pez-Morales}, Mercedes and {Valenti}, Jeff A. and {Alam}, Munazza K. and {Alderson}, Lili and {Fu}, Guangwei and {Gonzalez-Quiles}, Junellie and {Lothringer}, Joshua D. and {Rustamkulov}, Zafar and {Sotzen}, Kristin S.},
        title = "{Double Trouble: Two Transits of the Super-Earth GJ 1132 b Observed with JWST NIRSpec G395H}",
      journal = {\apjl},
         year = 2023,
        month = dec,
       volume = {959},
       number = {1},
          eid = {L9},
        pages = {L9},
          doi = {10.3847/2041-8213/ad054f},
archivePrefix = {arXiv},
       eprint = {2310.10711},
 primaryClass = {astro-ph.EP},
       adsurl = {https://ui.adsabs.harvard.edu/abs/2023ApJ...959L...9M}
}

@ARTICLE{Lim2023,
       author = {{Lim}, Olivia and {Benneke}, Bj{\"o}rn and {Doyon}, Ren{\'e} and {MacDonald}, Ryan J. and {Piaulet}, Caroline and {Artigau}, {\'E}tienne and {Coulombe}, Louis-Philippe and {Radica}, Michael and {L'Heureux}, Alexandrine and {Albert}, Lo{\"\i}c and {Rackham}, Benjamin V. and {de Wit}, Julien and {Salhi}, Salma and {Roy}, Pierre-Alexis and {Flagg}, Laura and {Fournier-Tondreau}, Marylou and {Taylor}, Jake and {Cook}, Neil J. and {Lafreni{\`e}re}, David and {Cowan}, Nicolas B. and {Kaltenegger}, Lisa and {Rowe}, Jason F. and {Espinoza}, N{\'e}stor and {Dang}, Lisa and {Darveau-Bernier}, Antoine},
        title = "{Atmospheric Reconnaissance of TRAPPIST-1 b with JWST/NIRISS: Evidence for Strong Stellar Contamination in the Transmission Spectra}",
      journal = {\apjl},
         year = 2023,
        month = sep,
       volume = {955},
       number = {1},
          eid = {L22},
        pages = {L22},
          doi = {10.3847/2041-8213/acf7c4},
archivePrefix = {arXiv},
       eprint = {2309.07047},
 primaryClass = {astro-ph.EP},
       adsurl = {https://ui.adsabs.harvard.edu/abs/2023ApJ...955L..22L}
}

@ARTICLE{Zhang2024,
       author = {{Zhang}, Michael and {Hu}, Renyu and {Inglis}, Julie and {Dai}, Fei and {Bean}, Jacob L. and {Knutson}, Heather A. and {Lam}, Kristine and {Goffo}, Elisa and {Gandolfi}, Davide},
        title = "{GJ 367b Is a Dark, Hot, Airless Sub-Earth}",
      journal = {\apjl},
         year = 2024,
        month = feb,
       volume = {961},
       number = {2},
          eid = {L44},
        pages = {L44},
          doi = {10.3847/2041-8213/ad1a07},
archivePrefix = {arXiv},
       eprint = {2401.01400},
 primaryClass = {astro-ph.EP},
       adsurl = {https://ui.adsabs.harvard.edu/abs/2024ApJ...961L..44Z}
}

@ARTICLE{Xue2025,
       author = {{Xue}, Qiao and {Zhang}, Michael and {Coy}, Brandon Park and {Brady}, Madison and {Ji}, Xuan and {Bean}, Jacob L. and {Radica}, Michael and {Seifahrt}, Andreas and {St{\"u}rmer}, Julian and {Luque}, Rafael and {Basant}, Ritvik and {Brown}, Nina and {Das}, Tanya and {Kasper}, David and {Piaulet-Ghorayeb}, Caroline and {Kempton}, Eliza M.-R. and {Kite}, Edwin},
        title = "{The JWST Rocky Worlds DDT Program Reveals GJ 3929b to Likely Be a Bare Rock}",
      journal = {\apjl},
         year = 2025,
        month = dec,
       volume = {995},
       number = {2},
          eid = {L52},
        pages = {L52},
          doi = {10.3847/2041-8213/ae2098},
archivePrefix = {arXiv},
       eprint = {2508.12516},
 primaryClass = {astro-ph.EP},
       adsurl = {https://ui.adsabs.harvard.edu/abs/2025ApJ...995L..52X}
}

@ARTICLE{Xue2024,
       author = {{Xue}, Qiao and {Bean}, Jacob L. and {Zhang}, Michael and {Mahajan}, Alexandra and {Ih}, Jegug and {Eastman}, Jason D. and {Lunine}, Jonathan and {Mansfield}, Megan Weiner and {Coy}, Brandon Park and {Kempton}, Eliza M.-R. and {Koll}, Daniel and {Kite}, Edwin},
        title = "{JWST Thermal Emission of the Terrestrial Exoplanet GJ 1132b}",
      journal = {\apjl},
         year = 2024,
        month = sep,
       volume = {973},
       number = {1},
          eid = {L8},
        pages = {L8},
          doi = {10.3847/2041-8213/ad72e9},
archivePrefix = {arXiv},
       eprint = {2408.13340},
 primaryClass = {astro-ph.EP},
       adsurl = {https://ui.adsabs.harvard.edu/abs/2024ApJ...973L...8X}
}

@ARTICLE{WeinerMansfield2024,
       author = {{Weiner Mansfield}, Megan and {Xue}, Qiao and {Zhang}, Michael and {Mahajan}, Alexandra S. and {Ih}, Jegug and {Koll}, Daniel and {Bean}, Jacob L. and {Coy}, Brandon Park and {Eastman}, Jason D. and {Kempton}, Eliza M.-R. and {Kite}, Edwin S.},
        title = "{No Thick Atmosphere on the Terrestrial Exoplanet Gl 486b}",
      journal = {\apjl},
         year = 2024,
        month = nov,
       volume = {975},
       number = {1},
          eid = {L22},
        pages = {L22},
          doi = {10.3847/2041-8213/ad8161},
archivePrefix = {arXiv},
       eprint = {2408.15123},
 primaryClass = {astro-ph.EP},
       adsurl = {https://ui.adsabs.harvard.edu/abs/2024ApJ...975L..22W}
}

@ARTICLE{Zieba2023,
       author = {{Zieba}, Sebastian and {Kreidberg}, Laura and {Ducrot}, Elsa and {Gillon}, Micha{\"e}l and {Morley}, Caroline and {Schaefer}, Laura and {Tamburo}, Patrick and {Koll}, Daniel D.~B. and {Lyu}, Xintong and {Acu{\~n}a}, Lorena and {Agol}, Eric and {Iyer}, Aishwarya R. and {Hu}, Renyu and {Lincowski}, Andrew P. and {Meadows}, Victoria S. and {Selsis}, Franck and {Bolmont}, Emeline and {Mandell}, Avi M. and {Suissa}, Gabrielle},
        title = "{No thick carbon dioxide atmosphere on the rocky exoplanet TRAPPIST-1 c}",
      journal = {\nat},
         year = 2023,
        month = aug,
       volume = {620},
       number = {7975},
        pages = {746-749},
          doi = {10.1038/s41586-023-06232-z},
archivePrefix = {arXiv},
       eprint = {2306.10150},
 primaryClass = {astro-ph.EP},
       adsurl = {https://ui.adsabs.harvard.edu/abs/2023Natur.620..746Z}
}

@ARTICLE{Greene2023,
       author = {{Greene}, Thomas P. and {Bell}, Taylor J. and {Ducrot}, Elsa and {Dyrek}, Achr{\`e}ne and {Lagage}, Pierre-Olivier and {Fortney}, Jonathan J.},
        title = "{Thermal emission from the Earth-sized exoplanet TRAPPIST-1 b using JWST}",
      journal = {\nat},
         year = 2023,
        month = jun,
       volume = {618},
       number = {7963},
        pages = {39-42},
          doi = {10.1038/s41586-023-05951-7},
archivePrefix = {arXiv},
       eprint = {2303.14849},
 primaryClass = {astro-ph.EP},
       adsurl = {https://ui.adsabs.harvard.edu/abs/2023Natur.618...39G}
}

@ARTICLE{Crossfield2022,
       author = {{Crossfield}, Ian J.~M. and {Malik}, Matej and {Hill}, Michelle L. and {Kane}, Stephen R. and {Foley}, Bradford and {Polanski}, Alex S. and {Coria}, David and {Brande}, Jonathan and {Zhang}, Yanzhe and {Wienke}, Katherine and {Kreidberg}, Laura and {Cowan}, Nicolas B. and {Dragomir}, Diana and {Gorjian}, Varoujan and {Mikal-Evans}, Thomas and {Benneke}, Bj{\"o}rn and {Christiansen}, Jessie L. and {Deming}, Drake and {Morales}, Farisa Y.},
        title = "{GJ 1252b: A Hot Terrestrial Super-Earth with No Atmosphere}",
      journal = {\apjl},
         year = 2022,
        month = sep,
       volume = {937},
       number = {1},
          eid = {L17},
        pages = {L17},
          doi = {10.3847/2041-8213/ac886b},
archivePrefix = {arXiv},
       eprint = {2208.09479},
 primaryClass = {astro-ph.EP},
       adsurl = {https://ui.adsabs.harvard.edu/abs/2022ApJ...937L..17C}
}

@ARTICLE{Kreidberg2019,
       author = {{Kreidberg}, Laura and {Koll}, Daniel D.~B. and {Morley}, Caroline and {Hu}, Renyu and {Schaefer}, Laura and {Deming}, Drake and {Stevenson}, Kevin B. and {Dittmann}, Jason and {Vanderburg}, Andrew and {Berardo}, David and {Guo}, Xueying and {Stassun}, Keivan and {Crossfield}, Ian and {Charbonneau}, David and {Latham}, David W. and {Loeb}, Abraham and {Ricker}, George and {Seager}, Sara and {Vanderspek}, Roland},
        title = "{Absence of a thick atmosphere on the terrestrial exoplanet LHS 3844b}",
      journal = {\nat},
         year = 2019,
        month = aug,
       volume = {573},
       number = {7772},
        pages = {87-90},
          doi = {10.1038/s41586-019-1497-4},
archivePrefix = {arXiv},
       eprint = {1908.06834},
 primaryClass = {astro-ph.EP},
       adsurl = {https://ui.adsabs.harvard.edu/abs/2019Natur.573...87K}
}

@ARTICLE{Kreidberg2025,
       author = {{Kreidberg}, Laura and {Stevenson}, Kevin B.},
        title = "{A first look at rocky exoplanets with JWST}",
      journal = {Proceedings of the National Academy of Science},
         year = 2025,
        month = sep,
       volume = {122},
       number = {39},
          eid = {e2416190122},
        pages = {e2416190122},
          doi = {10.1073/pnas.2416190122},
archivePrefix = {arXiv},
       eprint = {2507.00933},
 primaryClass = {astro-ph.EP},
       adsurl = {https://ui.adsabs.harvard.edu/abs/2025PNAS..12216190K}
}

@ARTICLE{Wordsworth2022,
       author = {{Wordsworth}, Robin and {Kreidberg}, Laura},
        title = "{Atmospheres of Rocky Exoplanets}",
      journal = {\araa},
         year = 2022,
        month = aug,
       volume = {60},
        pages = {159-201},
          doi = {10.1146/annurev-astro-052920-125632},
archivePrefix = {arXiv},
       eprint = {2112.04663},
 primaryClass = {astro-ph.EP},
       adsurl = {https://ui.adsabs.harvard.edu/abs/2022ARA&A..60..159W}
}

@ARTICLE{Espinoza2025,
       author = {{Espinoza}, N{\'e}stor and {Perrin}, Marshall D.},
        title = "{Highlights from Exoplanet Observations by the James Webb Space Telescope}",
      journal = {arXiv e-prints},
         year = 2025,
        month = may,
          eid = {arXiv:2505.20520},
        pages = {arXiv:2505.20520},
          doi = {10.48550/arXiv.2505.20520},
archivePrefix = {arXiv},
       eprint = {2505.20520},
 primaryClass = {astro-ph.EP},
       adsurl = {https://ui.adsabs.harvard.edu/abs/2025arXiv250520520E}
}

@ARTICLE{Emissionpaper,
       author = {{Teske}, Johanna K. and {Wallack}, Nicole L. and {Piette}, Anjali A.~A. and {Dang}, Lisa and {Lichtenberg}, Tim and {Plotnykov}, Mykhaylo and {Pierrehumbert}, Raymond and {Postolec}, Emma and {Boucher}, Samuel and {McGinty}, Alex and {Peng}, Bo and {Valencia}, Diana and {Hammond}, Mark},
        title = "{A Thick Volatile Atmosphere on the Ultrahot Super-Earth TOI-561 b}",
      journal = {\apjl},
         year = 2025,
        month = dec,
       volume = {995},
       number = {2},
          eid = {L39},
        pages = {L39},
          doi = {10.3847/2041-8213/ae0a4c},
archivePrefix = {arXiv},
       eprint = {2509.17231},
 primaryClass = {astro-ph.EP},
       adsurl = {https://ui.adsabs.harvard.edu/abs/2025ApJ...995L..39T}
}

@article{lacedelli21,
  title = {An Unusually Low Density Ultra-Short Period Super-{{Earth}} and Three Mini-{{Neptunes}} around the Old Star {{TOI-561}}},
  author = {Lacedelli, G. and Malavolta, L. and Borsato, L. and Piotto, G. and Nardiello, D. and Mortier, A. and Stalport, M. and Cameron, A. Collier and Poretti, E. and Buchhave, L. A. and {L{\'o}pez-Morales}, M. and Nascimbeni, V. and Wilson, T. G. and Udry, S. and Latham, D. W. and Bonomo, A. S. and Damasso, M. and Dumusque, X. and Jenkins, J. M. and Lovis, C. and Rice, K. and Sasselov, D. and Winn, J. N. and Andreuzzi, G. and Cosentino, R. and Charbonneau, D. and Fabrizio, L. Di and Fiorenzano, A. F. Martinez and Ghedina, A. and Harutyunyan, A. and Lienhard, F. and Micela, G. and Molinari, E. and Pagano, I. and Pepe, F. and Phillips, D. F. and Pinamonti, M. and Ricker, G. and Scandariato, G. and Sozzetti, A. and Watson, C. A.},
  year = {2021},
  month = jan,
  journal = {Monthly Notices of the Royal Astronomical Society},
  volume = {501},
  number = {3},
  eprint = {2009.02332},
  primaryclass = {astro-ph},
  pages = {4148--4166},
  issn = {0035-8711, 1365-2966},
  doi = {10.1093/mnras/staa3728},
  urldate = {2025-06-12},
  archiveprefix = {arXiv}
}

@ARTICLE{Boer2025ApJ,
       author = {{Boer}, Iris D. and {Nicholls}, Harrison and {Lichtenberg}, Tim},
        title = "{Absence of a Runaway Greenhouse Limit on Lava Planets}",
      journal = {\apj},
         year = 2025,
        month = jul,
       volume = {987},
       number = {2},
          eid = {172},
        pages = {172},
          doi = {10.3847/1538-4357/add69f},
archivePrefix = {arXiv},
       eprint = {2505.11149},
 primaryClass = {astro-ph.EP},
       adsurl = {https://ui.adsabs.harvard.edu/abs/2025ApJ...987..172B}
}

@ARTICLE{Yoshida2025A&A,
       author = {{Yoshida}, Tatsuya and {Gaidos}, Eric},
        title = "{Water-cooled (sub)-Neptunes get better gas mileage}",
      journal = {\aap},
         year = 2025,
        month = apr,
       volume = {696},
          eid = {L13},
        pages = {L13},
          doi = {10.1051/0004-6361/202553667},
archivePrefix = {arXiv},
       eprint = {2503.23020},
 primaryClass = {astro-ph.EP},
       adsurl = {https://ui.adsabs.harvard.edu/abs/2025A&A...696L..13Y}
}

@ARTICLE{Burn2024NatAs,
       author = {{Burn}, Remo and {Mordasini}, Christoph and {Mishra}, Lokesh and {Haldemann}, Jonas and {Venturini}, Julia and {Emsenhuber}, Alexandre and {Henning}, Thomas},
        title = "{A radius valley between migrated steam worlds and evaporated rocky cores}",
      journal = {Nature Astronomy},
         year = 2024,
        month = apr,
       volume = {8},
        pages = {463-471},
          doi = {10.1038/s41550-023-02183-7},
archivePrefix = {arXiv},
       eprint = {2401.04380},
 primaryClass = {astro-ph.EP},
       adsurl = {https://ui.adsabs.harvard.edu/abs/2024NatAs...8..463B}
}

@ARTICLE{Venturini2024A&A,
       author = {{Venturini}, J. and {Ronco}, M.~P. and {Guilera}, O.~M. and {Haldemann}, J. and {Mordasini}, C. and {Miller Bertolami}, M.},
        title = "{A fading radius valley towards M dwarfs, a persistent density valley across stellar types}",
      journal = {\aap},
         year = 2024,
        month = jun,
       volume = {686},
          eid = {L9},
        pages = {L9},
          doi = {10.1051/0004-6361/202349088},
archivePrefix = {arXiv},
       eprint = {2404.01967},
 primaryClass = {astro-ph.EP},
       adsurl = {https://ui.adsabs.harvard.edu/abs/2024A&A...686L...9V}
}

@article{emcee,
  title = {Emcee: {{The MCMC Hammer}}},
  shorttitle = {Emcee},
  author = {{Foreman-Mackey}, Daniel and Conley, Alex and Meierjurgen Farr, Will and Hogg, David W. and Lang, Dustin and Marshall, Phil and {Price-Whelan}, Adrian and Sanders, Jeremy and Zuntz, Joe},
  year = {2013},
  month = mar,
  journal = {Astrophysics Source Code Library},
  pages = {ascl:1303.002},
  urldate = {2025-08-17}
}

@article{batman,
  title = {Batman: {{BAsic Transit Model cAlculatioN}} in {{Python}}},
  shorttitle = {Batman},
  author = {Kreidberg, Laura},
  year = {2015},
  month = nov,
  journal = {Publications of the Astronomical Society of the Pacific},
  volume = {127},
  number = {957},
  pages = {1161--1165},
  publisher = {IOP Publishing},
  issn = {0004-6280, 1538-3873},
  doi = {10.1086/683602},
  urldate = {2025-07-23},
  langid = {english}
}

@article{celerite,
  title = {Fast and {{Scalable Gaussian Process Modeling}} with {{Applications}} to {{Astronomical Time Series}}},
  author = {{Foreman-Mackey}, Daniel and Agol, Eric and Ambikasaran, Sivaram and Angus, Ruth},
  year = {2017},
  month = dec,
  journal = {The Astronomical Journal},
  volume = {154},
  number = {6},
  pages = {220},
  issn = {0004-6256, 1538-3881},
  doi = {10.3847/1538-3881/aa9332},
  urldate = {2025-08-06}
}

@article{dang,
  title = {Detection of a {{Westward Hotspot Offset}} in the {{Atmosphere}} of a {{Hot Gas Giant CoRoT-2b}}},
  author = {Dang, Lisa and Cowan, Nicolas B. and Schwartz, Joel C. and Rauscher, Emily and Zhang, Michael and Knutson, Heather A. and Line, Michael and {Dobbs-Dixon}, Ian and Deming, Drake and Sundararajan, Sudarsan and Fortney, Jonathan J. and Zhao, Ming},
  year = {2018},
  month = jan,
  journal = {Nature Astronomy},
  volume = {2},
  number = {3},
  eprint = {1801.06548},
  primaryclass = {astro-ph},
  pages = {220--227},
  issn = {2397-3366},
  doi = {10.1038/s41550-017-0351-6},
  urldate = {2025-05-15},
  archiveprefix = {arXiv}
}

@article{cowanalbedo,
  title = {{{THE STATISTICS OF ALBEDO AND HEAT RECIRCULATION ON HOT EXOPLANETS}}},
  author = {Cowan, Nicolas B. and Agol, Eric},
  year = {2011},
  month = mar,
  journal = {The Astrophysical Journal},
  volume = {729},
  number = {1},
  pages = {54},
  issn = {0004-637X, 1538-4357},
  doi = {10.1088/0004-637X/729/1/54},
  urldate = {2025-05-28}
}

@article{nguyen_clouds_2024,
	author = {Nguyen, T. Giang and Cowan, Nicolas B. and Dang, Lisa},
	title = {{Clouds in Partial Atmospheres of Lava Planets and Where to Find Them}},
	journal = {Astron. J.},
	volume = {168},
	number = {6},
	pages = {287},
	year = {2024},
	month = nov,
	issn = {1538-3881},
	publisher = {The American Astronomical Society},
	doi = {10.3847/1538-3881/ad85e0}
}

@article{gao_aerosols_2021,
	author = {Gao, Peter and Wakeford, Hannah R. and Moran, Sarah E. and Parmentier, Vivien},
	title = {{Aerosols in Exoplanet Atmospheres}},
	journal = {J. Geophys. Res. Planets},
	volume = {126},
	number = {4},
	pages = {e2020JE006655},
	year = {2021},
	month = apr,
	issn = {2169-9097},
	publisher = {John Wiley {\&} Sons, Ltd},
	doi = {10.1029/2020JE006655}
}

@article{husser_a_2013,
	author = {Husser, T.-O. and Berg, S. Wende-von and Dreizler, S. and Homeier, D. and Reiners, A. and Barman, T. and Hauschildt, P. H.},
	title = {{A new extensive library of PHOENIX stellar atmospheres and synthetic spectra}},
	journal = {Astron. Astrophys.},
	volume = {553},
	pages = {A6},
	year = {2013},
	month = may,
	issn = {0004-6361},
	publisher = {EDP Sciences},
	doi = {10.1051/0004-6361/201219058}
}

@article{lacedelli,
  title = {Investigating the Architecture and Internal Structure of the {{TOI-561}} System Planets with {{CHEOPS}}, {{HARPS-N}}, and {{TESS}}},
  author = {Lacedelli, G. and Wilson, T. G. and Malavolta, L. and Hooton, M. J. and Collier Cameron, A. and Alibert, Y. and Mortier, A. and Bonfanti, A. and Haywood, R. D. and Hoyer, S. and Piotto, G. and Bekkelien, A. and Vanderburg, A. M. and Benz, W. and Dumusque, X. and Deline, A. and {L{\'o}pez-Morales}, M. and Borsato, L. and Rice, K. and Fossati, L. and Latham, D. W. and Brandeker, A. and Poretti, E. and Sousa, S. G. and Sozzetti, A. and Salmon, S. and Burke, C. J. and Van Grootel, V. and Fausnaugh, M. M. and Adibekyan, V. and Huang, C. X. and Osborn, H. P. and Mustill, A. J. and Pall{\'e}, E. and Bourrier, V. and Nascimbeni, V. and Alonso, R. and Anglada, G. and B{\'a}rczy, T. and {Barrado y Navascues}, D. and Barros, S. C. C. and Baumjohann, W. and Beck, M. and Beck, T. and Billot, N. and Bonfils, X. and Broeg, C. and Buchhave, L. A. and Cabrera, J. and Charnoz, S. and Cosentino, R. and Csizmadia, Sz and Davies, M. B. and Deleuil, M. and Delrez, L. and Demangeon, O. and Demory, B. -O. and Ehrenreich, D. and Erikson, A. and {Esparza-Borges}, E. and Flor{\'e}n, H. G. and Fortier, A. and Fridlund, M. and Futyan, D. and Gandolfi, D. and Ghedina, A. and Gillon, M. and G{\"u}del, M. and Guterman, P. and Harutyunyan, A. and Heng, K. and Isaak, K. G. and Jenkins, J. M. and Kiss, L. and Laskar, J. and {Lecavelier des Etangs}, A. and Lendl, M. and Lovis, C. and Magrin, D. and Marafatto, L. and Martinez Fiorenzano, A. F. and Maxted, P. F. L. and Mayor, M. and Micela, G. and Molinari, E. and Murgas, F. and Narita, N. and Olofsson, G. and Ottensamer, R. and Pagano, I. and Pasetti, A. and Pedani, M. and Pepe, F. A. and Peter, G. and Phillips, D. F. and Pollacco, D. and Queloz, D. and Ragazzoni, R. and Rando, N. and Ratti, F. and Rauer, H. and Ribas, I. and Santos, N. C. and Sasselov, D. and Scandariato, G. and Seager, S. and S{\'e}gransan, D. and Serrano, L. M. and Simon, A. E. and Smith, A. M. S. and Steinberger, M. and Steller, M. and Szab{\'o}, Gy and Thomas, N. and Twicken, J. D. and Udry, S. and Walton, N. and Winn, J. N.},
  year = {2022},
  month = apr,
  journal = {Monthly Notices of the Royal Astronomical Society},
  volume = {511},
  pages = {4551--4571},
  publisher = {OUP},
  issn = {0035-8711},
  doi = {10.1093/mnras/stac199},
  urldate = {2025-08-05}
}

@article{weiss2021,
  title = {The {{TESS-Keck Survey}}. {{II}}. {{An Ultra-short-period Rocky Planet}} and {{Its Siblings Transiting}} the {{Galactic Thick-disk Star TOI-561}}},
  author = {Weiss, Lauren M. and Dai, Fei and Huber, Daniel and Brewer, John M. and Collins, Karen A. and Ciardi, David R. and Matthews, Elisabeth C. and Ziegler, Carl and Howell, Steve B. and Batalha, Natalie M. and Crossfield, Ian J. M. and Dressing, Courtney and Fulton, Benjamin and Howard, Andrew W. and Isaacson, Howard and Kane, Stephen R. and Petigura, Erik A. and Robertson, Paul and Roy, Arpita and Rubenzahl, Ryan A. and Twicken, Joseph D. and Claytor, Zachary R. and Stassun, Keivan G. and MacDougall, Mason G. and Chontos, Ashley and Giacalone, Steven and Dalba, Paul A. and Mocnik, Teo and Hill, Michelle L. and Beard, Corey and Akana Murphy, Joseph M. and Rosenthal, Lee J. and Behmard, Aida and Van Zandt, Judah and Lubin, Jack and Kosiarek, Molly R. and Lund, Michael B. and Christiansen, Jessie L. and Matson, Rachel A. and Beichman, Charles A. and Schlieder, Joshua E. and Gonzales, Erica J. and Brice{\~n}o, C{\'e}sar and Law, Nicholas and Mann, Andrew W. and Collins, Kevin I. and Evans, Phil and Fukui, Akihiko and Jensen, Eric L. N. and Murgas, Felipe and Narita, Norio and Palle, Enric and Parviainen, Hannu and Schwarz, Richard P. and Tan, Thiam-Guan and Acton, Jack S. and Bryant, Edward M. and Chaushev, Alexander and Gill, Sam and Eigm{\"u}ller, Philipp and Jenkins, Jon and Ricker, George and Seager, Sara and Winn, Joshua N.},
  year = {2021},
  month = feb,
  journal = {The Astronomical Journal},
  volume = {161},
  pages = {56},
  publisher = {IOP},
  issn = {0004-6256},
  doi = {10.3847/1538-3881/abd409},
  urldate = {2025-05-15}
}

@article{patel2023,
  title = {{{CHEOPS}} and {{TESS}} View of the Ultra-Short-Period Super-{{Earth TOI-561}} b},
  author = {Patel, J. A. and Egger, J. A. and Wilson, T. G. and Bourrier, V. and Carone, L. and Beck, M. and Ehrenreich, D. and Sousa, S. G. and Benz, W. and Brandeker, A. and Deline, A. and Alibert, Y. and Lam, K. W. F. and Lendl, M. and Alonso, R. and Anglada, G. and B{\'a}rczy, T. and Barrado, D. and Barros, S. C. C. and Baumjohann, W. and Beck, T. and Billot, N. and Bonfils, X. and Broeg, C. and Busch, M.-D. and Cabrera, J. and Charnoz, S. and Collier Cameron, A. and Csizmadia, {\relax Sz}. and Davies, M. B. and Deleuil, M. and Delrez, L. and Demangeon, O. D. S. and Demory, B.-O. and Erikson, A. and Fortier, A. and Fossati, L. and Fridlund, M. and Gandolfi, D. and Gillon, M. and G{\"u}del, M. and Heng, K. and Hoyer, S. and Isaak, K. G. and Kiss, L. L. and Kopp, E. and Laskar, J. and Lecavelier Des Etangs, A. and Lovis, C. and Magrin, D. and Maxted, P. F. L. and Nascimbeni, V. and Olofsson, G. and Ottensamer, R. and Pagano, I. and Pall{\'e}, E. and Peter, G. and Piotto, G. and Pollacco, D. and Queloz, D. and Ragazzoni, R. and Rando, N. and Ratti, F. and Rauer, H. and Ribas, I. and Santos, N. C. and Scandariato, G. and S{\'e}gransan, D. and Simon, A. E. and Smith, A. M. S. and Steller, M. and Szab{\'o}, {\relax Gy}. M. and Thomas, N. and Udry, S. and Ulmer, B. and Van Grootel, V. and Viotto, V. and Walton, N. A.},
  year = {2023},
  month = nov,
  journal = {Astronomy \& Astrophysics},
  volume = {679},
  pages = {A92},
  issn = {0004-6361, 1432-0746},
  doi = {10.1051/0004-6361/202244946},
  urldate = {2025-08-07},
  copyright = {https://creativecommons.org/licenses/by/4.0}
}

@article{granguess,
  title = {{{{\emph{KEPLER}}}} {{MISSION STELLAR AND INSTRUMENT NOISE PROPERTIES}}},
  author = {Gilliland, Ronald L. and Chaplin, William J. and Dunham, Edward W. and Argabright, Vic S. and Borucki, William J. and Basri, Gibor and Bryson, Stephen T. and Buzasi, Derek L. and Caldwell, Douglas A. and Elsworth, Yvonne P. and Jenkins, Jon M. and Koch, David G. and Kolodziejczak, Jeffrey and Miglio, Andrea and Van Cleve, Jeffrey and Walkowicz, Lucianne M. and Welsh, William F.},
  year = {2011},
  month = nov,
  journal = {The Astrophysical Journal Supplement Series},
  volume = {197},
  number = {1},
  pages = {6},
  issn = {0067-0049, 1538-4365},
  doi = {10.1088/0067-0049/197/1/6},
  urldate = {2025-08-07}
}

@article{Grant2024, doi = {10.21105/joss.06816}, url = {https://doi.org/10.21105/joss.06816}, year = {2024}, publisher = {The Open Journal}, volume = {9}, number = {100}, pages = {6816}, author = {Grant, David and Wakeford, Hannah R.}, title = {ExoTiC-LD: thirty seconds to stellar limb-darkening coefficients}, journal = {Journal of Open Source Software} }

@ARTICLE{Lewis2022,
       author = {{Lewis}, Neil T. and {Hammond}, Mark},
        title = "{Temperature Structures Associated with Different Components of the Atmospheric Circulation on Tidally Locked Exoplanets}",
      journal = {\apj},
         year = 2022,
        month = dec,
       volume = {941},
       number = {2},
          eid = {171},
        pages = {171},
          doi = {10.3847/1538-4357/ac8fed},
archivePrefix = {arXiv},
       eprint = {2204.06503},
 primaryClass = {astro-ph.EP},
       adsurl = {https://ui.adsabs.harvard.edu/abs/2022ApJ...941..171L}
}

@ARTICLE{Hammond2017,
       author = {{Hammond}, Mark and {Pierrehumbert}, Raymond T.},
        title = "{Linking the Climate and Thermal Phase Curve of 55 Cancri e}",
      journal = {\apj},
         year = 2017,
        month = nov,
       volume = {849},
       number = {2},
          eid = {152},
        pages = {152},
          doi = {10.3847/1538-4357/aa9328},
archivePrefix = {arXiv},
       eprint = {1710.03556},
 primaryClass = {astro-ph.EP},
       adsurl = {https://ui.adsabs.harvard.edu/abs/2017ApJ...849..152H}
}

@article{fortin_lava_2024,
	author = {Fortin, Marc-Antoine and Gazel, Esteban and Williams, Daniel B. and Thompson, James O. and Kaltenegger, Lisa and Ramsey, Michael S.},
	title = {{Lava Worlds Surface Measurements at High Temperatures}},
	journal = {Astrophys. J. Lett.},
	volume = {974},
	number = {1},
	pages = {L7},
	year = {2024},
	month = oct,
	issn = {2041-8205},
	publisher = {The American Astronomical Society},
	doi = {10.3847/2041-8213/ad7d89}
}

@article{lai_three_2026,
	author = {Lai, Yanhong and Kang, Wanying and Yang, Jun and Tan, Xianyu},
	title = {{Three-Dimensional Ocean Dynamics and Detectability of Tidally Locked Lava Worlds}},
	journal = {arXiv},
	year = {2026},
	month = apr,
	eprint = {2604.00535},
	doi = {10.48550/arXiv.2604.00535}
}

@article{pereira2019,
       author = {{Pereira}, Filipe and {Campante}, Tiago L. and {Cunha}, Margarida S. and {Faria}, Jo{\~a}o P. and {Santos}, Nuno C. and {Barros}, Susana C.~C. and {Demangeon}, Olivier and {Kuszlewicz}, James S. and {Corsaro}, Enrico},
        title = "{Gaussian process modelling of granulation and oscillations in red giant stars}",
      journal = {\mnras},
         year = 2019,
        month = nov,
       volume = {489},
       number = {4},
        pages = {5764-5774},
          doi = {10.1093/mnras/stz2405},
archivePrefix = {arXiv},
       eprint = {1908.10662},
 primaryClass = {astro-ph.SR},
       adsurl = {https://ui.adsabs.harvard.edu/abs/2019MNRAS.489.5764P}
}

@article{Hammond2025,
       author = {{Hammond}, Mark and {Guimond}, Claire Marie and {Lichtenberg}, Tim and {Nicholls}, Harrison and {Fisher}, Chloe and {Luque}, Rafael and {Meier}, Tobias G. and {Taylor}, Jake and {Changeat}, Quentin and {Dang}, Lisa and {Hay}, Hamish C.~F.~C. and {Herbort}, Oliver and {Teske}, Johanna},
        title = "{Reliable Detections of Atmospheres on Rocky Exoplanets with Photometric JWST Phase Curves}",
      journal = {\apjl},
         year = 2025,
        month = jan,
       volume = {978},
       number = {2},
          eid = {L40},
        pages = {L40},
          doi = {10.3847/2041-8213/ada0bc},
archivePrefix = {arXiv},
       eprint = {2409.04386},
 primaryClass = {astro-ph.EP},
       adsurl = {https://ui.adsabs.harvard.edu/abs/2025ApJ...978L..40H}
}

@ARTICLE{Kang2023,
       author = {{Kang}, Wanying and {Nimmo}, Francis and {Ding}, Feng},
        title = "{True Polar Wander of Lava Worlds}",
      journal = {\apjl},
         year = 2023,
        month = jun,
       volume = {949},
       number = {2},
          eid = {L20},
        pages = {L20},
          doi = {10.3847/2041-8213/acd691},
archivePrefix = {arXiv},
       eprint = {2306.06768},
 primaryClass = {astro-ph.EP},
       adsurl = {https://ui.adsabs.harvard.edu/abs/2023ApJ...949L..20K}
}

@ARTICLE{Vallis2018,
       author = {{Vallis}, Geoffrey K. and {Colyer}, Greg and {Geen}, Ruth and {Gerber}, Edwin and {Jucker}, Martin and {Maher}, Penelope and {Paterson}, Alexander and {Pietschnig}, Marianne and {Penn}, James and {Thomson}, Stephen I.},
        title = "{Isca, v1.0: a framelrk for the global modelling of the atmospheres of Earth and other planets at varying levels of complexity}",
      journal = {Geoscientific Model Development},
         year = 2018,
        month = mar,
       volume = {11},
       number = {3},
        pages = {843-859},
          doi = {10.5194/gmd-11-843-2018},
       adsurl = {https://ui.adsabs.harvard.edu/abs/2018GMD....11..843V}
}

@ARTICLE{Thomson2019,
       author = {{Thomson}, Stephen I. and {Vallis}, Geoffrey K.},
        title = "{The effects of gravity on the climate and circulation of a terrestrial planet}",
      journal = {Quarterly Journal of the Royal Meteorological Society},
         year = 2019,
        month = jul,
       volume = {145},
       number = {723},
        pages = {2627-2640},
          doi = {10.1002/qj.3582},
archivePrefix = {arXiv},
       eprint = {1901.11426},
 primaryClass = {astro-ph.EP},
       adsurl = {https://ui.adsabs.harvard.edu/abs/2019QJRMS.145.2627T}
}

@ARTICLE{Guillot2010,
       author = {{Guillot}, T.},
        title = "{On the radiative equilibrium of irradiated planetary atmospheres}",
      journal = {\aap},
         year = 2010,
        month = sep,
       volume = {520},
          eid = {A27},
        pages = {A27},
          doi = {10.1051/0004-6361/200913396},
archivePrefix = {arXiv},
       eprint = {1006.4702},
 primaryClass = {astro-ph.EP},
       adsurl = {https://ui.adsabs.harvard.edu/abs/2010A&A...520A..27G}
}

@article{splinter,
  title = {Precise {{Constraints}} on the {{Energy Budget}} of {{WASP-121}} b from {{Its JWST NIRISS}}/{{SOSS Phase Curve}}},
  author = {Splinter, Jared and Coulombe, Louis-Philippe and Frazier, Robert C. and Cowan, Nicolas B. and Rauscher, Emily and Dang, Lisa and Radica, Michael and Collins, Sean and Pelletier, Stefan and Allart, Romain and MacDonald, Ryan J. and Lafreni{\`e}re, David and {Lo\"ic Albert} and Benneke, Bj{\"o}rn and Doyon, Ren{\'e} and Jayawardhana, Ray and Johnstone, Doug and Krishnamurthy, Vigneshwaran and {Piaulet-Ghorayeb}, Caroline and Kaltenegger, Lisa and Meyer, Michael R. and Taylor, Jake and Turner, Jake D.},
  year = 2025,
  month = dec,
  journal = {The Astronomical Journal},
  volume = {170},
  number = {6},
  pages = {323},
  issn = {0004-6256, 1538-3881},
  doi = {10.3847/1538-3881/ae0e52},
  urldate = {2026-03-06}
}

@ARTICLE{Rogers2021MNRAS,
       author = {{Rogers}, James G. and {Owen}, James E.},
        title = "{Unveiling the planet population at birth}",
      journal = {\mnras},
         year = 2021,
        month = may,
       volume = {503},
       number = {1},
        pages = {1526-1542},
          doi = {10.1093/mnras/stab529},
archivePrefix = {arXiv},
       eprint = {2007.11006},
 primaryClass = {astro-ph.EP},
       adsurl = {https://ui.adsabs.harvard.edu/abs/2021MNRAS.503.1526R}
}

@ARTICLE{Arora2025arXiv,
       author = {{Arora}, R. and {Ranjan}, S. and {Moitra}, P. and {Mallik}, A.},
        title = "{Thin H$_2$-dominated Atmospheres as Signposts of Magmatic Outgassing on Tidally-Heated Terrestrial Exoplanets}",
      journal = {arXiv e-prints},
         year = 2025,
        month = oct,
          eid = {arXiv:2510.07378},
        pages = {arXiv:2510.07378},
          doi = {10.48550/arXiv.2510.07378},
archivePrefix = {arXiv},
       eprint = {2510.07378},
 primaryClass = {astro-ph.EP},
       adsurl = {https://ui.adsabs.harvard.edu/abs/2025arXiv251007378A}
}

@ARTICLE{Krissansen-Totton2022ApJ,
       author = {{Krissansen-Totton}, J. and {Fortney}, J.~J.},
        title = "{Predictions for Observable Atmospheres of Trappist-1 Planets from a Fully Coupled Atmosphere-Interior Evolution Model}",
      journal = {\apj},
         year = 2022,
        month = jul,
       volume = {933},
       number = {1},
          eid = {115},
        pages = {115},
          doi = {10.3847/1538-4357/ac69cb},
archivePrefix = {arXiv},
       eprint = {2207.04164},
 primaryClass = {astro-ph.EP},
       adsurl = {https://ui.adsabs.harvard.edu/abs/2022ApJ...933..115K}
}

@ARTICLE{Ito2025ApJ,
       author = {{Ito}, Yuichi and {Kimura}, Tadahiro and {Ohno}, Kazumasa and {Fujii}, Yuka and {Ikoma}, Masahiro},
        title = "{Monosilane Worlds: Sub-Neptunes with Atmospheres Shaped by Reduced Magma Oceans}",
      journal = {\apj},
         year = 2025,
        month = jul,
       volume = {987},
       number = {2},
          eid = {174},
        pages = {174},
          doi = {10.3847/1538-4357/add3fe},
archivePrefix = {arXiv},
       eprint = {2505.03200},
 primaryClass = {astro-ph.EP},
       adsurl = {https://ui.adsabs.harvard.edu/abs/2025ApJ...987..174I}
}

@ARTICLE{Seo2024ApJ,
       author = {{Seo}, Chanoul and {Ito}, Yuichi and {Fujii}, Yuka},
        title = "{Role of Magma Oceans in Controlling Carbon and Oxygen of Sub-Neptune Atmospheres}",
      journal = {\apj},
         year = 2024,
        month = nov,
       volume = {975},
       number = {1},
          eid = {14},
        pages = {14},
          doi = {10.3847/1538-4357/ad7461},
archivePrefix = {arXiv},
       eprint = {2408.17056},
 primaryClass = {astro-ph.EP},
       adsurl = {https://ui.adsabs.harvard.edu/abs/2024ApJ...975...14S}
}

@ARTICLE{Boukare2025NatAs,
       author = {{Boukar{\'e}}, Charles-{\'E}douard and {Lemasquerier}, Daphn{\'e} and {Cowan}, Nicolas B. and {Samuel}, Henri and {Badro}, James and {Dang}, Lisa and {Falco}, Aur{\'e}lien and {Charnoz}, S{\'e}bastien},
        title = "{The role of interior dynamics and differentiation on the surface and in the atmosphere of lava planets}",
      journal = {Nature Astronomy},
         year = 2025,
        month = jul,
       volume = {9},
        pages = {1511-1522},
          doi = {10.1038/s41550-025-02617-4},
archivePrefix = {arXiv},
       eprint = {2308.13614},
 primaryClass = {astro-ph.EP},
       adsurl = {https://ui.adsabs.harvard.edu/abs/2025NatAs...9.1511B}
}

@ARTICLE{Lichtenberg2021JGRE,
       author = {{Lichtenberg}, Tim and {Bower}, Dan J. and {Hammond}, Mark and {Boukrouche}, Ryan and {Sanan}, Patrick and {Tsai}, Shang-Min and {Pierrehumbert}, Raymond T.},
        title = "{Vertically Resolved Magma Ocean-Protoatmosphere Evolution: H$_{2}$, H$_{2}$O, CO$_{2}$, CH$_{4}$, CO, O$_{2}$, and N$_{2}$ as Primary Absorbers}",
      journal = {Journal of Geophysical Research (Planets)},
         year = 2021,
        month = feb,
       volume = {126},
       number = {2},
          eid = {e06711},
        pages = {e06711},
          doi = {10.1029/2020JE006711},
archivePrefix = {arXiv},
       eprint = {2101.10991},
 primaryClass = {astro-ph.EP},
       adsurl = {https://ui.adsabs.harvard.edu/abs/2021JGRE..12606711L}
}

@ARTICLE{Piette2023ApJ,
       author = {{Piette}, Anjali A.~A. and {Gao}, Peter and {Brugman}, Kara and {Shahar}, Anat and {Lichtenberg}, Tim and {Miozzi}, Francesca and {Driscoll}, Peter},
        title = "{Rocky Planet or Water World? Observability of Low-density Lava World Atmospheres}",
      journal = {\apj},
         year = 2023,
        month = sep,
       volume = {954},
       number = {1},
          eid = {29},
        pages = {29},
          doi = {10.3847/1538-4357/acdef2},
archivePrefix = {arXiv},
       eprint = {2306.10100},
 primaryClass = {astro-ph.EP},
       adsurl = {https://ui.adsabs.harvard.edu/abs/2023ApJ...954...29P}
}

@ARTICLE{Steinmeyer2026ApJ,
       author = {{Steinmeyer}, Marie-Luise and {Dorn}, Caroline and {Werlen}, Aaron and {Grimm}, Simon L.},
        title = "{Coupled Thermal─Chemical Evolution Models of Sub-Neptunes Reveal Atmospheric Signatures of Their Formation Location}",
      journal = {\apj},
         year = 2026,
        month = apr,
       volume = {1001},
       number = {1},
          eid = {36},
        pages = {36},
          doi = {10.3847/1538-4357/ae4c47},
archivePrefix = {arXiv},
       eprint = {2601.21377},
 primaryClass = {astro-ph.EP},
       adsurl = {https://ui.adsabs.harvard.edu/abs/2026ApJ..1001...36S}
}

@ARTICLE{Nicholls2025MNRAS,
       author = {{Nicholls}, Harrison and {Guimond}, Claire Marie and {Hay}, Hamish C.~F.~C. and {Chatterjee}, Richard D. and {Lichtenberg}, Tim and {Pierrehumbert}, Raymond T.},
        title = "{Self-limited tidal heating and prolonged magma oceans in the L 98-59 system}",
      journal = {\mnras},
         year = 2025,
        month = aug,
       volume = {541},
       number = {3},
        pages = {2566-2584},
          doi = {10.1093/mnras/staf1167},
archivePrefix = {arXiv},
       eprint = {2505.03604},
 primaryClass = {astro-ph.EP},
       adsurl = {https://ui.adsabs.harvard.edu/abs/2025MNRAS.541.2566N}
}

@ARTICLE{Schaefer2016ApJ,
       author = {{Schaefer}, Laura and {Wordsworth}, Robin D. and {Berta-Thompson}, Zachory and {Sasselov}, Dimitar},
        title = "{Predictions of the Atmospheric Composition of GJ 1132b}",
      journal = {\apj},
         year = 2016,
        month = oct,
       volume = {829},
       number = {2},
          eid = {63},
        pages = {63},
          doi = {10.3847/0004-637X/829/2/63},
archivePrefix = {arXiv},
       eprint = {1607.03906},
 primaryClass = {astro-ph.EP},
       adsurl = {https://ui.adsabs.harvard.edu/abs/2016ApJ...829...63S}
}

@ARTICLE{Bower2019A&A,
       author = {{Bower}, Dan J. and {Kitzmann}, Daniel and {Wolf}, Aaron S. and {Sanan}, Patrick and {Dorn}, Caroline and {Oza}, Apurva V.},
        title = "{Linking the evolution of terrestrial interiors and an early outgassed atmosphere to astrophysical observations}",
      journal = {\aap},
         year = 2019,
        month = nov,
       volume = {631},
          eid = {A103},
        pages = {A103},
          doi = {10.1051/0004-6361/20193571010.31223/osf.io/ctqe3},
       adsurl = {https://ui.adsabs.harvard.edu/abs/2019A&A...631A.103B}
}

@ARTICLE{ulrich1986,
       author = {{Ulrich}, R.~K.},
        title = "{Determination of Stellar Ages from Asteroseismology}",
      journal = {\apjl},
         year = 1986,
        month = jul,
       volume = {306},
        pages = {L37},
          doi = {10.1086/184700},
       adsurl = {https://ui.adsabs.harvard.edu/abs/1986ApJ...306L..37U}
}

@ARTICLE{brown1991,
       author = {{Brown}, Timothy M. and {Gilliland}, Ronald L. and {Noyes}, Robert W. and {Ramsey}, Lawrence W.},
        title = "{Detection of Possible p-Mode Oscillations on Procyon}",
      journal = {\apj},
         year = 1991,
        month = feb,
       volume = {368},
        pages = {599},
          doi = {10.1086/169725},
       adsurl = {https://ui.adsabs.harvard.edu/abs/1991ApJ...368..599B}
}

@ARTICLE{KrissansenTotton2024NatCo,
       author = {{Krissansen-Totton}, Joshua and {Wogan}, Nicholas and {Thompson}, Maggie and {Fortney}, Jonathan J.},
        title = "{The erosion of large primary atmospheres typically leaves behind substantial secondary atmospheres on temperate rocky planets}",
      journal = {Nature Communications},
         year = 2024,
        month = dec,
       volume = {15},
       number = {1},
          eid = {8374},
        pages = {8374},
          doi = {10.1038/s41467-024-52642-6},
archivePrefix = {arXiv},
       eprint = {2409.18940},
 primaryClass = {astro-ph.EP},
       adsurl = {https://ui.adsabs.harvard.edu/abs/2024NatCo..15.8374K}
}

@ARTICLE{Ji2025ApJ,
       author = {{Ji}, Xuan and {Chatterjee}, Richard D. and {Coy}, Brandon Park and {Kite}, Edwin},
        title = "{The Cosmic Shoreline Revisited: A Metric for Atmospheric Retention Informed by Hydrodynamic Escape}",
      journal = {\apj},
         year = 2025,
        month = oct,
       volume = {992},
       number = {2},
          eid = {198},
        pages = {198},
          doi = {10.3847/1538-4357/adfe69},
archivePrefix = {arXiv},
       eprint = {2504.19872},
 primaryClass = {astro-ph.EP},
       adsurl = {https://ui.adsabs.harvard.edu/abs/2025ApJ...992..198J}
}

@ARTICLE{Chatterjee2026ApJ,
       author = {{Chatterjee}, Richard D. and {Pierrehumbert}, Raymond T.},
        title = "{Novel Physics of Escaping Secondary Atmospheres May Shape the Cosmic Shoreline}",
      journal = {\apj},
         year = 2026,
        month = feb,
       volume = {998},
       number = {2},
          eid = {236},
        pages = {236},
          doi = {10.3847/1538-4357/ae2ffa},
archivePrefix = {arXiv},
       eprint = {2412.05188},
 primaryClass = {astro-ph.EP},
       adsurl = {https://ui.adsabs.harvard.edu/abs/2026ApJ...998..236C}
}

@ARTICLE{Owen2019AREPS,
       author = {{Owen}, James E.},
        title = "{Atmospheric Escape and the Evolution of Close-In Exoplanets}",
      journal = {Annual Review of Earth and Planetary Sciences},
         year = 2019,
        month = may,
       volume = {47},
        pages = {67-90},
          doi = {10.1146/annurev-earth-053018-060246},
archivePrefix = {arXiv},
       eprint = {1807.07609},
 primaryClass = {astro-ph.EP},
       adsurl = {https://ui.adsabs.harvard.edu/abs/2019AREPS..47...67O}
}

@ARTICLE{Lichtenberg2025TrGeo,
       author = {{Lichtenberg}, Tim and {Miguel}, Yamila},
        title = "{Super-Earths and Earth-like Exoplanets}",
      journal = {Treatise on Geochemistry},
         year = 2025,
        month = jan,
       volume = {7},
        pages = {51-112},
          doi = {10.1016/B978-0-323-99762-1.00122-4},
archivePrefix = {arXiv},
       eprint = {2405.04057},
 primaryClass = {astro-ph.EP},
       adsurl = {https://ui.adsabs.harvard.edu/abs/2025TrGeo...7...51L}
}

@ARTICLE{Kallinger2014AAP,
       author = {{Kallinger}, T. and {De Ridder}, J. and {Hekker}, S. and {Mathur}, S. and {Mosser}, B. and {Gruberbauer}, M. and {Garc{\'\i}a}, R.~A. and {Karoff}, C. and {Ballot}, J.},
        title = "{The connection between stellar granulation and oscillation as seen by the Kepler mission}",
      journal = {\aap},
         year = 2014,
        month = oct,
       volume = {570},
          eid = {A41},
        pages = {A41},
          doi = {10.1051/0004-6361/201424313},
archivePrefix = {arXiv},
       eprint = {1408.0817},
 primaryClass = {astro-ph.SR},
       adsurl = {https://ui.adsabs.harvard.edu/abs/2014A&A...570A..41K}
}

@ARTICLE{Rasr2003APJ,
       author = {{Rast}, Mark Peter},
        title = "{The Scales of Granulation, Mesogranulation, and Supergranulation}",
      journal = {\apj},
         year = 2003,
        month = nov,
       volume = {597},
       number = {2},
        pages = {1200-1210},
          doi = {10.1086/381221},
       adsurl = {https://ui.adsabs.harvard.edu/abs/2003ApJ...597.1200R}
}

@ARTICLE{Lund2019mnras,
       author = {{Lund}, Mikkel N.},
        title = "{Bolometric corrections of stellar oscillation amplitudes as observed by the Kepler, CoRoT, and TESS missions}",
      journal = {\mnras},
         year = 2019,
        month = oct,
       volume = {489},
       number = {1},
        pages = {1072-1081},
          doi = {10.1093/mnras/stz2010},
archivePrefix = {arXiv},
       eprint = {1907.12557},
 primaryClass = {astro-ph.SR},
       adsurl = {https://ui.adsabs.harvard.edu/abs/2019MNRAS.489.1072L}
}

@ARTICLE{KjeldsenBedding1995,
       author = {{Kjeldsen}, H. and {Bedding}, T.~R.},
        title = "{Amplitudes of stellar oscillations: the implications for asteroseismology.}",
      journal = {\aap},
         year = 1995,
        month = jan,
       volume = {293},
        pages = {87-106},
          doi = {10.48550/arXiv.astro-ph/9403015},
archivePrefix = {arXiv},
       eprint = {astro-ph/9403015},
 primaryClass = {astro-ph},
       adsurl = {https://ui.adsabs.harvard.edu/abs/1995A&A...293...87K}
}

@ARTICLE{Mathur2011,
       author = {{Mathur}, S. and {Hekker}, S. and {Trampedach}, R. and {Ballot}, J. and {Kallinger}, T. and {Buzasi}, D. and {Garc{\'\i}a}, R.~A. and {Huber}, D. and {Jim{\'e}nez}, A. and {Mosser}, B. and {Bedding}, T.~R. and {Elsworth}, Y. and {R{\'e}gulo}, C. and {Stello}, D. and {Chaplin}, W.~J. and {De Ridder}, J. and {Hale}, S.~J. and {Kinemuchi}, K. and {Kjeldsen}, H. and {Mullally}, F. and {Thompson}, S.~E.},
        title = "{Granulation in Red Giants: Observations by the Kepler Mission and Three-dimensional Convection Simulations}",
      journal = {\apj},
         year = 2011,
        month = nov,
       volume = {741},
       number = {2},
          eid = {119},
        pages = {119},
          doi = {10.1088/0004-637X/741/2/119},
archivePrefix = {arXiv},
       eprint = {1109.1194},
 primaryClass = {astro-ph.SR},
       adsurl = {https://ui.adsabs.harvard.edu/abs/2011ApJ...741..119M}
}

@ARTICLE{Kitzmann2018,
       author = {{Kitzmann}, Daniel and {Heng}, Kevin},
        title = "{Optical properties of potential condensates in exoplanetary atmospheres}",
      journal = {\mnras},
         year = 2018,
        month = mar,
       volume = {475},
       number = {1},
        pages = {94-107},
          doi = {10.1093/mnras/stx3141},
archivePrefix = {arXiv},
       eprint = {1710.04946},
 primaryClass = {astro-ph.EP},
       adsurl = {https://ui.adsabs.harvard.edu/abs/2018MNRAS.475...94K}
}

@article{chase1998j,
  title={J. Phys. Chem. Ref. Data Monogr.},
  author={Chase, MW},
  journal={NIST-JANAF Thermochemical Tables},
  pages={1952},
  year={1998},
  publisher={Monograph}
}

@article{garvin1986codata,
  title={CODATA thermodynamic tables},
  author={Garvin, David and Parker, Vivian B and White Jr, HJ},
  year={1986},
  publisher={Springer-Verlag New York Inc., New York, NY}
}

@ARTICLE{2010JQSRT.111.2139R,
       author = {{Rothman}, L.~S. and {Gordon}, I.~E. and {Barber}, R.~J. and {Dothe}, H. and {Gamache}, R.~R. and {Goldman}, A. and {Perevalov}, V.~I. and {Tashkun}, S.~A. and {Tennyson}, J.},
        title = "{HITEMP, the high-temperature molecular spectroscopic database}",
      journal = {\jqsrt},
         year = 2010,
        month = oct,
       volume = {111},
        pages = {2139-2150},
          doi = {10.1016/j.jqsrt.2010.05.001},
       adsurl = {https://ui.adsabs.harvard.edu/abs/2010JQSRT.111.2139R}
}

@ARTICLE{2020MNRAS.496.5282Y,
       author = {{Yurchenko}, S.~N. and {Mellor}, Thomas M. and {Freedman}, Richard S. and {Tennyson}, J.},
        title = "{ExoMol line lists - XXXIX. Ro-vibrational molecular line list for CO$_{2}$}",
      journal = {\mnras},
         year = 2020,
        month = aug,
       volume = {496},
       number = {4},
        pages = {5282-5291},
          doi = {10.1093/mnras/staa1874},
archivePrefix = {arXiv},
       eprint = {2007.02122},
 primaryClass = {astro-ph.EP},
       adsurl = {https://ui.adsabs.harvard.edu/abs/2020MNRAS.496.5282Y}
}

@ARTICLE{2022JQSRT.27707949G,
       author = {{Gordon}, I.~E. and {Rothman}, L.~S. and {Hargreaves}, R.~J. and {Hashemi}, R. and {Karlovets}, E.~V. and {Skinner}, F.~M. and {Conway}, E.~K. and {Hill}, C. and {Kochanov}, R.~V. and {Tan}, Y. and {Wcis{\l}o}, P. and {Finenko}, A.~A. and {Nelson}, K. and {Bernath}, P.~F. and {Birk}, M. and {Boudon}, V. and {Campargue}, A. and {Chance}, K.~V. and {Coustenis}, A. and {Drouin}, B.~J. and {Flaud}, J.-M. and {Gamache}, R.~R. and {Hodges}, J.~T. and {Jacquemart}, D. and {Mlawer}, E.~J. and {Nikitin}, A.~V. and {Perevalov}, V.~I. and {Rotger}, M. and {Tennyson}, J. and {Toon}, G.~C. and {Tran}, H. and {Tyuterev}, V.~G. and {Adkins}, E.~M. and {Baker}, A. and {Barbe}, A. and {Can{\`e}}, E. and {Cs{\'a}sz{\'a}r}, A.~G. and {Dudaryonok}, A. and {Egorov}, O. and {Fleisher}, A.~J. and {Fleurbaey}, H. and {Foltynowicz}, A. and {Furtenbacher}, T. and {Harrison}, J.~J. and {Hartmann}, J.-M. and {Horneman}, V.-M. and {Huang}, X. and {Karman}, T. and {Karns}, J. and {Kassi}, S. and {Kleiner}, I. and {Kofman}, V. and {Kwabia-Tchana}, F. and {Lavrentieva}, N.~N. and {Lee}, T.~J. and {Long}, D.~A. and {Lukashevskaya}, A.~A. and {Lyulin}, O.~M. and {Makhnev}, V. Yu. and {Matt}, W. and {Massie}, S.~T. and {Melosso}, M. and {Mikhailenko}, S.~N. and {Mondelain}, D. and {M{\"u}ller}, H.~S.~P. and {Naumenko}, O.~V. and {Perrin}, A. and {Polyansky}, O.~L. and {Raddaoui}, E. and {Raston}, P.~L. and {Reed}, Z.~D. and {Rey}, M. and {Richard}, C. and {T{\'o}bi{\'a}s}, R. and {Sadiek}, I. and {Schwenke}, D.~W. and {Starikova}, E. and {Sung}, K. and {Tamassia}, F. and {Tashkun}, S.~A. and {Vander Auwera}, J. and {Vasilenko}, I.~A. and {Vigasin}, A.~A. and {Villanueva}, G.~L. and {Vispoel}, B. and {Wagner}, G. and {Yachmenev}, A. and {Yurchenko}, S.~N.},
        title = "{The HITRAN2020 molecular spectroscopic database}",
      journal = {\jqsrt},
         year = 2022,
        month = jan,
       volume = {277},
          eid = {107949},
        pages = {107949},
          doi = {10.1016/j.jqsrt.2021.107949},
       adsurl = {https://ui.adsabs.harvard.edu/abs/2022JQSRT.27707949G}
}

@ARTICLE{Plotnykov2024,
       author = {{Plotnykov}, Mykhaylo and {Valencia}, Diana},
        title = "{Observation uncertainty effects on the precision of interior planetary parameters}",
      journal = {\mnras},
         year = 2024,
        month = may,
       volume = {530},
       number = {3},
        pages = {3488-3499},
          doi = {10.1093/mnras/stae993},
archivePrefix = {arXiv},
       eprint = {2405.03860},
 primaryClass = {astro-ph.EP},
       adsurl = {https://ui.adsabs.harvard.edu/abs/2024MNRAS.530.3488P}
}
\bibliographystyle{aasjournalv7}


\end{document}